\documentclass{aastex63}

\usepackage{natbib}
\usepackage{graphicx}
\usepackage{epstopdf}
\usepackage[caption=false]{subfig}
\usepackage{url}

\newcommand{\epeak}{$E_{\rm peak}$}

\shorttitle{GBM Spectroscopy Catalog}
\shortauthors{Poolakkil, Preece, Goldstein et al.}
\begin{document}

\title{The Fermi GBM Gamma-Ray Burst Spectral Catalog: 10 Years of Data}

%
%
\author[0000-0002-6269-0452]{S.~Poolakkil}
\affiliation{Department of Space Science, University of Alabama in Huntsville, Huntsville, AL 35899, USA}
\affiliation{Center for Space Plasma and Aeronomic Research, University of Alabama in Huntsville, Huntsville, AL 35899, USA}
\author[0000-0003-1626-7335]{R.~Preece}
\affiliation{Department of Space Science, University of Alabama in Huntsville, Huntsville, AL 35899, USA}
\author{C.~Fletcher}
\affiliation{Science and Technology Institute, Universities Space Research Association, Huntsville, AL 35805, USA}
\author[0000-0002-0587-7042]{A.~Goldstein}
\affiliation{Science and Technology Institute, Universities Space Research Association, Huntsville, AL 35805, USA}

%
%
\author[0000-0001-7916-2923]{P.N.~Bhat}
\affiliation{Center for Space Plasma and Aeronomic Research, University of Alabama in Huntsville, Huntsville, AL 35899, USA}
\affiliation{Department of Space Science, University of Alabama in Huntsville, Huntsville, AL 35899, USA}
\author[0000-0001-9935-8106]{E. Bissaldi}
\affil{Dipartimento Interateneo di Fisica dell'Università e Politecnico di Bari, Via E. Orabona 4, 70125, Bari, Italy}
\affil{Istituto Nazionale di Fisica Nucleare - Sezione di Bari, Via E. Orabona 4, 70125, Bari, Italy}
\author[0000-0003-2105-7711]{M.~S.~Briggs}
\affiliation{Department of Space Science, University of Alabama in Huntsville, Huntsville, AL 35899, USA}
\affiliation{Center for Space Plasma and Aeronomic Research, University of Alabama in Huntsville, Huntsville, AL 35899, USA}
\author[0000-0002-2942-3379]{E.~Burns}
\affiliation{Department of Physics and Astronomy, Louisiana State University, Baton Rouge, LA 70803 USA}
\author{W.~H.~Cleveland}
\affiliation{Science and Technology Institute, Universities Space Research Association, Huntsville, AL 35805, USA}
\author{M.~M.~Giles}
\affiliation{Jacobs Space Exploration Group, Huntsville, AL 35806, USA}
\author[0000-0002-0468-6025]{C.~M.~Hui}
\affiliation{NASA Marshall Space Flight Center, Huntsville, AL 35812, USA}
\author{D.~Kocevski}
\affiliation{NASA Marshall Space Flight Center, Huntsville, AL 35812, USA}
\author[0000-0001-8058-9684]{S.~Lesage}
\affiliation{Department of Space Science, University of Alabama in Huntsville, Huntsville, AL 35899, USA}
\affiliation{Center for Space Plasma and Aeronomic Research, University of Alabama in Huntsville, Huntsville, AL 35899, USA}
\author[0000-0002-2531-3703]{B.~Mailyan}
\affiliation{Center for Astro, Particle, and Planetary Physics, New York University Abu Dhabi}
\author[0000-0002-0380-0041]{C.~Malacaria}
\affiliation{NASA Marshall Space Flight Center, NSSTC, 320 Sparkman Drive, Huntsville, AL 35805, USA}\thanks{NASA Postdoctoral Fellow}
\affiliation{Universities Space Research Association, Science and Technology Institute, 320 Sparkman Drive, Huntsville, AL 35805, USA}
\author[0000-0002-2481-5947]{W. S.~Paciesas}
\affil{Science and Technology Institute, Universities Space Research Association, Huntsville, AL 35805, USA}
\author[0000-0002-7150-9061]{O.J.~Roberts}
\affiliation{Science and Technology Institute, Universities Space Research Association, Huntsville, AL 35805, USA}
\author[0000-0002-2149-9846]{P.~Veres}
\affiliation{Center for Space Plasma and Aeronomic Research, University of Alabama in Huntsville, Huntsville, AL 35899, USA}
\author[0000-0002-0221-5916]{A.~von Kienlin}
\affil{Max-Planck-Institut f\"{u}r extraterrestrische Physik, Giessenbachstrasse 1, D-85748 Garching, Germany}
\author[0000-0002-8585-0084]{C. A. Wilson-Hodge}
\affil{ST12 Astrophysics Branch, NASA Marshall Space Flight Center, Huntsville, AL 35812, USA}

\begin{abstract}
We present the systematic spectral analyses of gamma-ray bursts (GRBs) detected by the
\textit{Fermi} Gamma-Ray Burst Monitor (GBM) during its first ten years of operation.
This catalog contains two types of spectra; time-integrated spectral fits and spectral 
fits at the brightest time bin, from 2297 GRBs, resulting in a compendium of over 18000 
spectra.
The four different spectral models used for fitting the spectra were selected based on 
their empirical importance to the shape of many GRBs.
We describe in detail our procedure and criteria for the analyses, and present the bulk 
results in the form of parameter distributions both in the observer frame and in the GRB 
rest frame.
941 GRBs from the first four years have been re-fitted using the same methodology as 
that of the 1356 GRBs in years five through ten.
The data files containing the complete results are available from the High-Energy 
Astrophysics 
Science Archive Research Center (HEASARC).
\end{abstract}


\section{Introduction}
\label{sec:Intro}
Gamma-ray bursts (GRBs) have been studied extensively since their discovery in the 1960s,
but many aspects of their prompt emission remain a mystery. The bimodal distribution of 
GRB durations admits a natural division between short and long classes of 
bursts, with a dividing line at $T_{90}$ = 2 s \citep{Kouveliotou_1993}, where $T_{90}$ is the time between the 5\% and 95\% values of the total fluence. The prompt gamma-ray episode is known to be followed by 
radiation at all wavelengths, in the manner of an expanding fireball explosion, fading in 
time and energy. The observed flux originates in a relativistic jet, which is 
inferred by energetic and compactness constraints \citep{cavallo78}, as well as observations of achromatic 
jet breaks in the temporal power-law decay observed in some afterglow light curves. However, the composition of the 
jet is yet unknown and could be 
baryon or magnetic field dominated \citep{Veres+12fit,Burgess+14corr}. The 
mechanism that accelerates the emitting particles (usually assumed to be electrons and positrons) to 
their inferred power-law energy distributions is also not well understood. Finally, the 
process that can produce the enormous observed fluxes of gamma-rays extremely efficiently has yet to be determined.

As we are now entering the multi-messenger era of astronomy, new and exciting observations 
are just beginning to yield results. The 
nearly simultaneous observation of a short GRB with the gravitational signature of a 
coalescing binary neutron star system by the Laser Interferometer Gravitational-wave 
Observatory (LIGO)/Virgo gravitational-wave detectors \citep{Abbott+17BNS}, the \textit{Fermi} Gamma-Ray 
Burst Monitor (GBM) \citep{GBMLVC, Goldstein, INTEGRAL_170817A} and \textit{INTEGRAL} 
\citep{INTEGRAL_2017} on August 17, 2017, has confirmed that at least some short bursts originate in such 
mergers of compact objects. This observation kicked off one of the most extensive follow-up campaigns in 
astronomy, covering nearly all wavelengths. GRB 170817A was unusual in many respects, not the 
least of which that it was extraordinarily under-luminous. Comparison with the spectral properties of the ensemble of short bursts,
though, shows the otherwise ordinary nature of this extraordinary burst \citep{Goldstein}. 
Other than being extremely under-luminous, the spectral properties of 
GRB 170817A fall nearly in the median of the observed flux, fluence, and duration distributions. The 
distributions used in this comparison were drawn from the data set that comprises this 
Catalog, showing just one use of these data. The numerous discoveries still to come 
can only be judged relative to the expected properties of their cohorts. It is for this 
purpose that we have assembled these data; to serve as a benchmark for future discoveries in the studies of GRBs.

In 10 years, GBM has triggered on 2356 GRBs \citep{vonKienlin+20GBM10yrcat}, of which, 
2297 are useful for spectroscopy and are included in this Catalog. All of 
the data from these bursts are available online at the HEASARC
\footnote{\url{https://heasarc.gsfc.nasa.gov/W3Browse/fermi/fermigbrst.html}}
website. As with the 2-year \citep{Goldstein_2012} and 4-year \citep{Gruber_2014} Catalogs, the 
analyses presented herein are comprised of two spectra for each burst: a `fluence' 
spectrum that represents the entire duration of emission and a `peak flux' spectrum 
that depicts the brightest portion of each burst, on a fixed timescale of 1.024 s for 
long GRBs and 64 ms for short GRBs. The selection of fluence time bins for each of these 
two classes is made by including every (energy-integrated) time bin that has flux that is at 
least 3.5 sigma in excess of the background model for that bin. 
We fit 4 spectral functions to each spectrum: power law (PLAW), exponentially cut-off 
power law (COMP), the Band GRB function \citep{Band_1993} and smoothly-broken power law 
(SBPL), as described in the previous 
Catalogs. These form a set of empirical spectral functions that have 2, 3, 4, and 5 free 
parameters, respectively. The best fit of these should not by any means be considered the true spectral form 
of the incident photons, because they are not motivated by any theoretical guidance; but rather
they serve as a model-independent basis for inter-comparison between different bursts, even those 
observed by different instruments. For each spectral fit, we assign a rating (GOOD), based upon the uncertainties 
of the fitted functional parameters. In this Catalog, we introduce two-sided uncertainties 
for each fitted parameter, where these could be determined. Thus, assignment of a spectral 
fit to the GOOD category requires not-to-be-exceeded limits on both tails of the error 
distribution (Sec.~\ref{sec:Good_Best}). Finally, based upon goodness of fit criteria, 
we determine which function provided the BEST fit to the spectral data.

\section{Analysis Method}

\subsection{Instrument and Data}

\textit{Fermi} GBM consists of 14 detector modules: 12 Sodium Iodide (NaI) detectors, covering the energies 8 - 1000 keV,
and two Bismuth Germanate (BGO) detectors, covering 200 keV to 40 MeV \citep{Meegan_2009}. The NaI detectors are distributed in 
four clusters of three detectors on each corner of the spacecraft and oriented in such a way that 
enables the prime GRB scientific objectives of all-sky coverage and burst localization. The 
two BGO detectors are positioned on opposite sides of the spacecraft, also for full-sky coverage. The 
spectroscopy data products each have 128 channels of energy resolution, with CSPEC data being accumulated 
at fixed time intervals, and time-tagged event (TTE) data recording the time and energy of each count. 
The TTE data type is the most flexible, since it can be binned arbitrarily in time, and comprises the majority of 
data used for the analyses in this work. Since November 27, 2012, GBM has been operating in a mode in which TTE data 
are collected all the time. Previously, TTE data collection was initiated by a trigger, which occurs when the GBM flight software 
detects a significant rise above a preset threshold in the counting rates in two or more detectors in one of several energy and time 
ranges. When operating in this mode, TTE pre-burst data, which is being constantly accumulated in a ring buffer, 
is frozen and scheduled for transmission to accompany the triggered TTE data. Before the transition to continuous 
TTE data collection, if the burst was so bright as to fill up the finite-sized TTE ring buffer or the
pre-burst TTE were somehow corrupted, it would be difficult to reconstruct the 
entire time history and in such cases, CSPEC or CTIME data could be used.

\subsection{Data Selection}

Data selection is identical to that as described in \citet{Gruber_2014}. In brief, up to three NaI detectors 
with observing angles to the source less than $60^{\circ}$ are selected, along with the BGO detector that has 
the smallest observing angle of the burst. For each of these, standard energy ranges that avoid 
unmodeled effects, such as an electronic roll-off at 
low energies and high-energy overflow bins are selected. Each data set is binned according to whether the burst is long 
(1.024 s binning) or short (0.064 s binning), where the boundary between the long or short classes is defined by 
$T_{90}= 2.0$ s. Next, a background model (polynomial in time) is chosen to fit regions of the light curve that bracket
the emission interval. 
Although the individual energy channels are fitted separately, the background model shares the polynomial order 
(up to 4) chosen by the analyst to best represent the general trend of the non-burst portions of the data. 
The resulting model is then interpolated over the entire light curve, including the region(s) where the burst is active. 
The background uncertainty in each energy channel of each time bin is typically dominated by the uncertainties of 
the fitted temporal model parameters, except at the highest energies, where Poisson errors dominate.

In order to autonomously determine the time bins that comprise the burst source selections, we combine the 
individual NaI detector rate histories by summing over the selected detectors. This produces a single rate history 
(count s$^{-1}$) for each burst, where the source rates are added coherently and the background incoherently,  
thus improving the source statistics. We do the same with the interpolated 
background rate histories. We convert each of these into integrated light curves by multiplying each energy 
channel by the energy bin width and summing over the energy bins. The count rates are converted into counts 
by multiplying each time bin by the bin width. The signal-to-noise ratio (S/N) for each time bin is calculated 
by subtracting the background 
counts from the total counts, and dividing by the square root of the background counts. The source region is determined by 
those time bins that have a S/N in excess of 3.5 $\sigma$, relative to the background model. 
The sum of the rate histories over these (possibly discrete) time bins defines the spectrum for the `fluence' (F) sample. 
The time bin with the highest S/N selects the spectrum for the `peak flux' (P) sample. The source 
selections are propagated to each detector's light curve, including that of the BGOs.

The sum of the duration of the selected time bins is defined as the `accumulation time', which serves as a proxy for 
the duration of the burst, as seen in panel (a) of Figure~\ref{fig:Accum_times}. Notably, this can be modeled as the sum 
of two Gaussians, much like the more familiar $T_{90}$ distribution \citep{Koshut_1996}.
Panel (b) shows the accumulation times for the short GRBs ($T_{90} < 2.0$ s, \citealt{Kouveliotou_1993}) 
as a separate histogram. Interestingly, there is very little overlap of the short GRB distribution (in grey) into the accumulation time distribution of long GRBs. Finally,
panel (c) shows that there are substantial differences between $T_{90}$ and the accumulation time. One difference is that 
$10\%$ of the burst fluence is omitted in the $T_{90}$, by design, resulting in a considerable number of bursts that fall 
below the line of equality (dashed). The other main difference is that the accumulation 
time omits quiescent portions of the light curve, so many bursts fall above the equality line (accumulation time shorter than $T_{90}$).

The data is then joint fit with 
RMfit\footnote{\url{https://fermi.gsfc.nasa.gov/ssc/data/analysis/rmfit/}} (currently at version
4.3.2, available at the Fermi Science Support Center),
using a set of standard model functions (section \ref{sec:Models}).
For a fit statistic, we have chosen a variant of the Cash-statistic likelihood 
\citep{Cash_1979}, called C-Stat in RMfit and pstat in Xspec \citep{Arnaud_2011}. C-Stat 
assumes the background model uncertainty to be negligible, which is a good approximation for the propagated 
uncertainties of a background model that is interpolated to a time interval much shorter 
than the intervals the model is based upon. Since the background uncertainties would ordinarily be combined 
with signal uncertainties using quadratic sum, the signal always dominates the 
total uncertainties. One can account for the Gaussian uncertainties in the background 
correctly by using the pgstat statistic in Xspec. This statistic was not
provided by RMfit at the time of the publication of the most recent GBM Spectroscopy Catalog 
\citep{Gruber_2014}.

\subsection{Detector Response Matrices}

Performing spectral analysis successfully is highly dependent upon the correct modeling of the 
detector response matrices. The response matrices in turn are dependent upon the source 
position, relative to each detector normal. Since the \textit{Fermi} spacecraft is in constant motion 
while in sky survey mode (the default mode), the detector response is a function of time. 
GBM uses OGIP\footnote{The FITS standard is maintained by the NASA HEASARC Office of 
Guest Investigators Program FITS Working Group:\\ 
\url{https://heasarc.gsfc.nasa.gov/docs/heasarc/ofwg/ofwg_intro.html}} 
standard response 
Flexible Image Transport System (FITS) files for the response matrices. This standard 
allows for multiple RESPONSE extensions in a single file, to represent a time sequence. 
For long spectral accumulations ($> 20$ s), each matrix is weighted by the fraction of 
the total counts for the corresponding period of time covered by the matrix and then 
the weighted matrices are summed together.
The GBM response generator creates a new response matrix for every 2 degrees of slew. 
These files are distinguished from single-matrix files by the `.rsp2' filename extension. 
The standard GBM burst data product, however, has only contained the single-matrix files 
(with extension `.rsp') by default since launch. Only the bursts longer than about 20 s 
require rsp2 files, so these were generated on an as-needed basis. Several types of errors
were found while this Catalog was being generated. The first and most important of these 
was that the rsp2 files were not systematically updated whenever a burst localization was 
changed. The improvement in location accuracy could have come from other spacecrafts, days
after the trigger, for example. Or else, 
the GBM team may decide to refine the location analysis after discussion of the burst trigger. 
Given the inherent latency in obtaining a `final' location, it was possible to 
create an initial set of response 
files using a localization that was superseded by a refined analysis. We have gone through 
the entire set of bursts in this Catalog to fix this and other errors that had manifested. 
Where the new matrix files had significant differences, we have redone the spectral analyses. The updated response files are available at the HEASARC website.

\subsection{Models} \label{sec:Models}

For consistency between the several previous editions of GBM (and BATSE) 
Spectroscopy Catalogs, we chose four spectral models to fit the spectra of GRBs in our sample. All models 
are formulated in units of photon flux with energy (\textit{E}) in keV and multiplied by a 
normalization constant \textit{A} (photon\ s$^{-1}$ cm$^{-2}$ keV$^{-1}$). Both COMP and BAND 
are parameterized such that the characteristic energy is given as the energy of the peak in
the differential $\nu f_{\nu}$ spectrum ($E_{\rm peak}$). The pivot energy ($E_{piv}$) 
normalizes the model to the energy range under consideration and helps reduce 
cross-correlation of other parameters. In all cases, $E_{piv}$ is held fixed at 100 keV. 

\begin{itemize}
\item \textit{PLAW:} A single power-law, with two free parameters; normalization ($A$) and power law index ($\lambda$)
\begin{equation}
  f_{ {PLAW} }(E) = A \left(\frac{ E }{ E_{{ piv}} } \right)^{ \lambda }
\end{equation}

\item \textit{COMP:} An exponentially-attenuated power law (`comptonized'), with normalization ($A$), low-energy power-law index ($\alpha$) and characteristic energy ($E_{\rm peak}$)
\begin{equation}
  f_{COMP}(E) = A \left( \frac{E}{E_{piv}} \right)^{\alpha} \exp \left[ -\frac{(\alpha + 2) E}{E_{peak}}  \right] \label{eq:comp}
\end{equation}

\item \textit{BAND:} The Band GRB function, with normalization ($A$), low-energy power-law index ($\alpha$), high-energy power-law index ($\beta$) and characteristic energy ($E_{\rm peak}$)
\begin{equation}
   f_{BAND}(E) = A \left\{
               \begin{array}{ll}
                 \left( \frac{E}{100 \ \rm keV} \right)^{\alpha} \exp \left[ -\frac{(\alpha + 2) E}{E_{peak}}  \right], & E \geq \frac{(\alpha - \beta) E_{peak}}{\alpha +2}    \\
                 \left( \frac{E}{100 \ \rm keV} \right)^{\beta} \exp\left(\beta - \alpha \right)\left[ \frac{(\alpha - \beta) E_{peak}}{100 \ \rm keV (\alpha +2)}  \right]^{\alpha - \beta}, & E < \frac{(\alpha - \beta) E_{peak}}{\alpha +2}
               \end{array}
             \right.
\end{equation}

\item \textit{SBPL:} A smoothly broken power law, with normalization ($A$), low-energy power-law index ($\lambda_1$), high-energy power-law index ($\lambda_2$), a characteristic break energy ($E_{\rm break}$) and the break scale ($\Delta$), in decades of energy. As in \citet{Gruber_2014}, we keep the value of $\Delta$ fixed at 0.3.
\begin{equation}
  f_{{\rm SBPL}}(E)=A \biggl (\frac{E}{E_{{\rm piv}}} \biggr)^b \ 10^{(a - a_{{\rm piv}})},
\end{equation}
where:
\begin{eqnarray*}
    a=m\Delta\ln\left(\frac{e^q+e^{-q}}{2}\right),\quad & \quad a_{\rm piv}=m\Delta\ln\left(\frac{e^{q_{\rm piv}}+e^{-q_{\rm piv}}}{2}\right),\\
    q=\frac{\log (E/E_b)}{\Delta},\quad & \quad q_{\rm piv}=\frac{\log (E_{\rm piv}/E_b)}{\Delta},\\
    m=\frac{\lambda_2-\lambda_1}{2},\quad & {\rm and} \quad b=\frac{\lambda_2+\lambda_1}{2}
\end{eqnarray*}
\end{itemize}

\subsection{Probability Density Histogram} \label{sec:Probability Density Histograms}

The spectral parameter distributions presented in the following section are histogrammed 
probability density plots, that were created via Monte Carlo sampling from the probability density function (PDF) of each 
quantity from each GRB \citep{Goldstein_2016}. In brief, for a quantity of interest from a 
total of \textit{N} GRBs in our sample, we first determine the edges of our bins, then we take 
a sample from each of the \textit{N} PDFs and place them in corresponding bins. This is done 
for a number of iterations (typically $>$1000), randomly sampling from the PDFs and recording
the counts in each bin for each iteration. This process creates a PDF for each bin of the 
histogram, from which we choose the median as the centroid of the bin and the error bars 
represent the 68\% credible interval centered at the median. This Monte Carlo sampling method 
allows us to represent the underlying distribution more accurately, especially at the extremes 
of the distribution where a combination of several low probability densities can produce a
finite probability density in the histogram.

\begin{figure}
  \centering
  \subfloat[]{\includegraphics[width=0.5\textwidth]{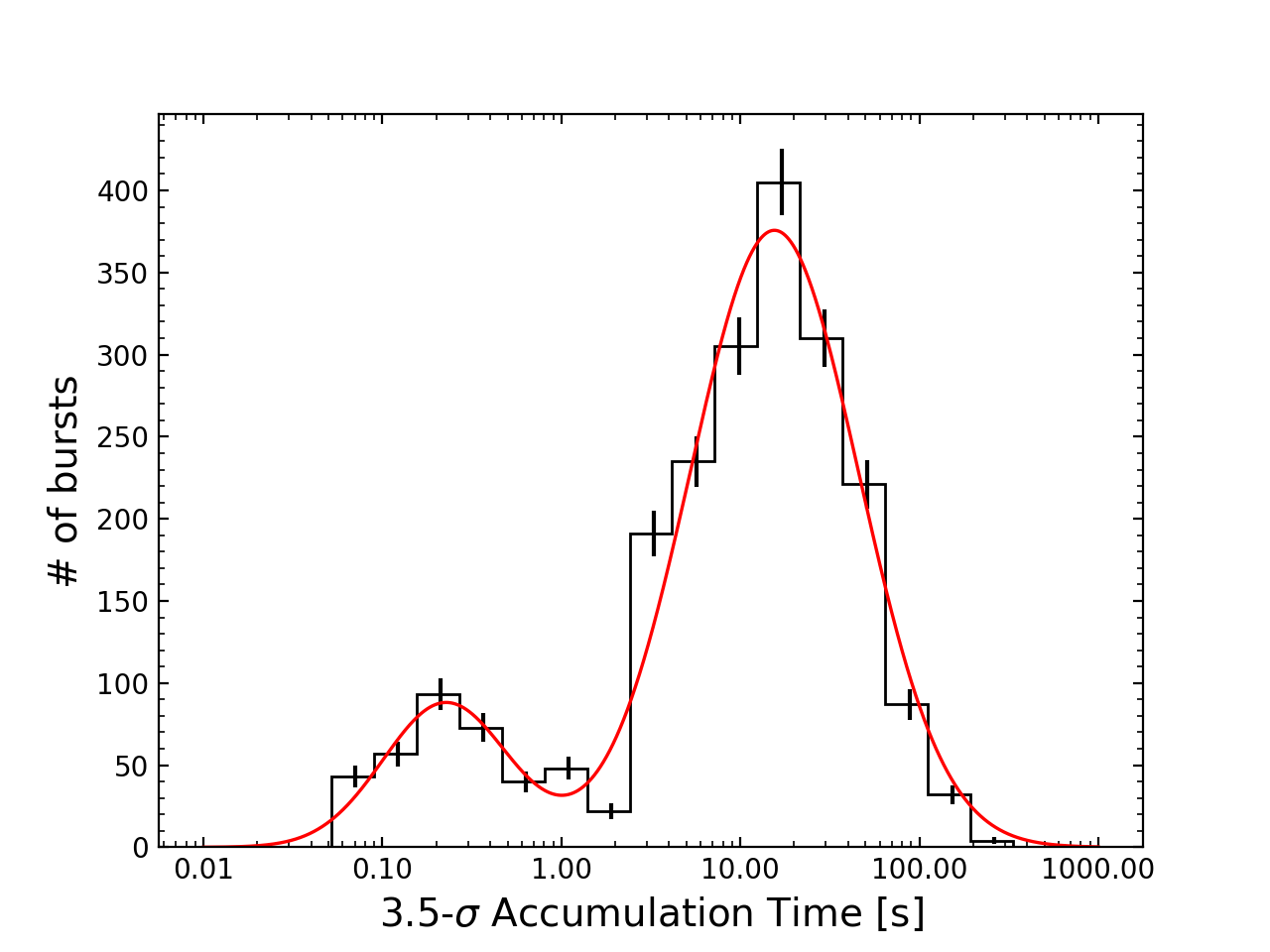}\label{fig:Accum_fit}}
  \subfloat[]{\includegraphics[width=0.48\textwidth]{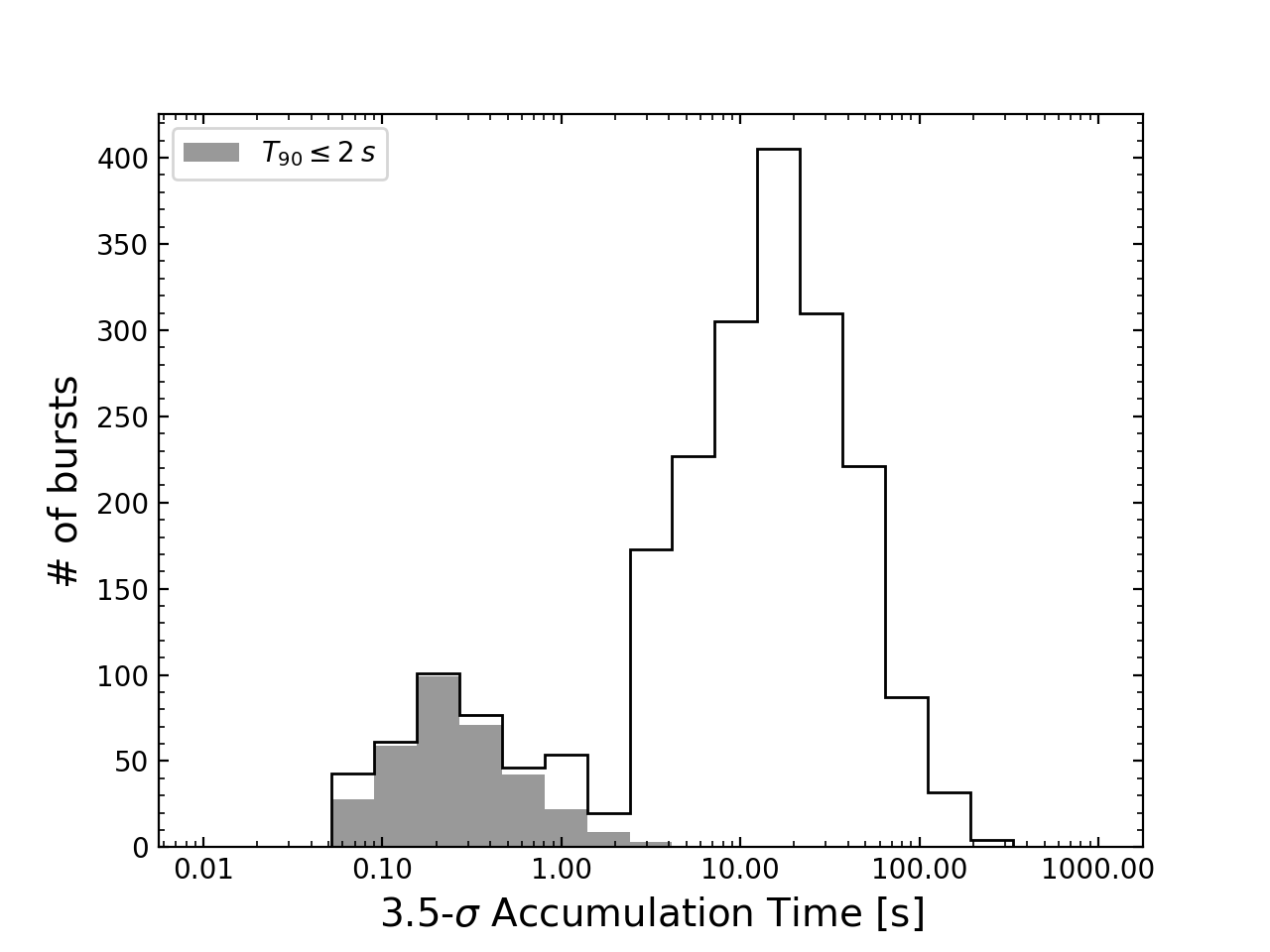}\label{fig:Accum_t90_overlay}}\hspace{1em}
  \subfloat[]{\includegraphics[width=0.48\textwidth]{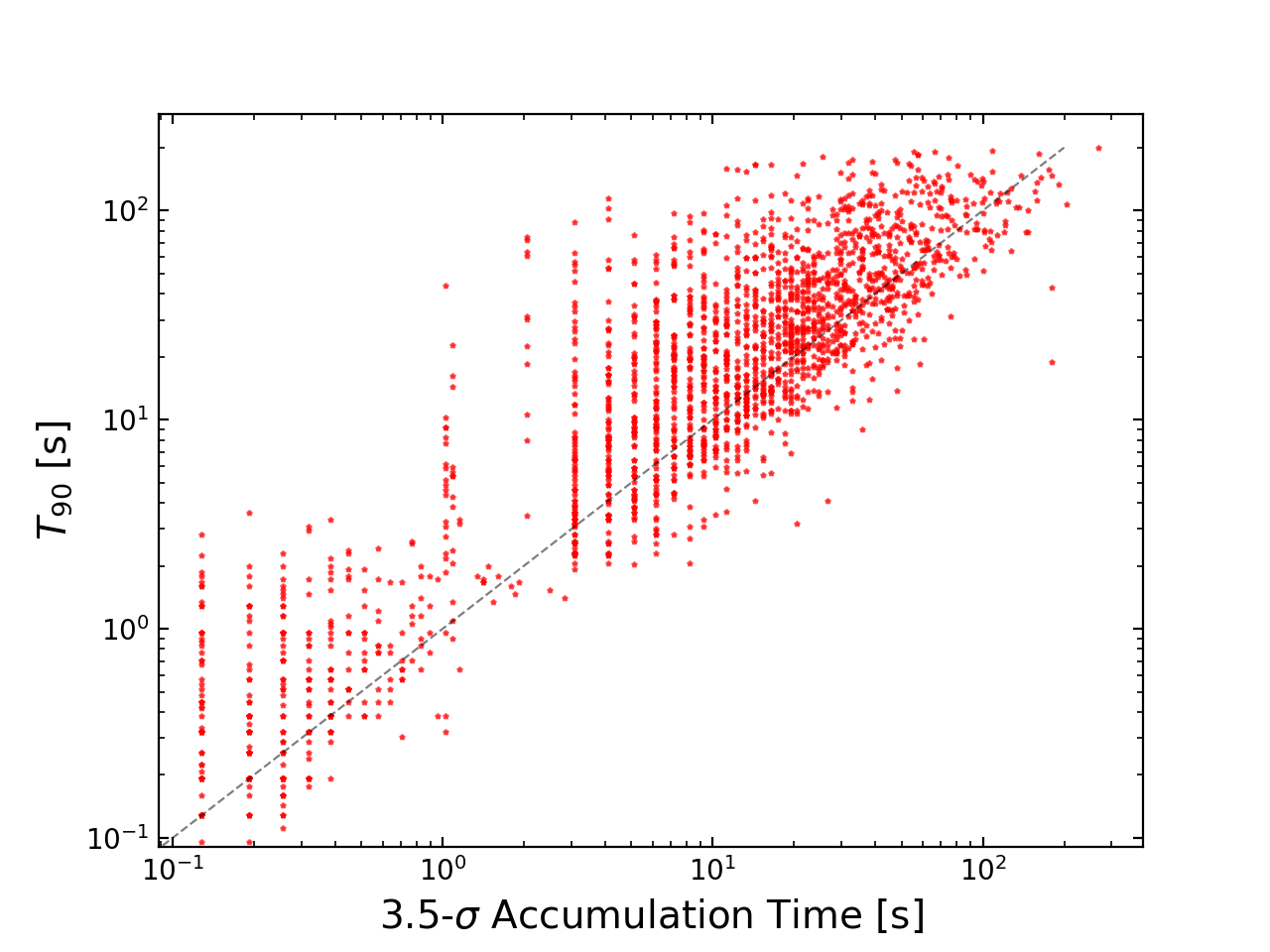}\label{fig:Accum_t90_scatterplot}}
  \caption{Panel (a) shows the distribution of the accumulation times based on the 3.5$\sigma$ S/N selections. In panel (b), the shaded region represents the accumulation times of short GRBs ($T_{90}\leq$ 2 s). Panel (c) shows the comparison between $T_{90}$ and accumulation times.}
  \label{fig:Accum_times}
\end{figure}

\section{Parameter Distributions} \label{sec:Parameter Distributions}

Distributions of the best-fit spectral parameters allow us to place each new burst in relation to the 
ensemble of all bursts. Comparisons may be made between the \textit{F} and \textit{P} spectral fits in general, 
where differences in the mean and FWHM values of the fitted parameters offer clues as to whether 
the peak in a burst is somehow special. Additionally, differences between spectral
parameter distributions of long and short GRBs may reveal something about the differences between merger versus collapsar 
jet environments. Many such analyses have been made in the previous Catalogs in this series and will not be discussed here. First, we will establish that this new iteration of the Catalog is 
not vastly different from the previous ones. Differences between the distributions of long and short bursts will be examined next. Finally, we will cover cases where there are differences that 
are derived from changes in our analysis protocol, such as the move to capture two-tailed uncertainties in the fitted parameters.

\subsection{The \textit{GOOD} and \textit{BEST} Sample} \label{sec:Good_Best}

\begin{figure}
    \centering
    \includegraphics[width=0.5\textwidth]{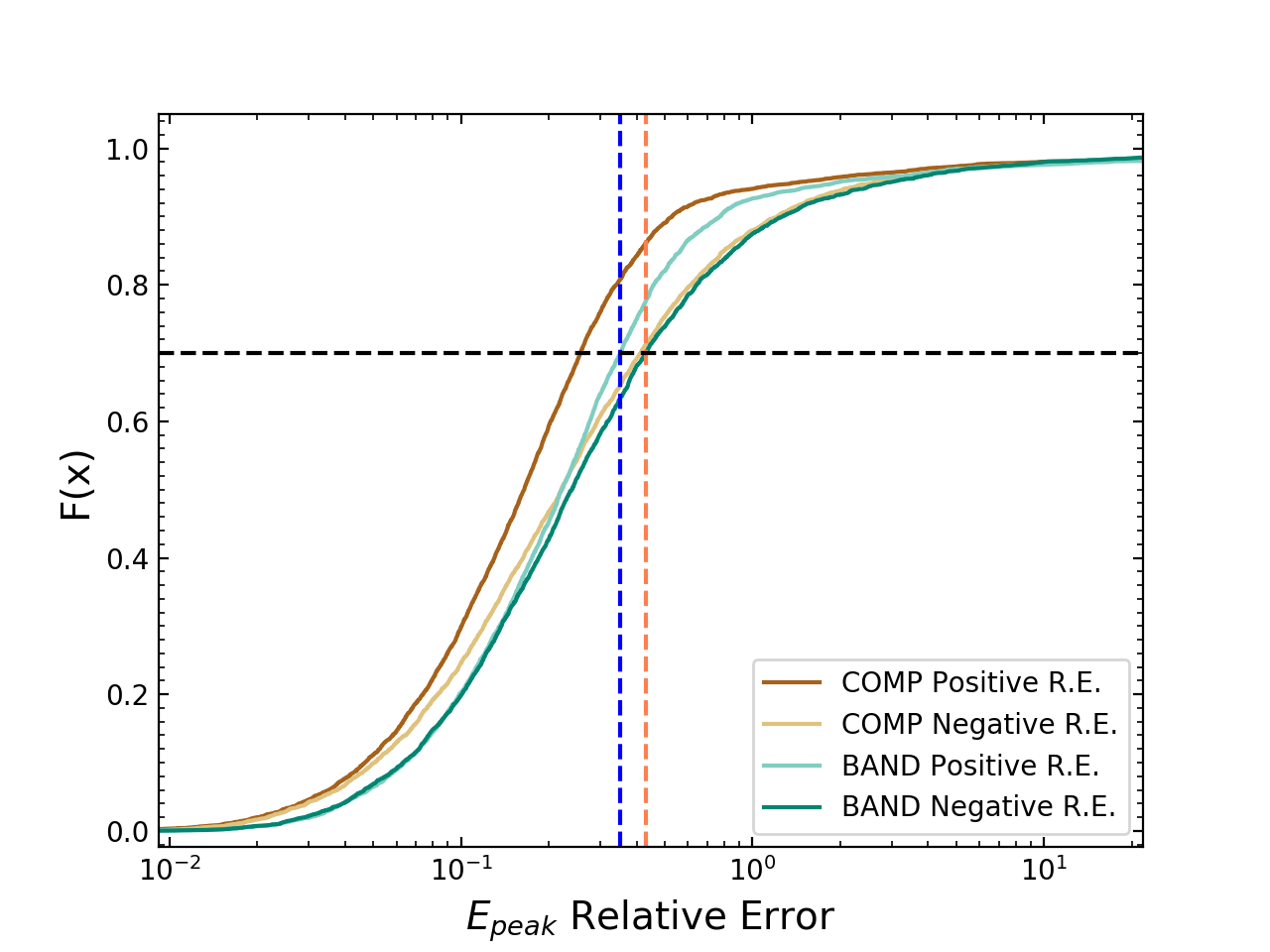}
    \caption{CDF of $E_{peak}$ relative errors obtained from GOOD \textit{F} and GOOD \textit{P} spectral fits. The blue and coral dashed lines indicate the positive and negative uncertainty cutoffs respectively for the GOOD criteria.}
    \label{fig:Epeak_Error}
\end{figure}

\begin{figure}
    \centering
    \includegraphics[width=0.5\textwidth]{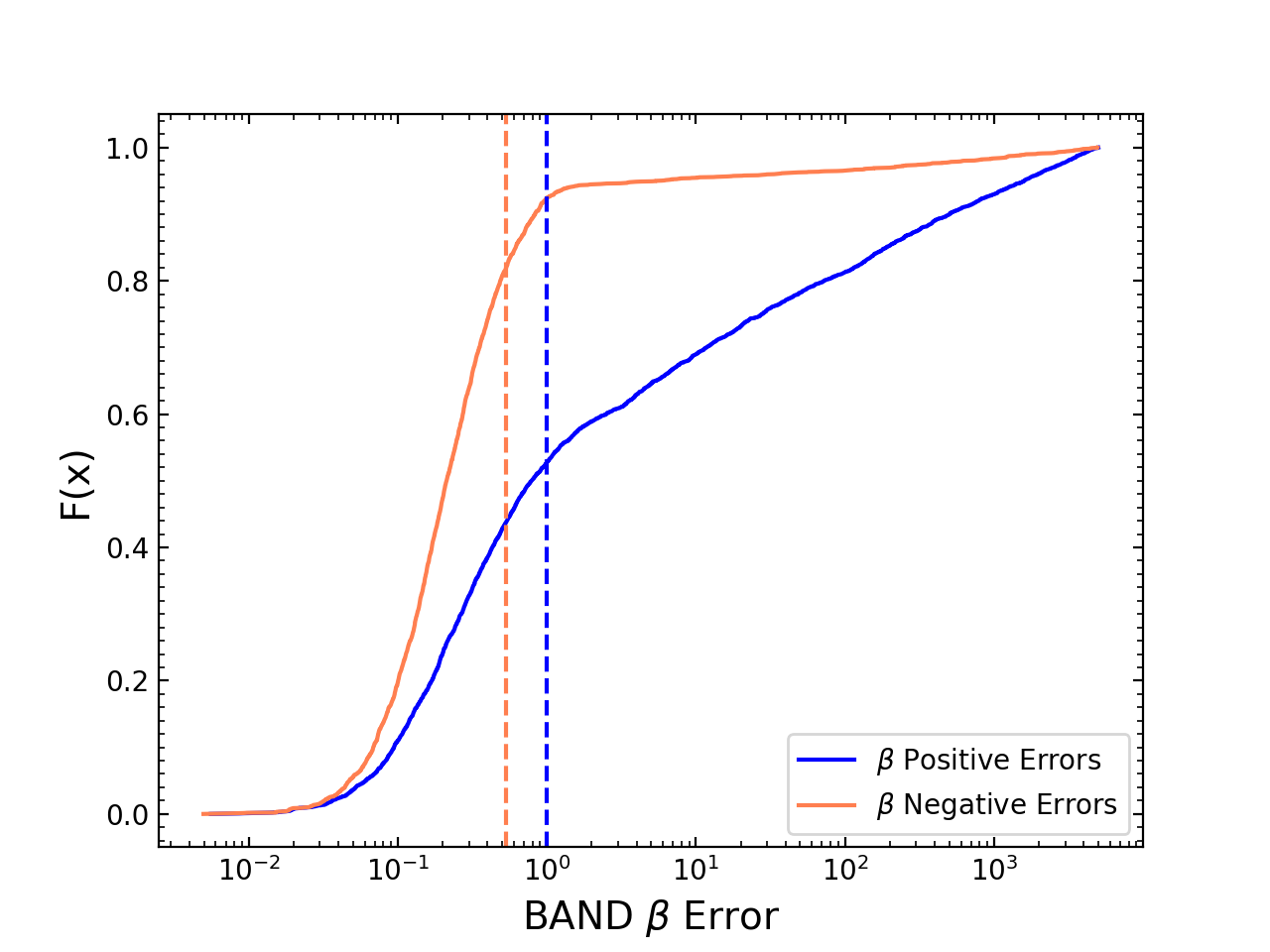}
    \caption{CDF of BAND $\beta$ errors obtained from GOOD \textit{F} and GOOD \textit{P} spectral fits.}
    \label{fig:Band_Beta_Error}
\end{figure}

\begin{figure}
    \centering
    \includegraphics[width=0.5\textwidth]{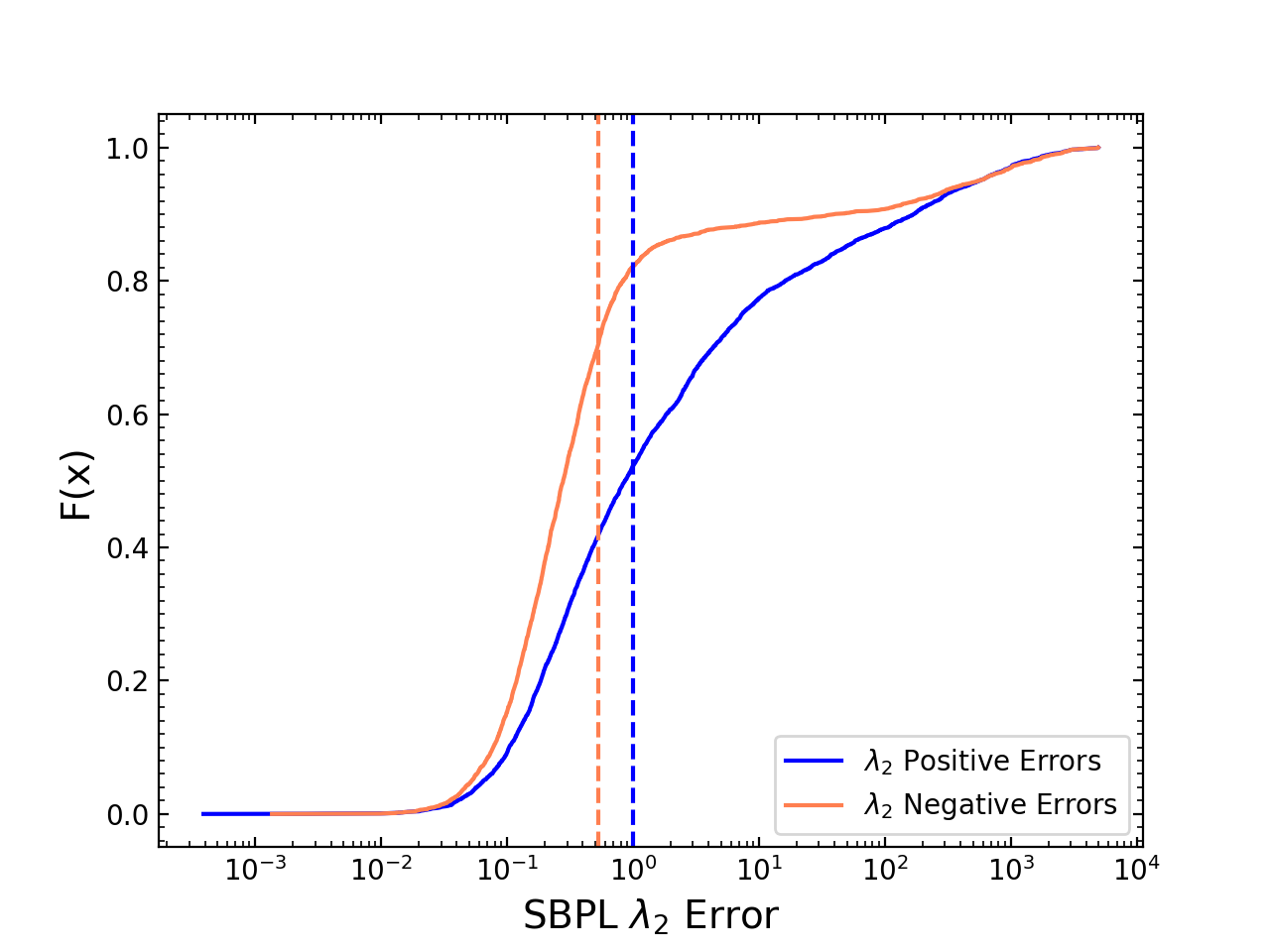}
    \caption{CDF of SBPL $\lambda_{2}$ errors obtained from GOOD \textit{F} and GOOD \textit{P} spectral fits.}
    \label{fig:SBPL_Index2_Error}
\end{figure}

We classify fitted burst models as \textit{GOOD} if the parameter error of \textit{all} model parameters are within
certain limits; we have chosen our threshold such that $\approx$70\% of the parameter uncertainties across the board
satisfy the cut-off. Figures \ref{fig:Epeak_Error}, \ref{fig:Band_Beta_Error} and \ref{fig:SBPL_Index2_Error},
depicting the cumulative distribution function (CDF) for errors of $E_{peak}$, $\beta$ and $\lambda_{2}$ respectively
were used as motivation for this approach. Note that for many GRBs there can be several models which qualify as
\textit{GOOD}.\\
We simultaneously require the following criteria to be satisfied in order for a model to be considered \textit{GOOD}:

\begin{itemize}
    \setlength\itemsep{0.005em}
	\item Amplitude: Positive relative error $<$ 0.44 \& Negative relative error $<$ 0.83
	\item Low-energy Index: Positive error $<$ 0.37 \& Negative error $<$ 0.51
	\item High-energy Index: Positive error $<$ 1.0 \& Negative error $<$ 0.53
	\item $E_{peak}$: Positive relative error $<$ 0.35 \& Negative relative error $<$ 0.43
	\item $E_{break}$: Positive relative error $<$ 1.0 \& Negative relative error $<$ 1.5
\end{itemize}

\indent We also identify a \textit{BEST} model in order to determine which of the \textit{GOOD} models is the best
representation of the burst emission. In addition to the aforesaid constraints for the \textit{GOOD}
sample, we compare the differences in \textit{C}-Stat ($\Delta\textit{C}$-Stat) per degree of freedom between the
various models. The idea of the \textit{BEST} parameter sample is to obtain the best estimate of the observed
properties of a GRB. 
Besides the necessity of having constrained parameters, already required for the \textit{GOOD} sample, we 
compare the difference in C-Stat ($\Delta$C-Stat) per degree of freedom between the various models in order to 
assess if a statistically more complex model, i.e., a model with more free fit parameters is preferred over a 
simpler model. If the $\Delta$C-Stat observed in the data exceeds a critical value ($\Delta$C-Stat$_{crit}$), 
then the statistically more complex model is preferred. Following the analysis done in \citet{Gruber_2014}, we 
use $\Delta$C-Stat$_{crit}$ = 8.58 for PLAW versus COMP and $\Delta$C-Stat$_{crit}$ = 11.83 for COMP versus BAND.
Since BAND and SBPL have the same number of degrees of freedom, the model with the lower C-Stat was preferred 
among them. Applying these criteria, the number of bursts that classify as \textit{GOOD} and \textit{BEST} for
each model can be seen in Table~\ref{tab:GOOD and BEST GRB Models.}, alongside a comparison with the previous spectral catalog \citep{Gruber_2014}.

\def\arraystretch{1.3}
\begin{table}[h]
    \movetableright=-0.75in
    \caption{GOOD and BEST GRB Models}
    \centering
    \begin{tabular}{lllll}
	\hline \hline
	
	~ & PLAW & COMP & BAND & SBPL \\ \hline
	
	~ & ~  & Fluence spectra & ~ & ~  \\ \hline
	This Catalog GOOD        & 2295 (99.9\%) & 1616 (70.3\%) & 666 (29.0\%) & 1013 (44.0\%)  \\
	\citet{Gruber_2014} GOOD & 941 (99.7\%)  & 684 (72.5\%)  & 342 (36.2\%) & 392 (41.5\%)  \\ \hline
	This Catalog BEST        & 693 (30.2\%)  & 1311 (57.0\%) & 209 (9.0\%)  & 82 (3.5\%)    \\
	\citet{Gruber_2014} BEST & 282 (29.9\%)  & 516 (54.7\%)  & 81 (8.6\%)   & 62 (6.6\%)    \\ \hline 
	
	~ & ~ & Peak flux spectra & ~ & ~  \\ \hline
	This Catalog GOOD        & 2287 (99.5\%) & 1047 (45.5\%) & 328 (14.2\%) & 522 (22.6\%)  \\ 
	\citet{Gruber_2014} GOOD & 932 (98.7\%)  & 430 (45.6\%) & 153 (16.2\%) & 196 (20.8\%)  \\ \hline
	This Catalog BEST        & 1248 (54.3\%) & 931 (40.5\%) & 79  (3.4\%)  & 29  (1.2\%)   \\
	\citet{Gruber_2014}BEST  & 514 (54.4\%)) & 375 (39.7\%) & 25 (2.6\%)   & 18 (1.9\%)    \\\hline
	
    \end{tabular}
    \label{tab:GOOD and BEST GRB Models.}
\end{table}

\begin{figure}
    \centering
    \includegraphics[width=0.5\textwidth]{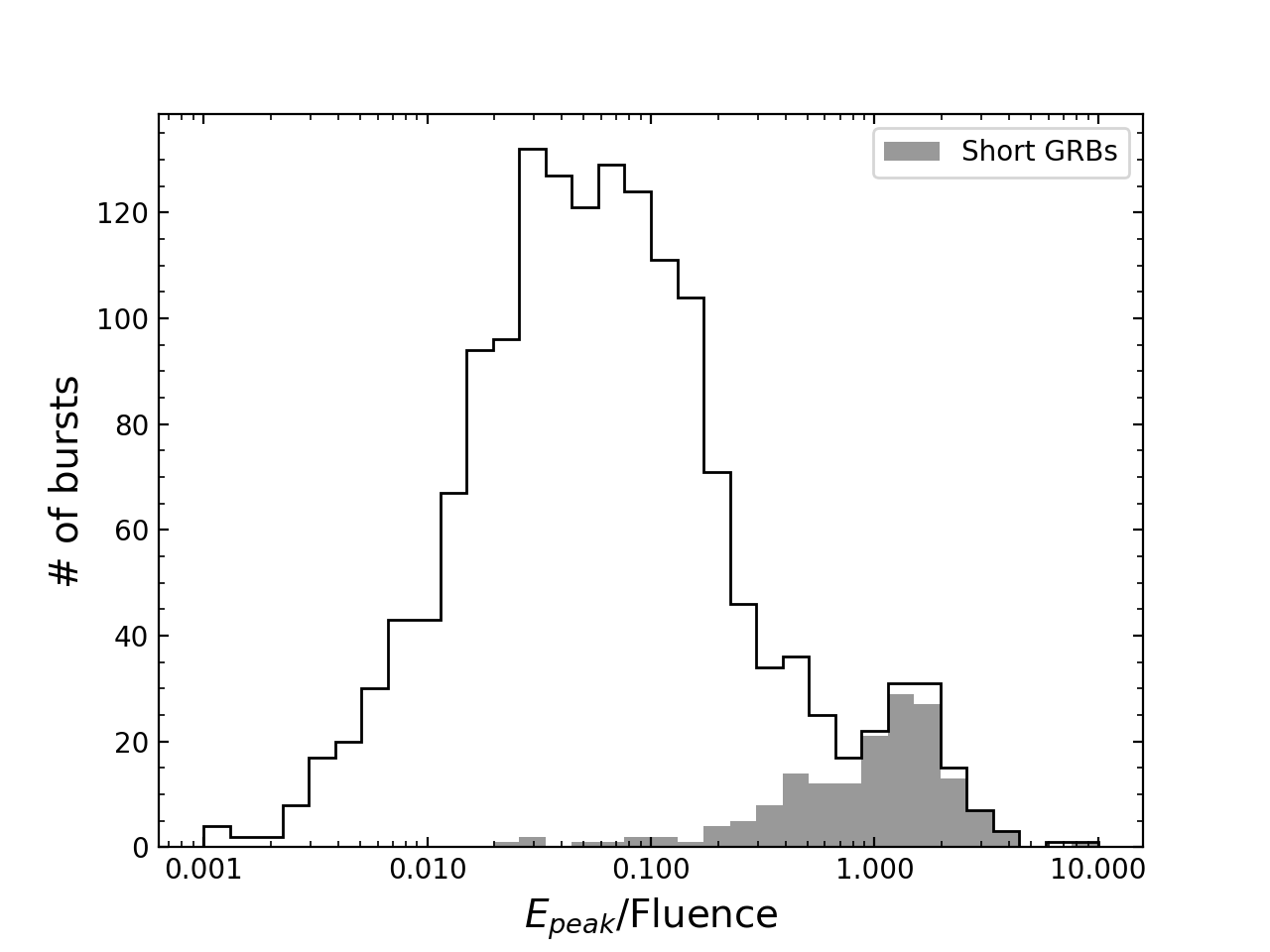}
    \caption{Distribution of $E_{peak}$/Fluence ratio for all COMP GOOD \textit{F} spectral fits. The solid 
    grey histogram contains bursts with $T_{90}\leq$ 2 s.}
    \label{fig:Epeak_Fluence_Comp}
\end{figure}

\subsection{Time-integrated (F) Spectral Fits}

\begin{figure}
    \centering
    \subfloat[]{\includegraphics[width=0.48\textwidth]{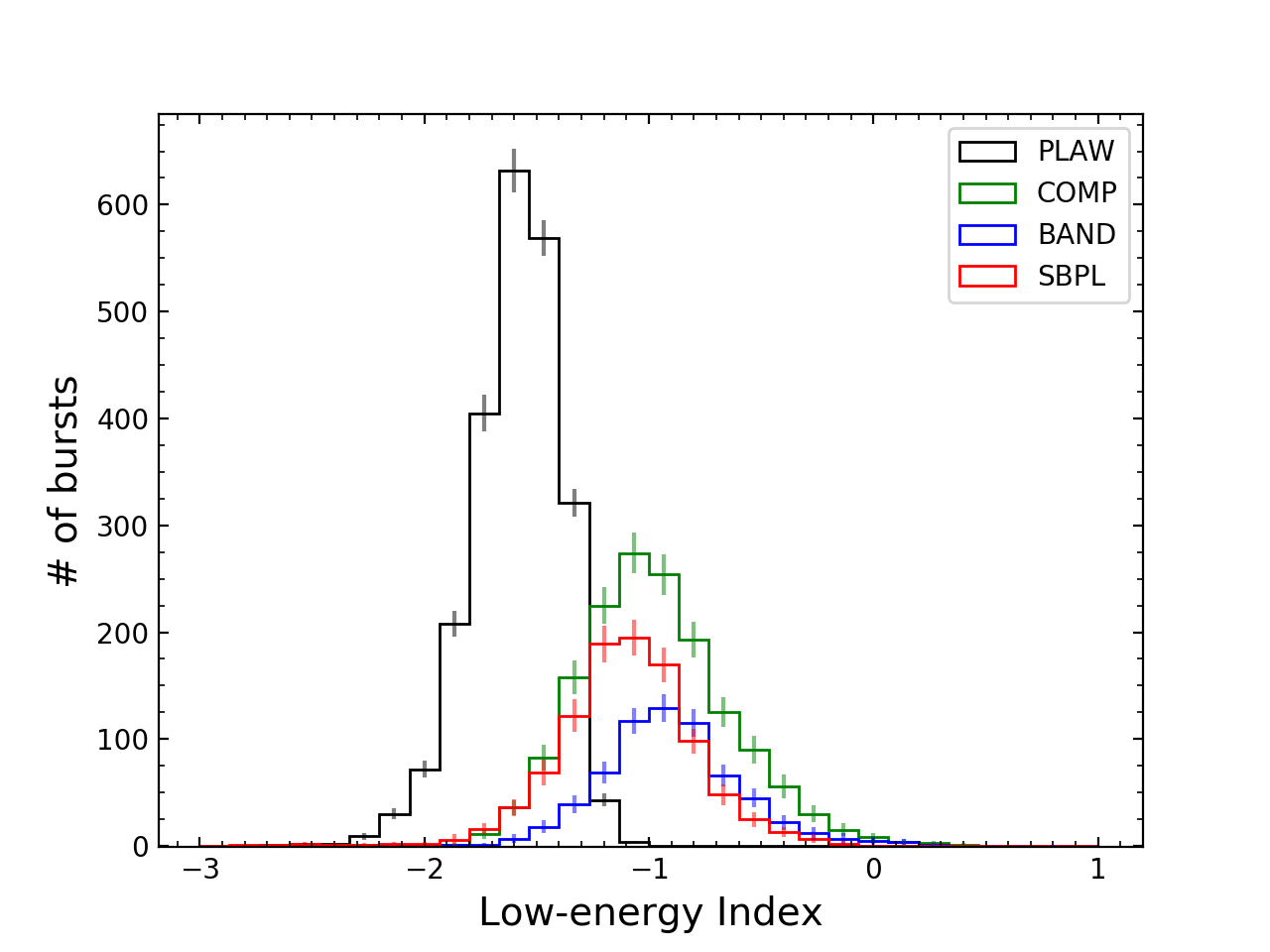}\label{fig:LE_Index_GOOD_F}}
    \subfloat[]{\includegraphics[width=0.48\textwidth]{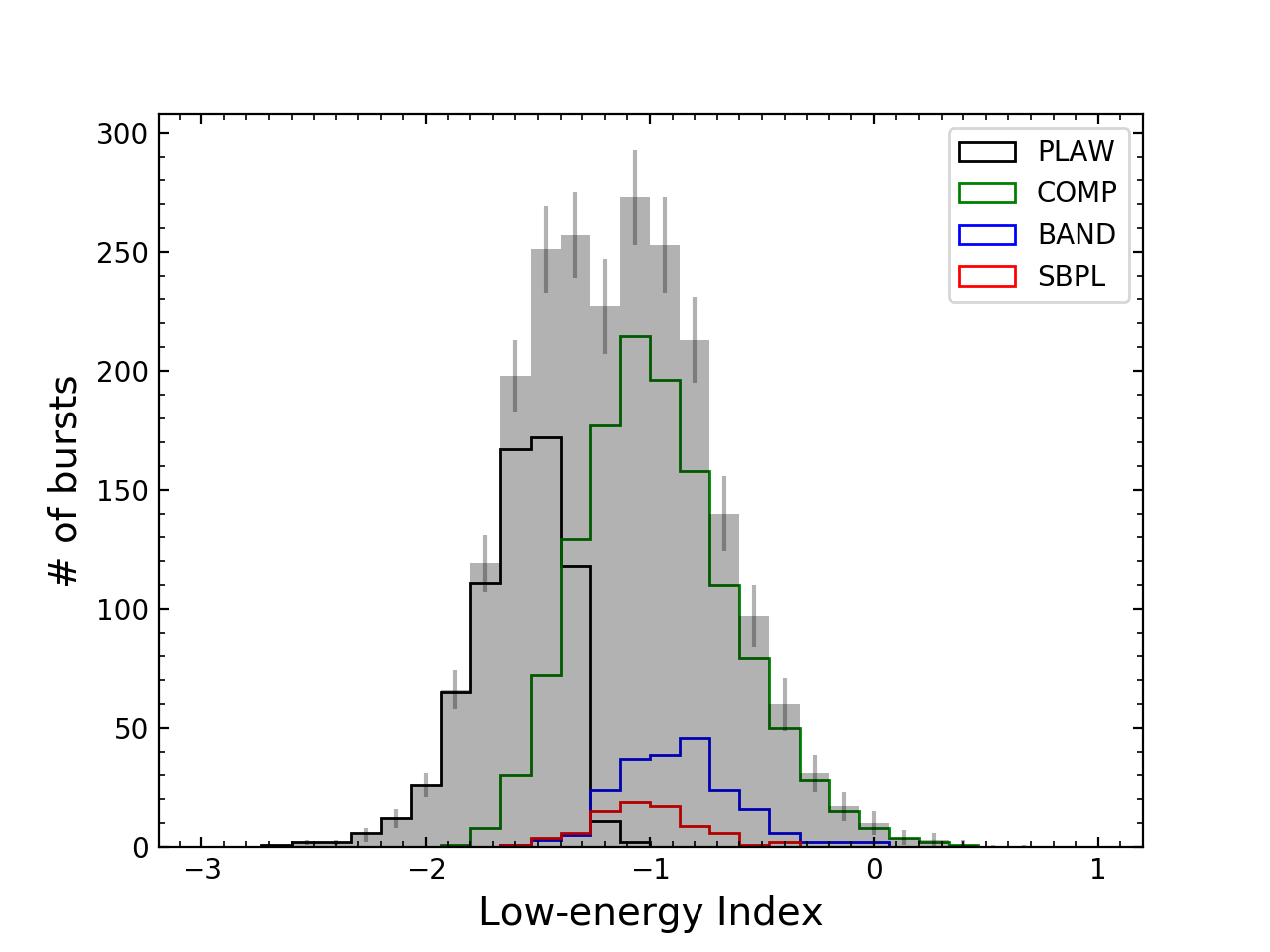}\label{fig:LE_Index_BEST_F}} \hspace{1em}
    \subfloat[]{\includegraphics[width=0.48\textwidth]{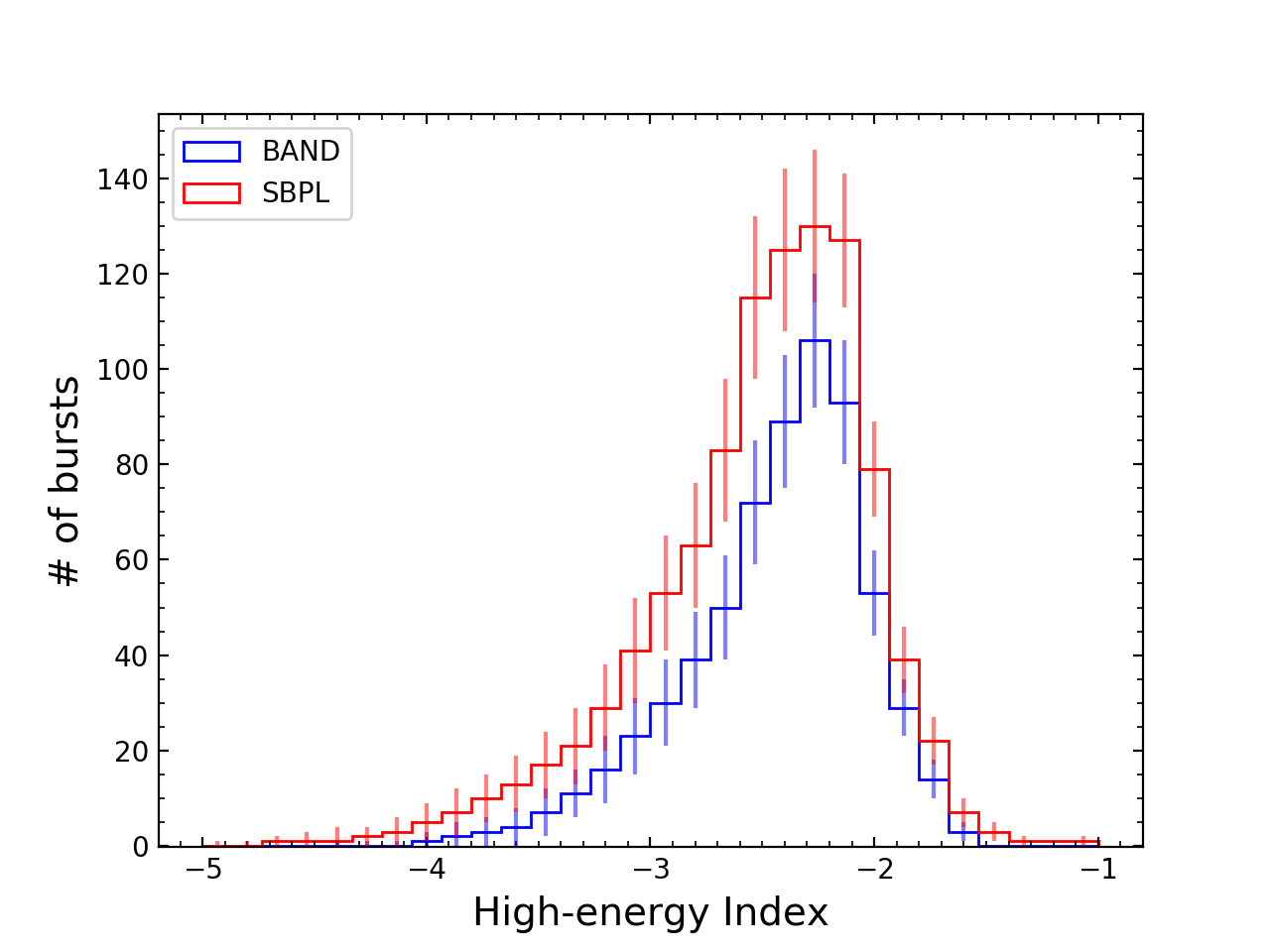}\label{fig:HE_Index_GOOD_F}}
    \subfloat[]{\includegraphics[width=0.48\textwidth]{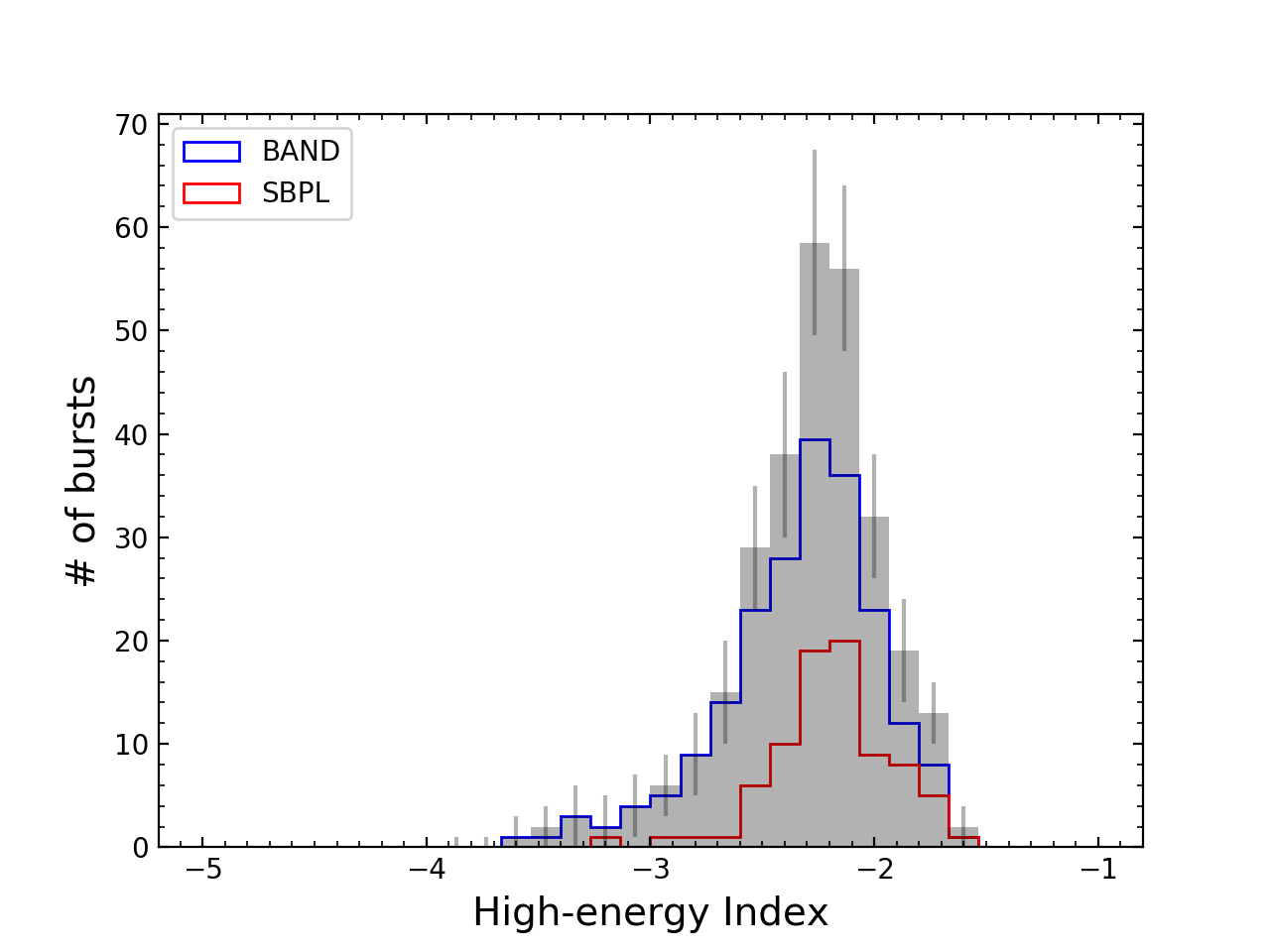}\label{fig:HE_Index_BEST_F}}\hspace{1em}
    \subfloat[]{\includegraphics[width=0.48\textwidth]{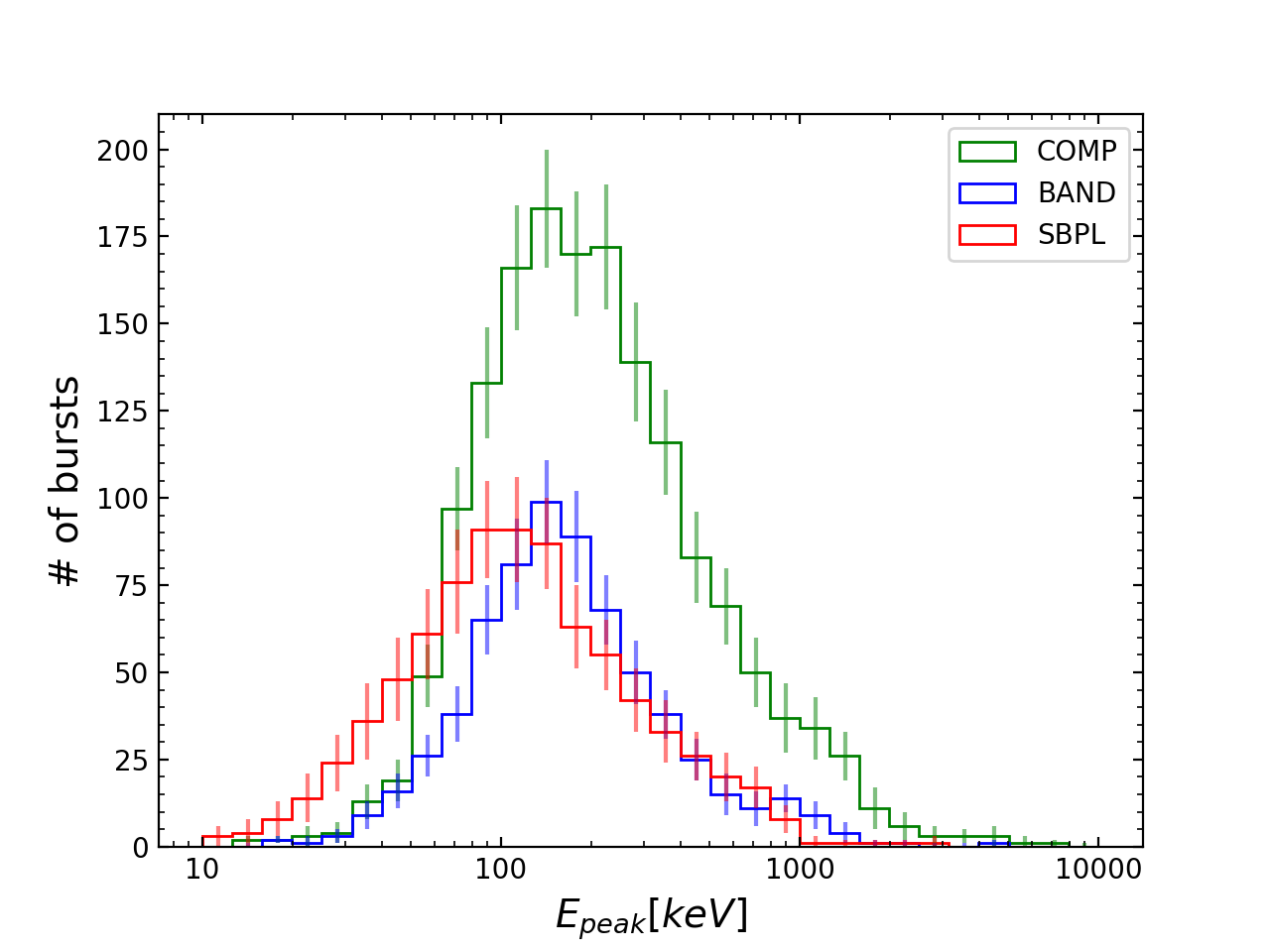}\label{fig:Epeak_GOOD_F}}
    \subfloat[]{\includegraphics[width=0.48\textwidth]{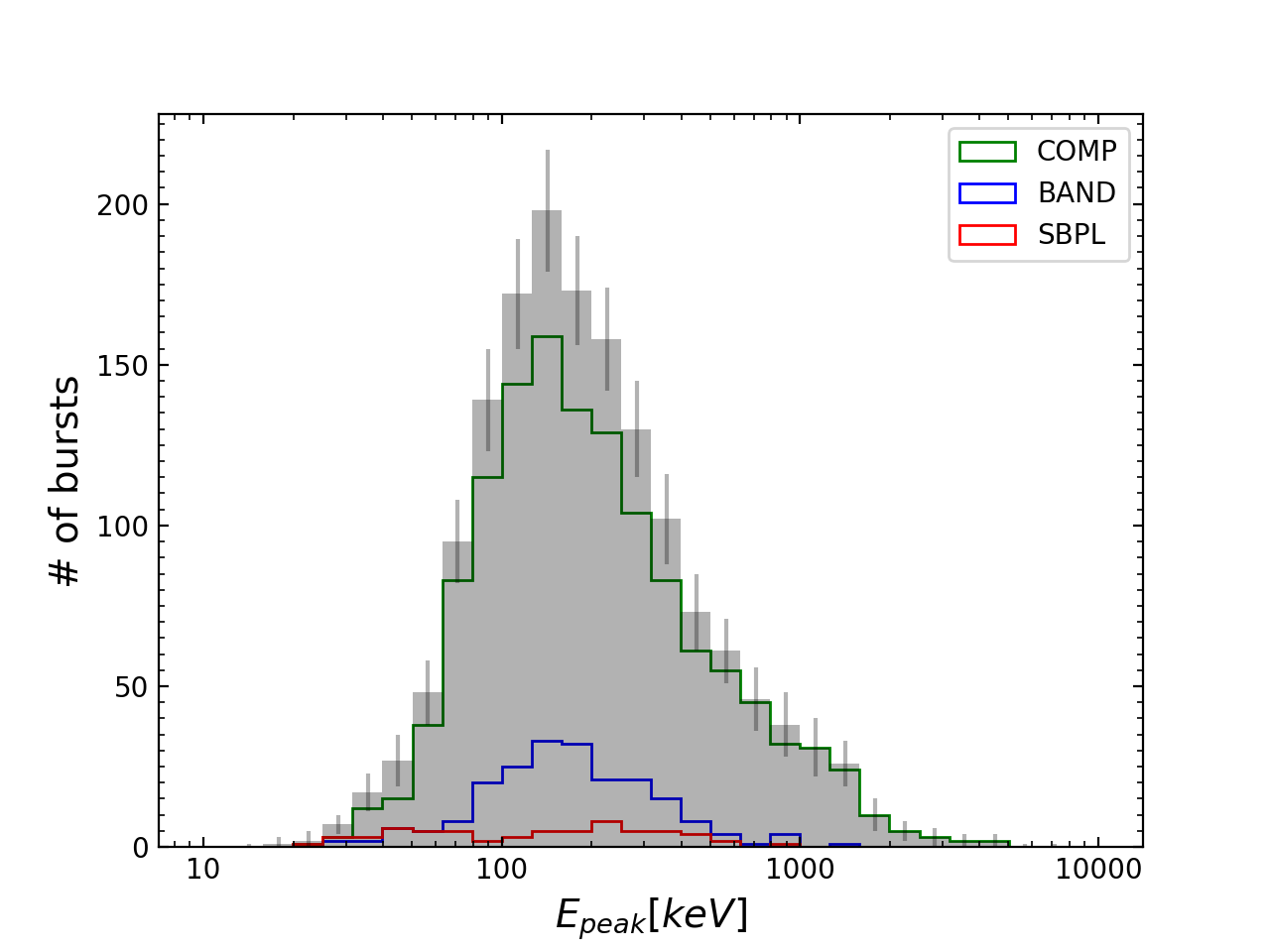}\label{fig:Epeak_BEST_F}}
    \caption{Distribution of the low-energy indices, high-energy indices and $E_{peak}$ obtained from the GOOD \textit{F} spectral fits are shown in (a), (c) and (e) respectively. The BEST parameter distribution (gray filled histogram) and its constituents are shown in (b), (d) and (f).
    \label{fig:GOOD_F}}
\end{figure}

\begin{figure}
    \centering
    \includegraphics[width=0.5\textwidth]{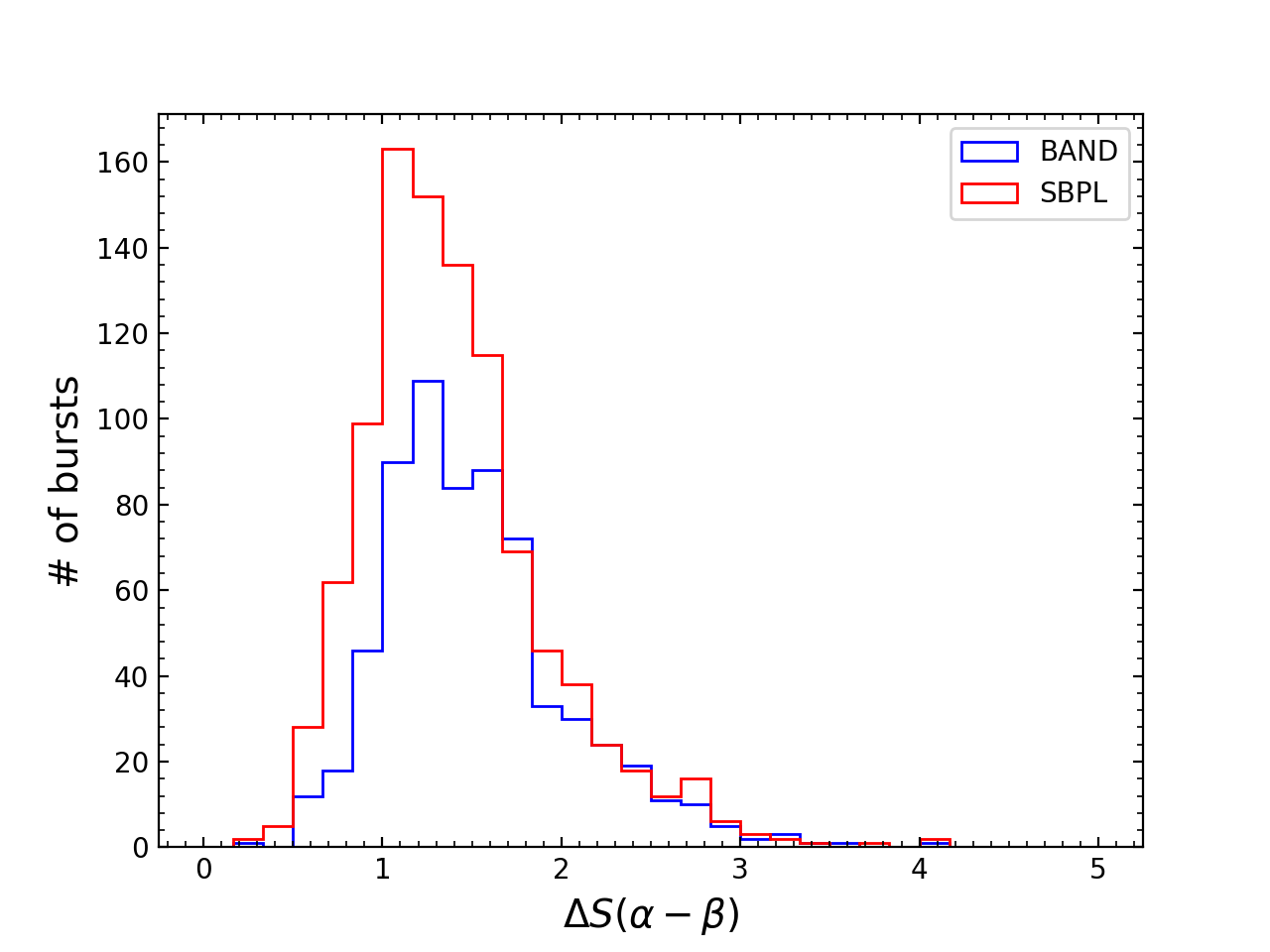}
    \caption{Distribution of $\Delta S$, the difference between low- and high-energy spectral indices $(\alpha - \beta)$ for the GOOD \textit{F} spectral fits.}
    \label{fig:DeltaS_F}
\end{figure}

\begin{figure}
    \subfloat[]{\includegraphics[width=0.33\textwidth]{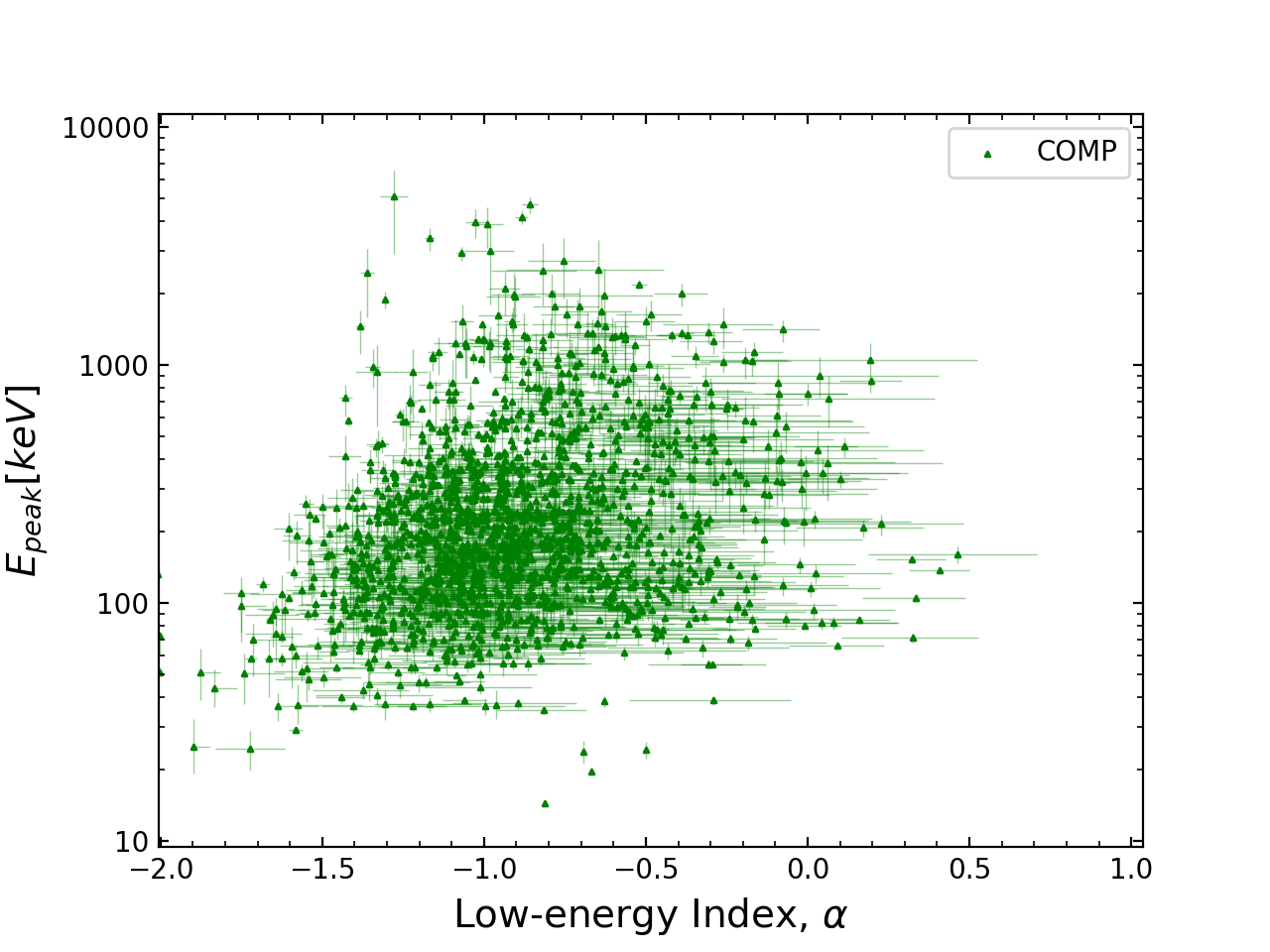}\label{fig:Epeak_Alpha_Comp}}
    \subfloat[]{\includegraphics[width=0.33\textwidth]{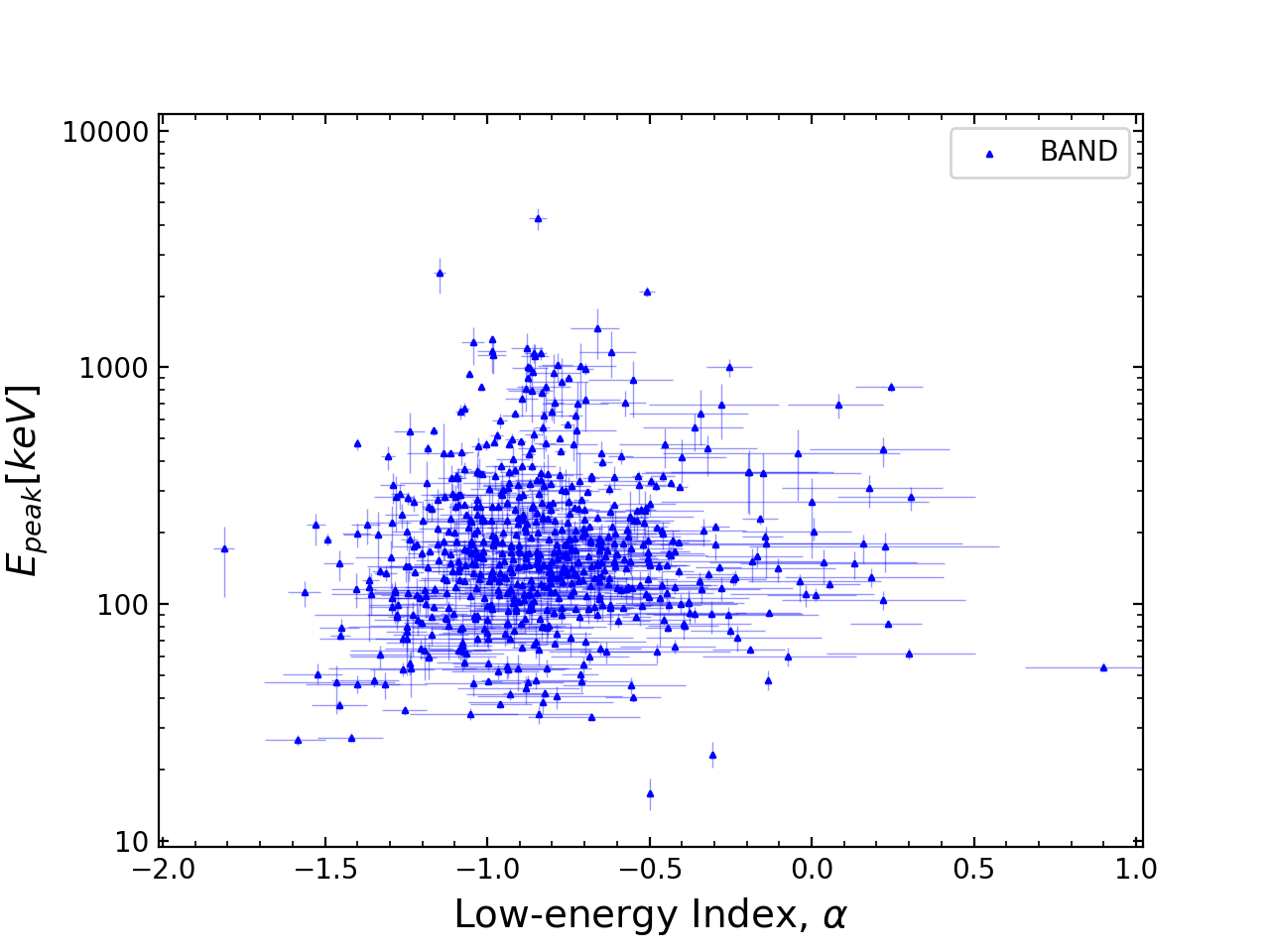}\label{fig:Epeak_Alpha_Band}}
    \subfloat[]{\includegraphics[width=0.33\textwidth]{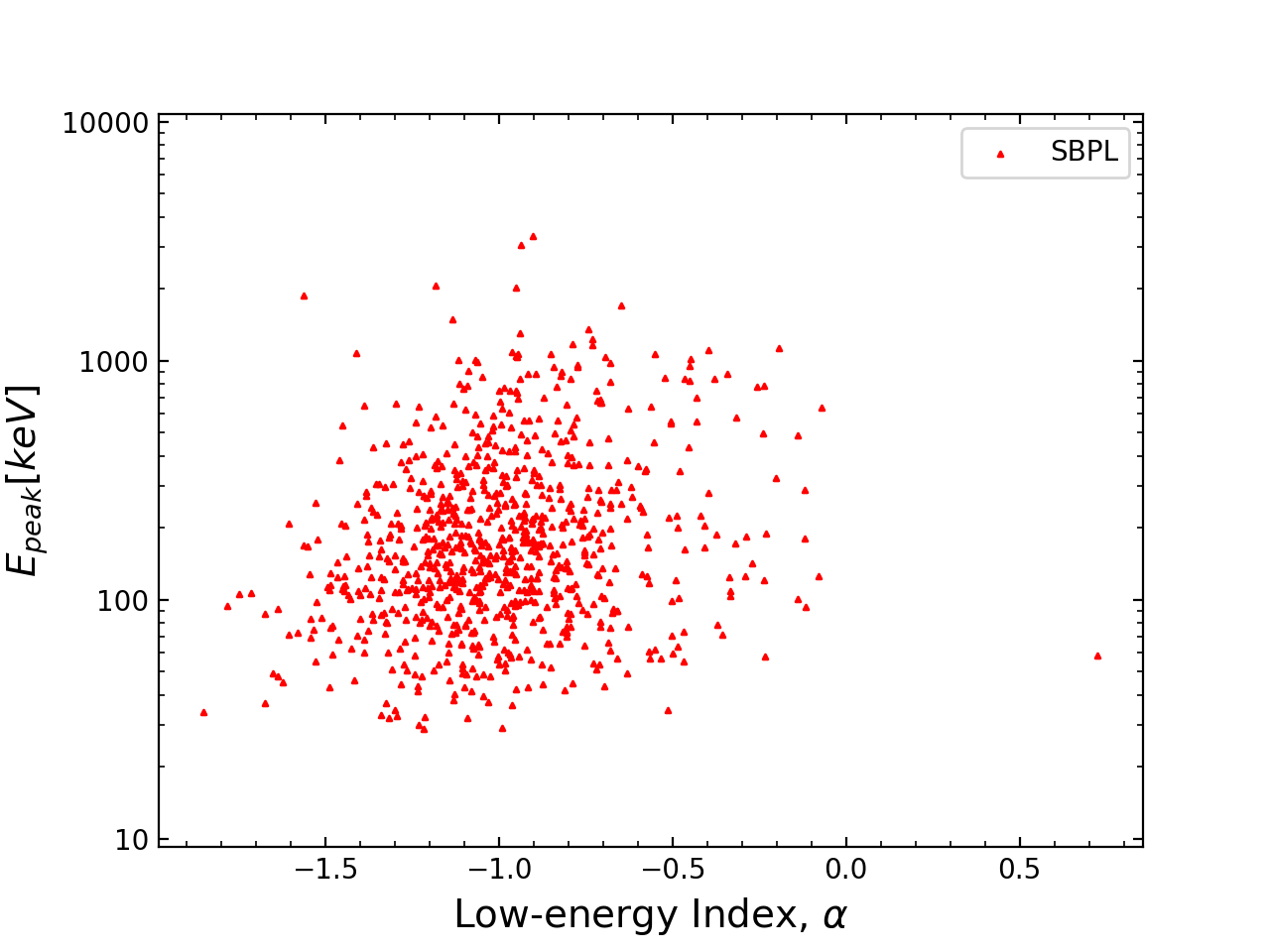}\label{fig:Epeak_Alpha_Sbpl}}
    \caption{Comparison of the low-energy index and $E_{peak}$ for three models from the GOOD \textit{F} spectral fits.
    \label{fig:Epeak_Alpha}}
\end{figure}

The time-integrated spectral distributions depict the overall emission properties and do not 
take into account any spectral evolution.
Figure~\ref{fig:Epeak_Fluence_Comp} shows the $E_{peak}$/Fluence ratio (in units of area) \citep{Goldstein_2010}
distribution for all COMP \textit{GOOD} \textit{F} spectral fits, 
with short GRBs highlighted in grey. This `energy ratio' plot further highlights the robustness 
of bimodality observed in GRB duration.
The low-energy indices $\alpha$ or $\lambda_1$ (from 
COMP, BAND or SBPL, as appropriate), as shown in Figure~\ref{fig:LE_Index_GOOD_F}, are distributed
about a -1.1 power law typical of most GRBs. About 17\% of the BEST low-energy indices 
(Figure~\ref{fig:LE_Index_BEST_F}) have
$\alpha > -2/3$, violating the synchrotron ``line-of-death'' \citep{Preece_1998a}, while 
82\% of the indices have $\alpha > -3/2$, violating the synchrotron slow-cooling 
limit \citep{Cohen_1997}. The distribution of high-energy indices $\beta$ or $\lambda_2$ (from 
BAND or SBPL) in Figure~\ref{fig:HE_Index_GOOD_F}, peak at a slope of about -2.1 and have a long 
tail toward steeper indices. The very steep high-energy indices indicate that the spectrum of 
these 
GRBs closely mimic a COMP model, which is equivalent to a BAND function with a high-energy 
index of $-\infty$. Figure~\ref{fig:DeltaS_F} shows the difference between the time-integrated 
low- and high-energy spectral indices, $\Delta S = (\alpha - \beta)$. This quantity is useful 
as the synchrotron shock model (SSM) \citep{baring06} makes predictions of this value in a 
number of cases and the power-law index, p, of the electron distribution can be inferred from 
$\Delta S$. \\
\indent The GOOD and BEST $E_{peak}$ distributions (shown in 
Figure~\ref{fig:Epeak_GOOD_F},~\ref{fig:Epeak_BEST_F}) 
generally peak around 150-200 keV and cover just over two orders of magnitude, which is 
consistent with previous findings \citep{Goldstein_2012, Gruber_2014}. As discussed 
in \citet{Kaneko_2006}, although the SBPL is parameterized with $E_{b}$, the $E_{peak}$ can be 
derived from the functional form. We have calculated the $E_{peak}$ for all bursts with a
low-energy index shallower than -2 and a high-energy index steeper than -2. The value of 
$E_{peak}$ can strongly affect the measurement of the low-energy index of the spectrum, as 
shown in Figure~\ref{fig:Epeak_Alpha}. A general trend appears to show that spectra with 
smaller $E_{peak}$ values also have smaller values of the low-energy power-law index. 
Asymmetric uncertainties for the SBPL $E_{peak}$ have not been calculated for this catalog.

\def\arraystretch{1.3}
\begin{table}[h]
    \movetableright=-0.75in
    \caption{The median parameter values and the 68\% CL of the distribution of the GOOD sample}
    \centering
    \begin{tabular}{lcccccc}
	\hline \hline
	
	Model & Low-energy & High-energy & $E_{peak}$ & $E_{break}$ & Photon Flux & Energy Flux \\
	~ & Index & Index & (keV) & (keV) & (photons s$^{-1}$cm$^{-2}$)& ($10^{-7}$erg s$^{-1}$cm$^{-2}$) \\ \hline
	
	~    & ~                       & ~                       & ~                   & Fluence spectra    & ~                      & ~                       \\ \hline
	PLAW   & $-1.55_{-0.20}^{+0.18}$ & ...                     & ...                 & ...                & $2.54_{-1.14}^{+3.98}$ & $3.42_{-1.51}^{+7.38}$  \\
	COMP & $-0.93_{-0.31}^{+0.39}$ & ...                     & $191_{-97}^{+309}$  & ...                & $2.62_{-1.19}^{+4.23}$ & $3.06_{-1.62}^{+8.61}$  \\ 
	BAND & $-0.84_{-0.26}^{+0.31}$ & $-2.29_{-0.42}^{+0.29}$ & $159_{-72}^{+178}$  & $111_{-41}^{+120}$ & $3.46_{-1.68}^{+4.98}$ & $4.76_{-2.56}^{+8.27}$  \\
	SBPL & $-1.00_{-0.26}^{+0.28}$ & $-2.32_{-0.48}^{+0.32}$ & $161_{-84}^{+237}$  & $112_{-58}^{+115}$ & $3.19_{-1.53}^{+4.41}$ & $4.15_{-2.12}^{+8.39}$  \\ 
	BEST & $-1.08_{-0.44}^{+0.45}$ & $-2.20_{-0.29}^{+0.26}$ & $180_{-88}^{+307}$  & $107_{-49}^{+88}$  & $2.37_{-1.05}^{+3.83}$ & $2.94_{-1.39}^{+7.90}$  \\\hline 
	
	~    & ~                       & ~                       & ~                    & Peak flux spectra  & ~                        & ~                          \\ \hline
	PLAW   & $-1.50_{-0.20}^{+0.17}$ & ...                     & ...                  & ...                & $4.81_{-2.73}^{+9.52}$   & $7.36_{-4.34}^{+16.29}$    \\
	COMP & $-0.69_{-0.30}^{+0.37}$ & ...                     & $242_{-127}^{+338}$  & ...                & $8.68_{-4.76}^{+13.79}$  & $13.25_{-8.17}^{+36.61}$   \\ 
	BAND & $-0.57_{-0.27}^{+0.33}$ & $-2.39_{-0.37}^{+0.29}$ & $222_{-100}^{+248}$  & $149_{-55}^{+176}$ & $16.31_{-8.43}^{+27.66}$ & $27.37_{-17.09}^{+60.25}$  \\
	SBPL & $-0.78_{-0.24}^{+0.27}$ & $-2.40_{-0.49}^{+0.31}$ & $214_{-105}^{+255}$  & $140_{-58}^{+140}$ & $13.49_{-7.69}^{+21.35}$ & $22.56_{-12.75}^{+50.34}$   \\ 
	BEST & $-1.30_{-0.33}^{+0.77}$ & $-2.34_{-0.36}^{+0.28}$ & $233_{-117}^{+316}$  & $163_{-65}^{+156}$ & $4.62_{-2.55}^{+8.90}$   & $6.46_{-3.52}^{+17.82}$    \\ \hline 
	
    \end{tabular}
\end{table}

\subsection{Peak Flux (P) Spectral Fits}

\begin{figure}
    \centering
    \subfloat[]{\includegraphics[width=0.48\textwidth]{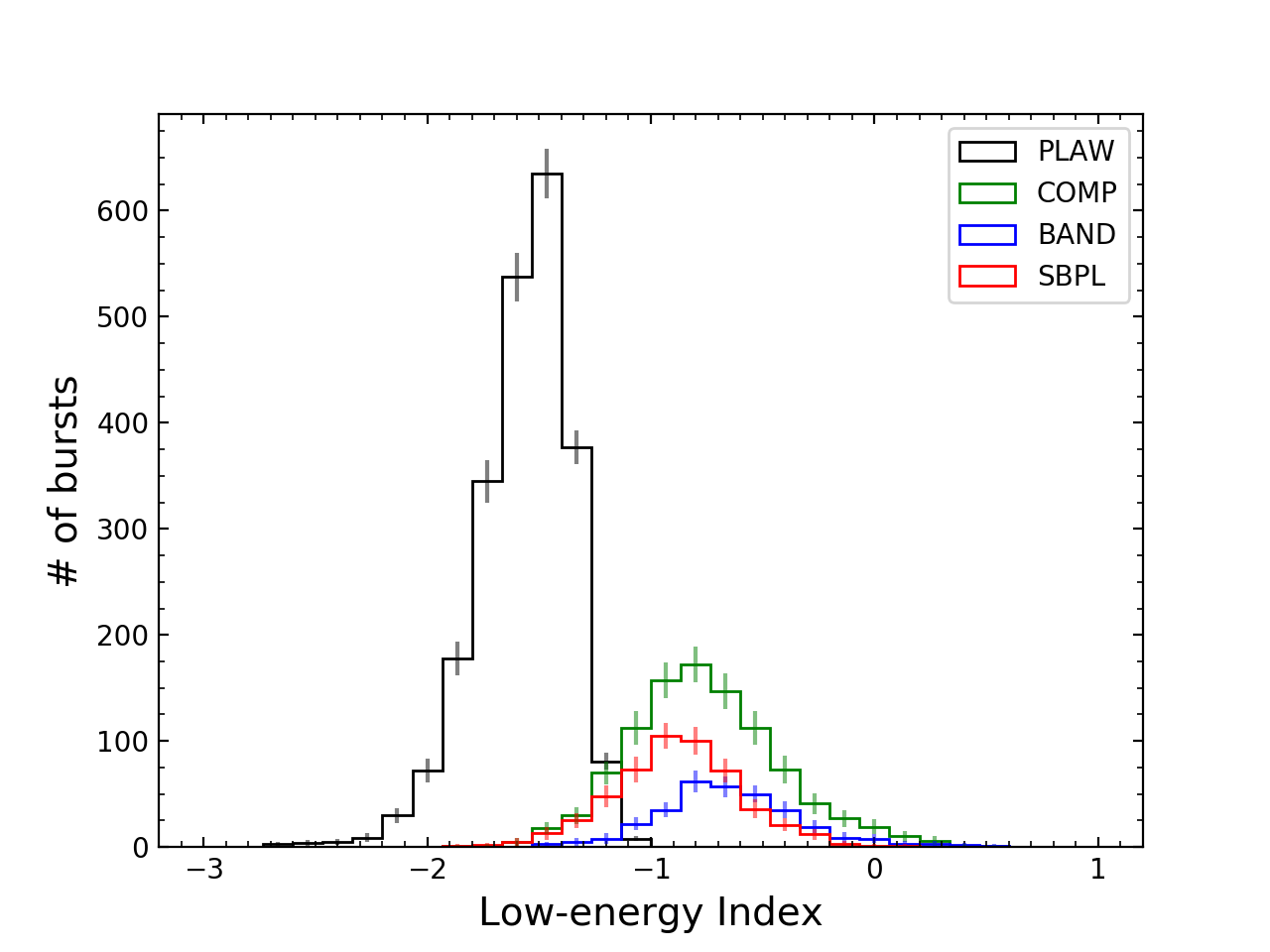}\label{fig:LE_Index_GOOD_P}}
    \subfloat[]{\includegraphics[width=0.48\textwidth]{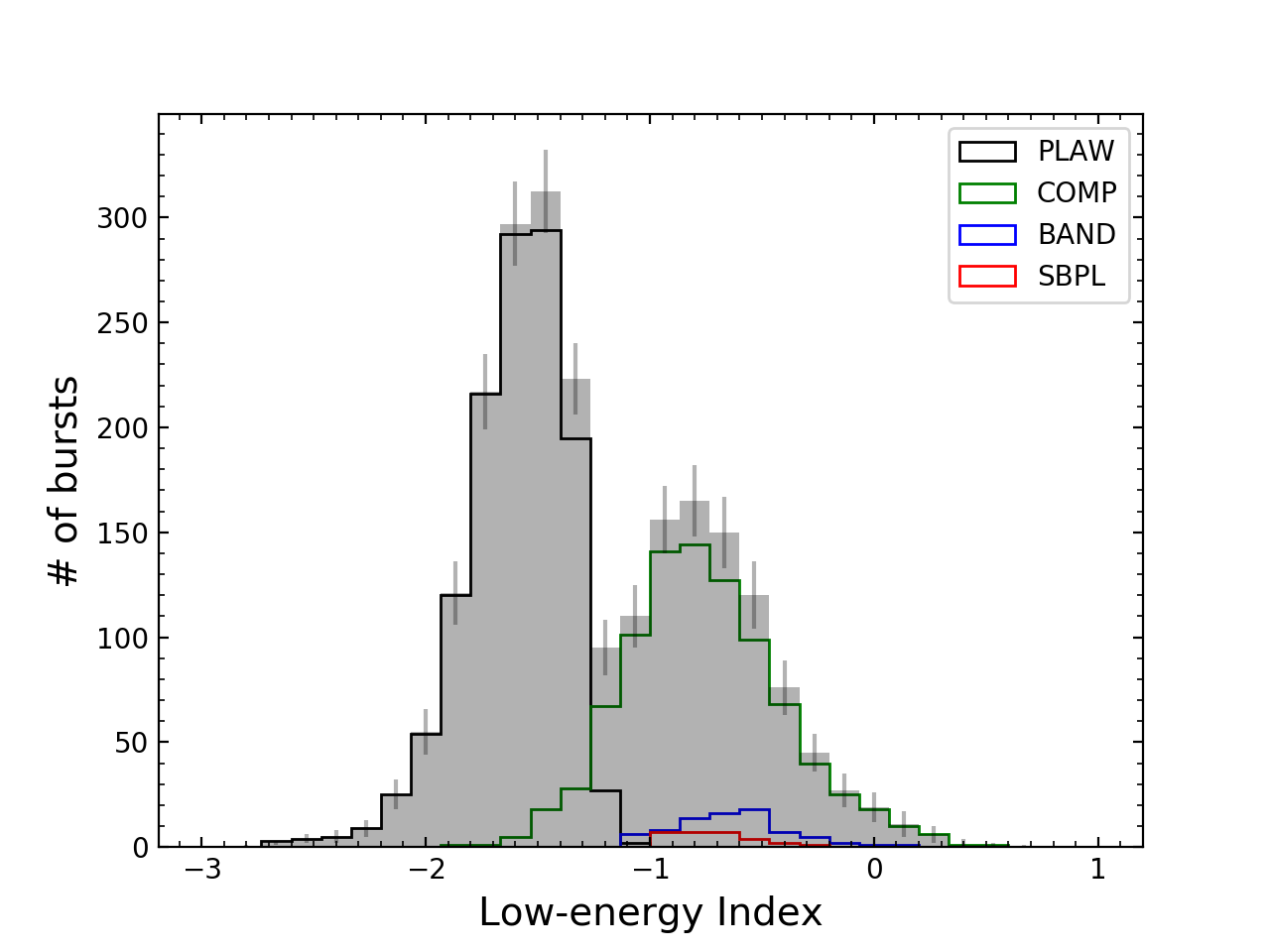}\label{fig:LE_Index_BEST_P}} \hspace{1em}
    \subfloat[]{\includegraphics[width=0.48\textwidth]{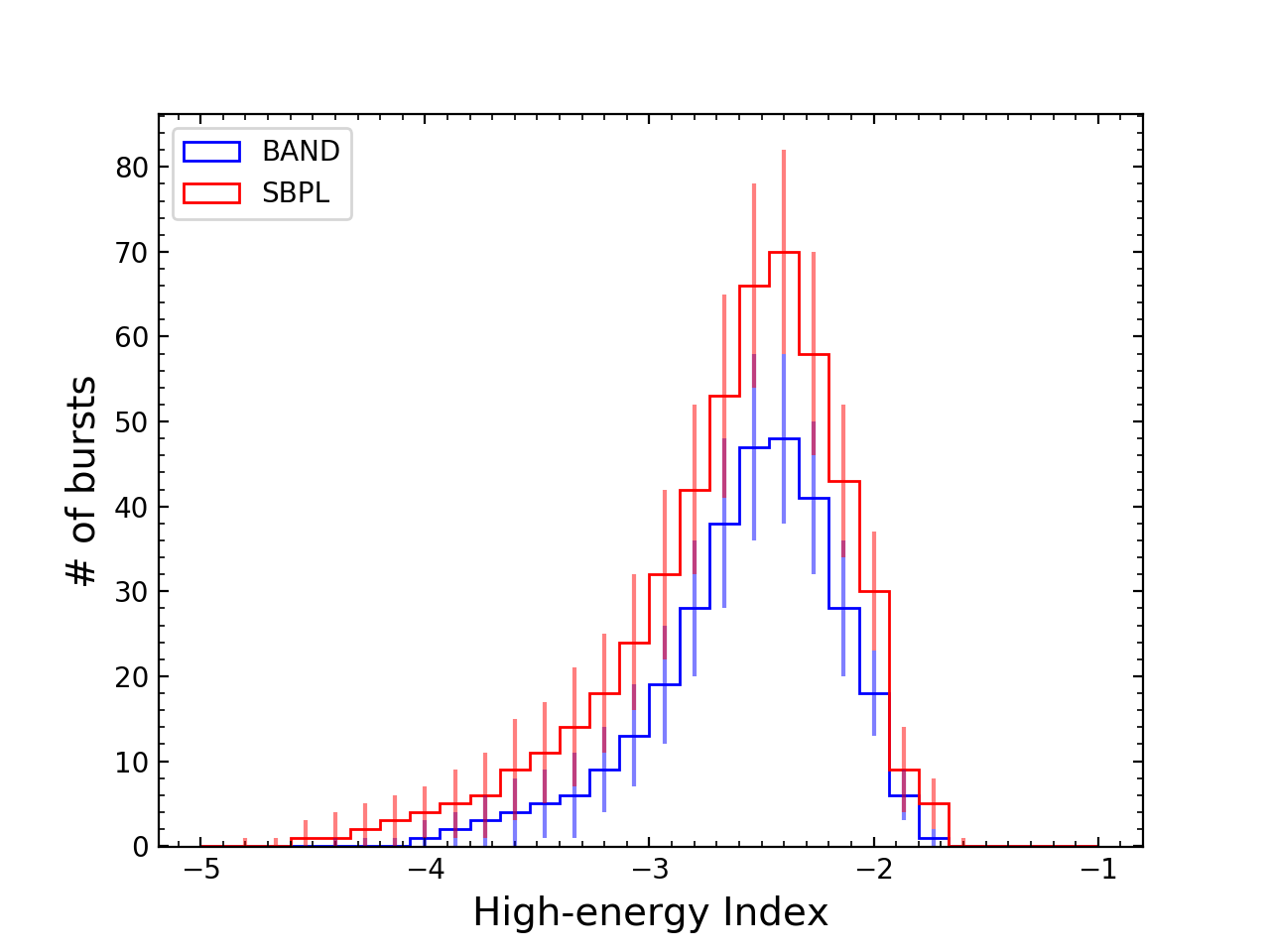}\label{fig:HE_Index_GOOD_P}}
    \subfloat[]{\includegraphics[width=0.48\textwidth]{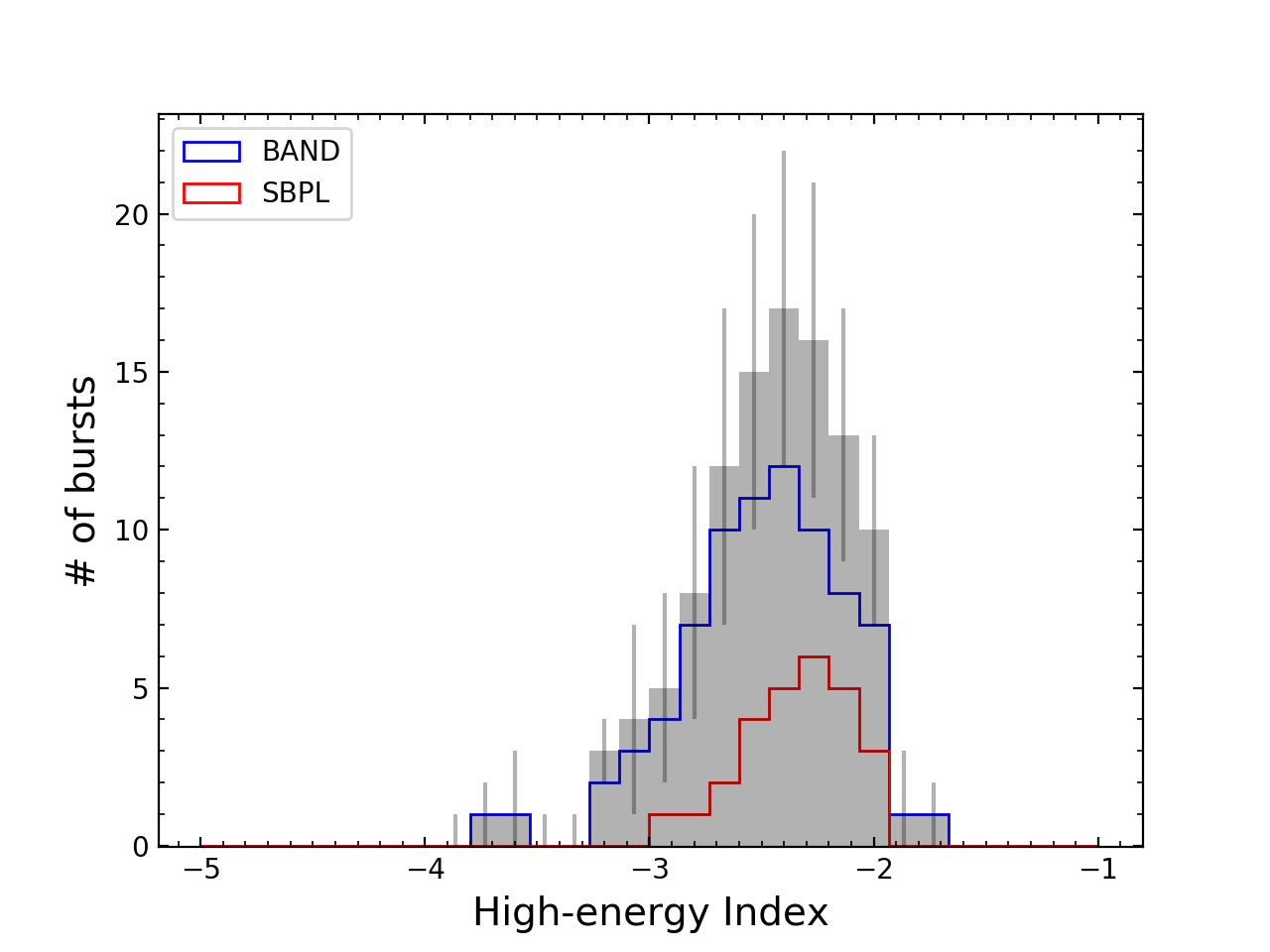}\label{fig:HE_Index_BEST_P}}\hspace{1em}
    \subfloat[]{\includegraphics[width=0.48\textwidth]{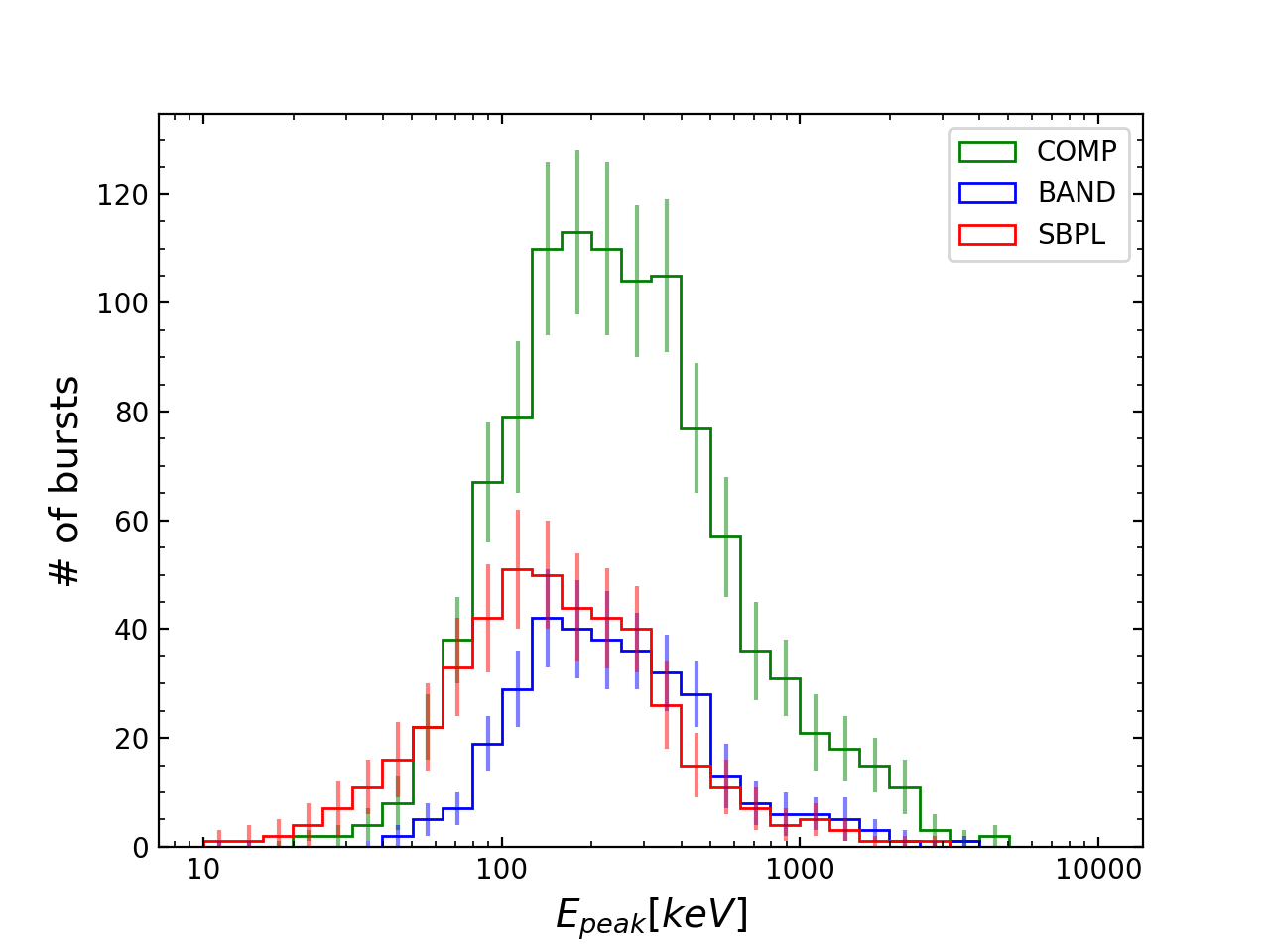}\label{fig:Epeak_GOOD_P}}
    \subfloat[]{\includegraphics[width=0.48\textwidth]{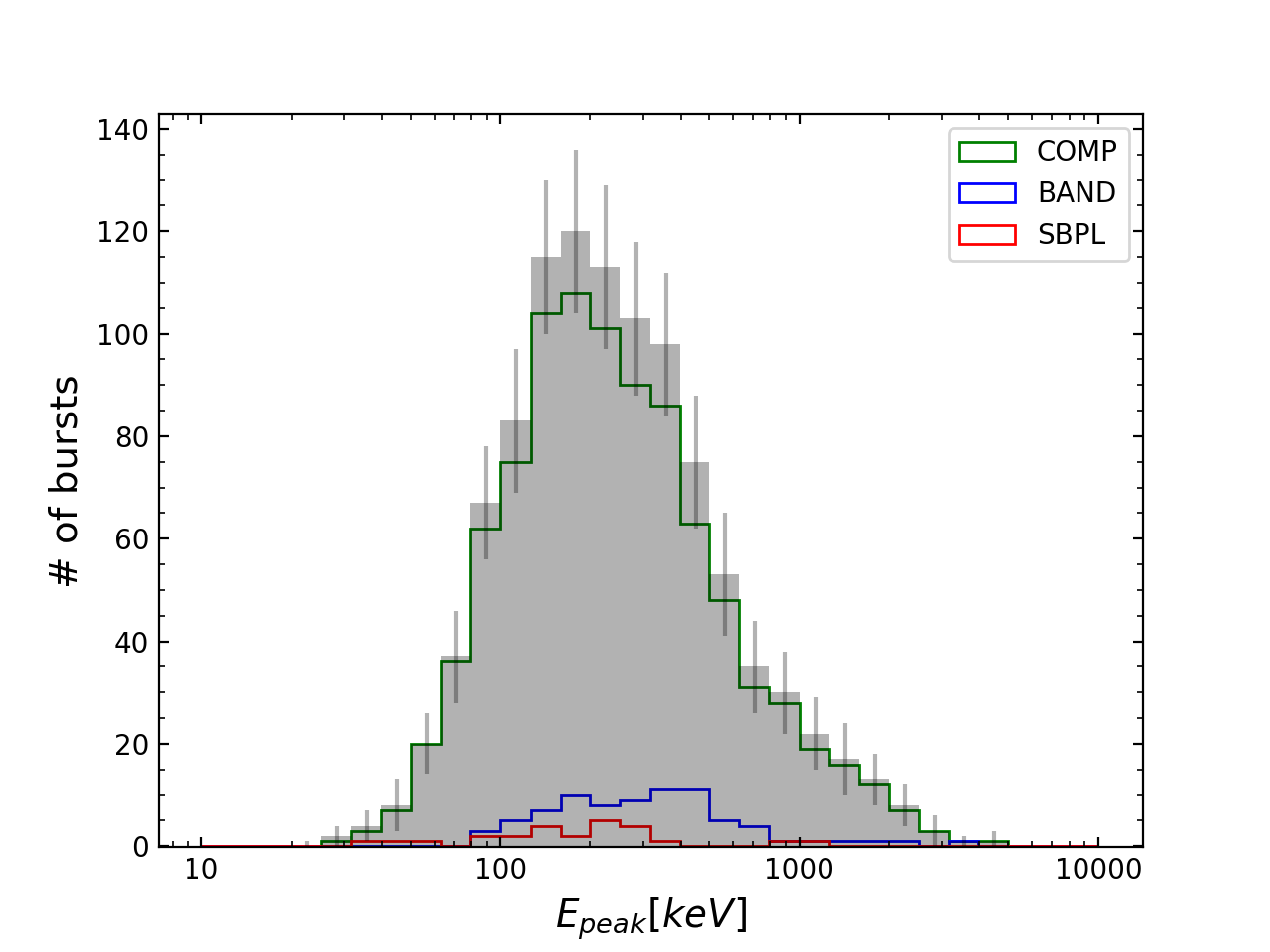}\label{fig:Epeak_BEST_P}}
    \caption{Distribution of the low-energy indices, high-energy indices and $E_{peak}$ obtained from the GOOD \textit{F} spectral fits are shown in (a), (c) and (e) respectively. The BEST parameter distribution (gray filled histogram) and its constituents are shown in (b), (d) and (f).
    \label{fig:GOOD_P}}
\end{figure}

\begin{figure}
    \centering
    \includegraphics[width=0.5\textwidth]{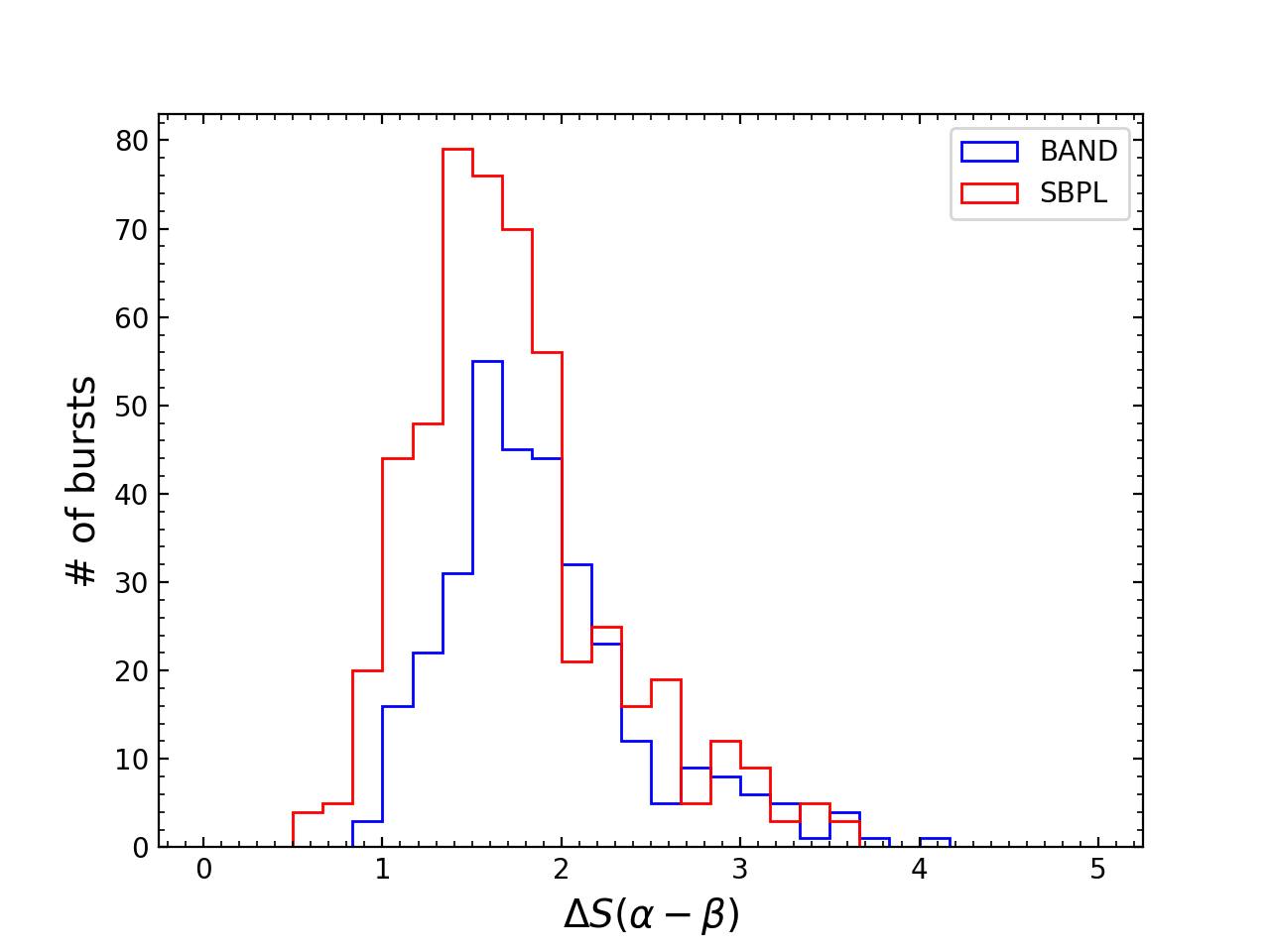}
    \caption{Distribution of $\Delta S$, the difference between low- and high-energy spectral indices $(\alpha - \beta)$ for the GOOD \textit{P} spectral fits.}
    \label{fig:DeltaS_P}
\end{figure}

The peak flux spectrum depicts the brightest portion of each burst, on a fixed 
timescale of 1.024 s for long GRBs and 64 ms for short GRBs. The time bin with the
highest significance is chosen for the ‘peak flux’ (\textit{P}) sample. The 
low-energy indices, shown in  Figure~\ref{fig:LE_Index_GOOD_P}, have a median value of 
about -1.3 and show a bimodal distribution. This is due to the fact that most GRBs
of the \textit{P} sample are best fit by the PLAW function, as less photon fluence
accumulation leads to a decrease in S/N. About 22\% of the BEST low-energy indices 
(Figure~\ref{fig:LE_Index_BEST_P})
have $\alpha > -2/3$, violating the synchrotron ``line-of-death" , while
70\% of the indices have $\alpha > -3/2$, violating the synchrotron
cooling limit; both of these are significantly larger percentages than those from
the \textit{F} spectra. The high-energy indices in Figure~\ref{fig:HE_Index_GOOD_P}
peak at about -2.3 and again have a long tail towards steeper indices. The lower
number of GOOD spectral fits compared to the \textit{F} spectra is likely due to
the poorer statistics resulting from shorter integration times. Shown in
Figure~\ref{fig:DeltaS_P} is the $\Delta S$ distribution for the \textit{P}
spectra, which suffers a deficit in values compared to the \textit{F} spectral fits largely
due to the inability of the data to sufficiently constrain the high-energy
power-law index. \\

\indent In Figure~\ref{fig:Epeak_GOOD_P},~\ref{fig:Epeak_BEST_P}, we show the $E_{peak}$ distribution 
for the \textit{P} spectra.
The $E_{peak}$ distribution for the BEST sample peaks at around 250 keV and covers
just over two orders of magnitude, which is consistent with previous findings
\citep{Goldstein_2012, Gruber_2014}. It should be noted that the data over the
short timescales in the \textit{P} spectra do not often favor either the BAND or
the SBPL model, resulting in large errors on parameters. 

\def\arraystretch{1.3}
\begin{table}[h]
    \movetableright=-0.75in
    \caption{The median parameter values and the 68\% CL of the BEST model fits}
    \centering
    \begin{tabular}{lcccccc}
	\hline \hline
	
	Data set & Low-energy & High-energy & $E_{peak}$ & $E_{break}$ & Photon Flux & Energy Flux \\
	~ & Index & Index & (keV) & (keV) & (photons s$^{-1}$cm$^{-2}$)& ($10^{-7}$erg s$^{-1}$cm$^{-2}$)  \\ \hline
	
	~                               & ~                       & ~                       & Fluence spectra     & ~                  & ~                      & ~                       \\ \hline
	This Catalog BEST               & $-1.08_{-0.44}^{+0.45}$ & $-2.20_{-0.29}^{+0.26}$ & $180_{-88}^{+307}$  & $107_{-49}^{+88}$  & $2.37_{-1.05}^{+3.83}$ & $2.94_{-1.39}^{+7.90}$  \\
	\citet{Gruber_2014}             & $-1.08_{-0.44}^{+0.43}$ & $-2.14_{-0.37}^{+0.27}$ & $196_{-100}^{+336}$ & $103_{-63}^{+129}$ & $2.38_{-1.05}^{+3.68}$ & $3.03_{-1.40}^{+7.41}$  \\
	\citet{Goldstein_2012}          & $-1.05_{-0.45}^{+0.44}$ & $-2.25_{-0.73}^{+0.34}$ & $205_{-121}^{+359}$ & $123_{-80}^{+240}$ & $2.92_{-1.31}^{+3.96}$ & $4.03_{-2.13}^{+9.38}$  \\
	\citet{Kaneko_2006}             & $-1.14_{-0.22}^{+0.20}$ & $-2.33_{-0.26}^{+0.24}$ & $251_{-68}^{+122}$  & $204_{-56}^{+76}$  &  ...                   & ...                     \\ \hline 
	
	~                               & ~                       & ~                       &  Peak flux spectra  & ~                   & ~                       & ~                        \\ \hline
	This Catalog BEST               & $-1.30_{-0.33}^{+0.77}$ & $-2.34_{-0.36}^{+0.28}$ & $233_{-117}^{+316}$ & $163_{-65}^{+156}$  & $4.62_{-2.55}^{+8.90}$  & $6.46_{-3.52}^{+17.82}$  \\
	\citet{Gruber_2014}             & $-1.32_{-0.33}^{+0.74}$ & $-2.24_{-0.38}^{+0.26}$ & $261_{-130}^{+364}$ & $133_{-39}^{+349}$  & $4.57_{-2.49}^{+8.82}$  & $6.49_{-3.46}^{+17.52}$  \\
	\citet{Goldstein_2012}          & $-1.12_{-0.50}^{+0.61}$ & $-2.27_{-0.50}^{+0.44}$ & $223_{-126}^{+352}$ & $172_{-100}^{+254}$ & $5.39_{-2.87}^{+10.18}$ & $8.35_{-4.98}^{+22.61}$  \\
	\citet{Nava_2011}               & $(-0.56_{-0.37}^{+0.40})^a$ & $-2.39_{-0.62}^{+0.23}$ & $225_{-122}^{+391}$ & ...             &  ...                    & $13.5_{-10.1}^{+79.8}$   \\
	\citet{Kaneko_2006}             & $-1.02_{-0.28}^{+0.26}$ & $-2.33_{-0.31}^{+0.26}$ & $281_{-99}^{+139}$  & $205_{-55}^{+72}$   &  ...                    & ...                      \\ \hline 
	
    \end{tabular}
    \textbf{Note}: $^a$ Low-energy index of the peak-flux spectra with curved function only.
\end{table}

\begin{figure}
    \subfloat[]{\includegraphics[width=0.33\textwidth]{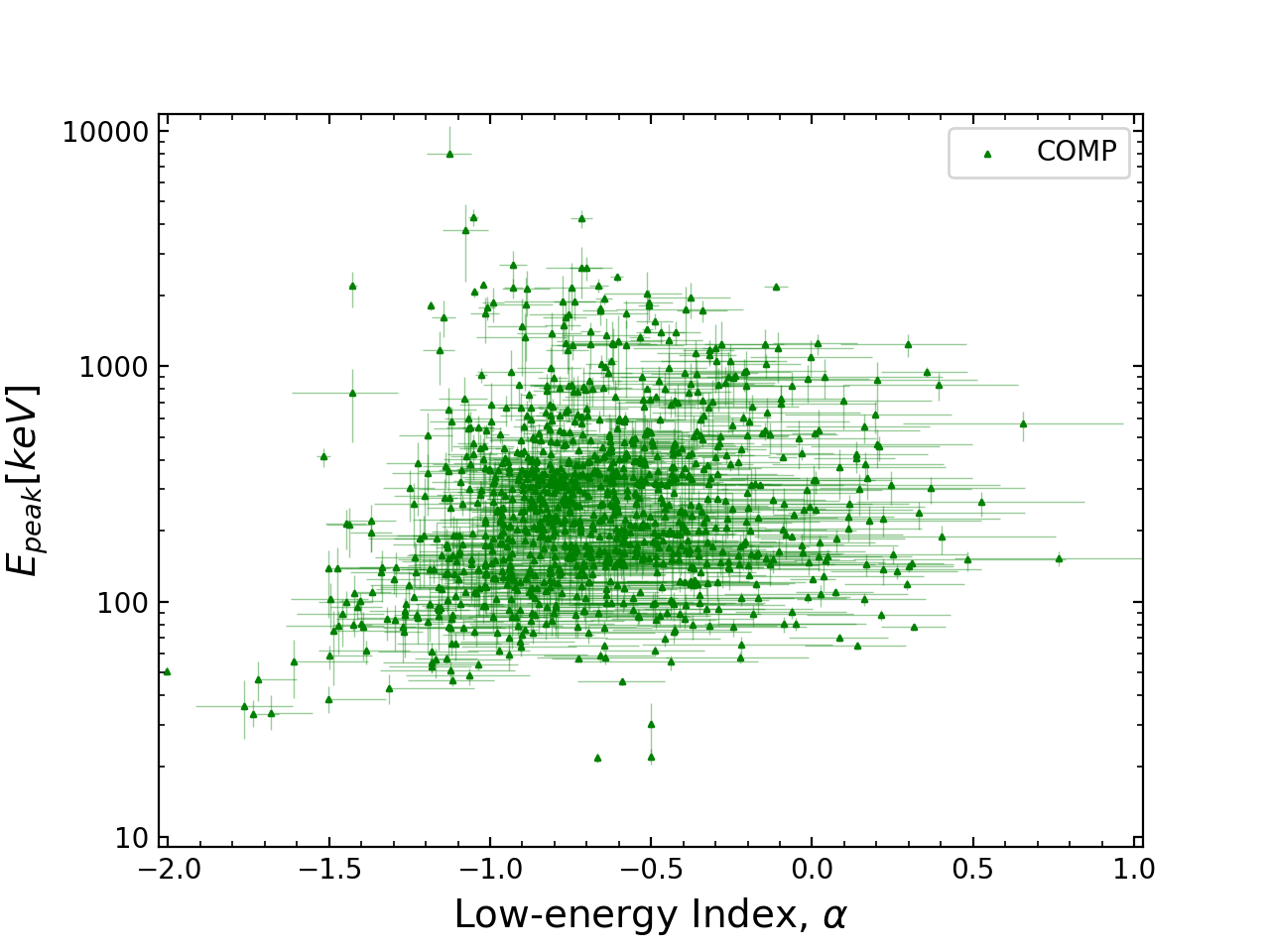}\label{fig:Epeak_Alpha_Comp_P}}
    \subfloat[]{\includegraphics[width=0.33\textwidth]{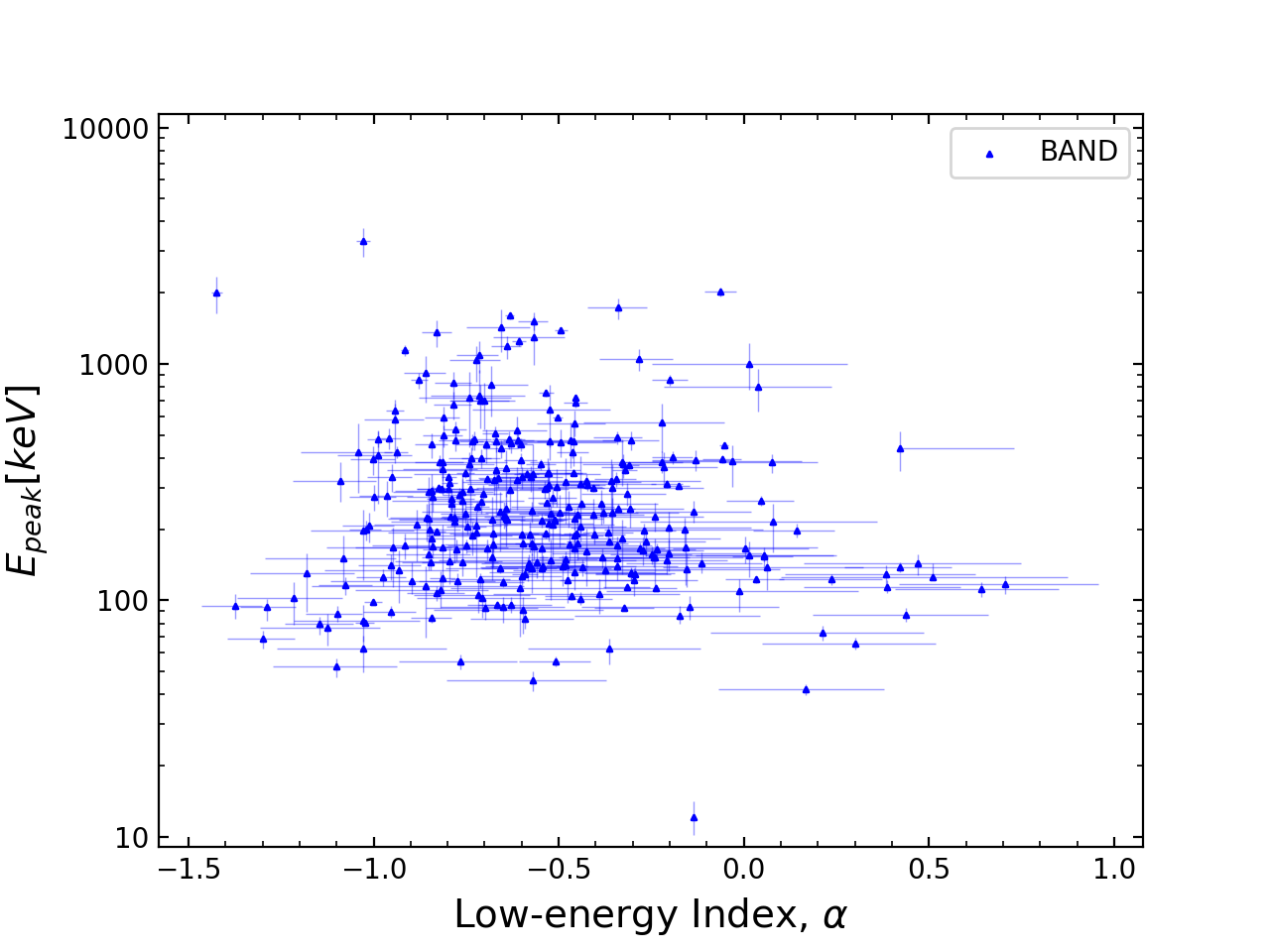}\label{fig:Epeak_Alpha_Band_P}}
    \subfloat[]{\includegraphics[width=0.33\textwidth]{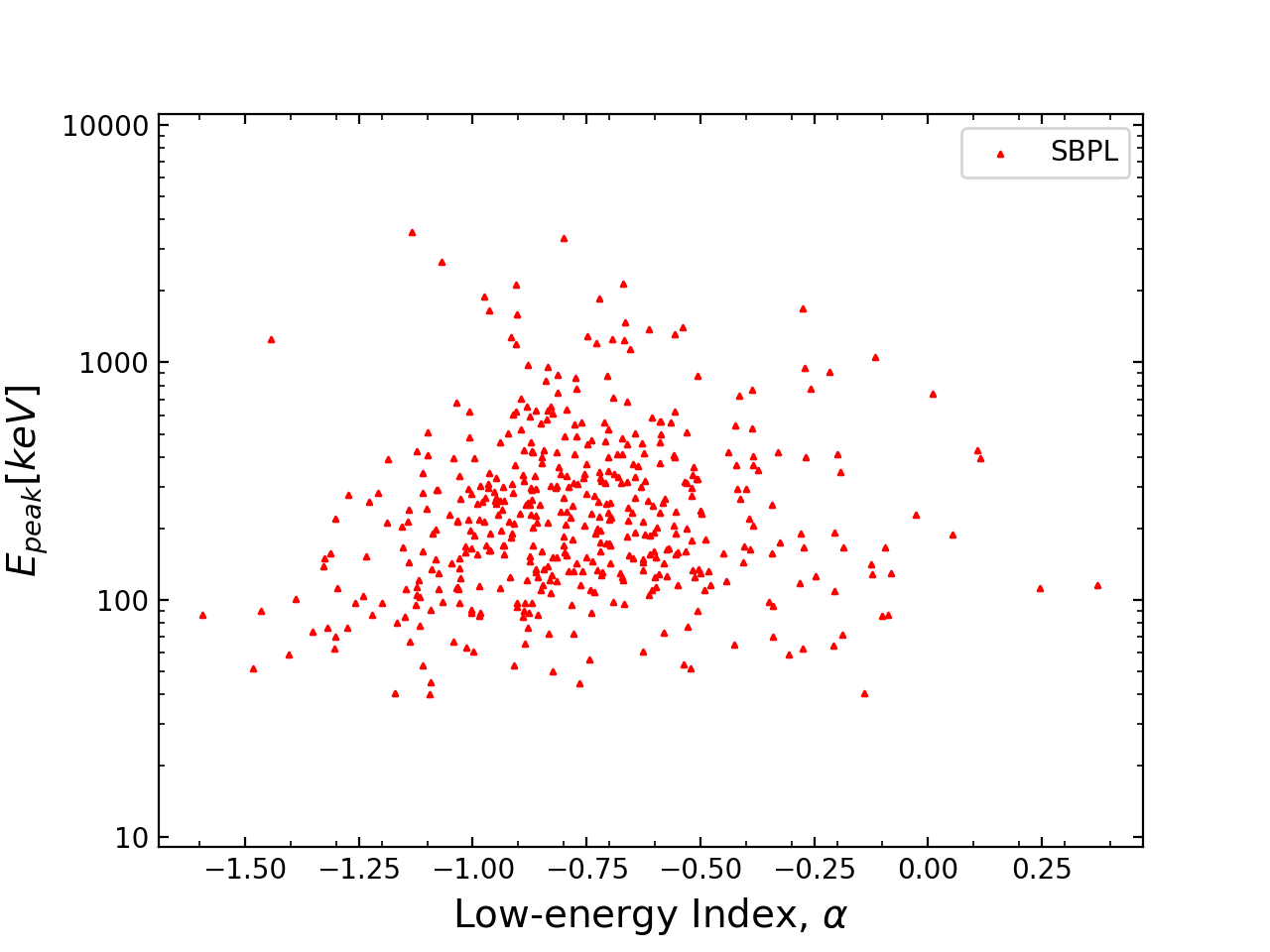}\label{fig:Epeak_Alpha_Sbpl_P}}
    \caption{Comparison of the low-energy index and $E_{peak}$ for three models from the GOOD \textit{P} spectral fits.
    \label{fig:Epeak_Alpha_P}}
\end{figure}

\subsection{Long vs.\ Short GRBs}

Over the ten years of operations covered in this Catalog, GBM triggered on 
395 short GRBs, $17\%$ of the total number of bursts. The idea that short GRBs and long GRBs represent two distinct populations was 
bolstered by the comparison between their hardness ratios \citep{Kouveliotou_1993, Bhat+16cat}. Short GRBs are significantly harder, as determined by the ratio of 
the counts in two broad energy bands (25 -- 100, 100 -- 300 keV) \citep{Kouveliotou_1993}. Spectral fit 
parameters should reflect this dichotomy in hardness in two ways. First, the median values for $E_{\rm peak}$ should be 
significantly different, with the higher value being associated with short bursts. Secondly, a low-energy power law 
index that is higher than another (e.g.$-1$ vs.$-2$) is said to be harder, as a positive uptick requires an 
increase in higher-energy photons, all other things being equal. Here, we can verify both of these by 
comparing the median fitted spectral parameters between short and long bursts in Table \ref{tab:Short and Long.}. This is 
in agreement with results from early on in the mission \citep{Nava_2011}.

\begin{figure}
    \centering
    \includegraphics[width=1.0\textwidth]{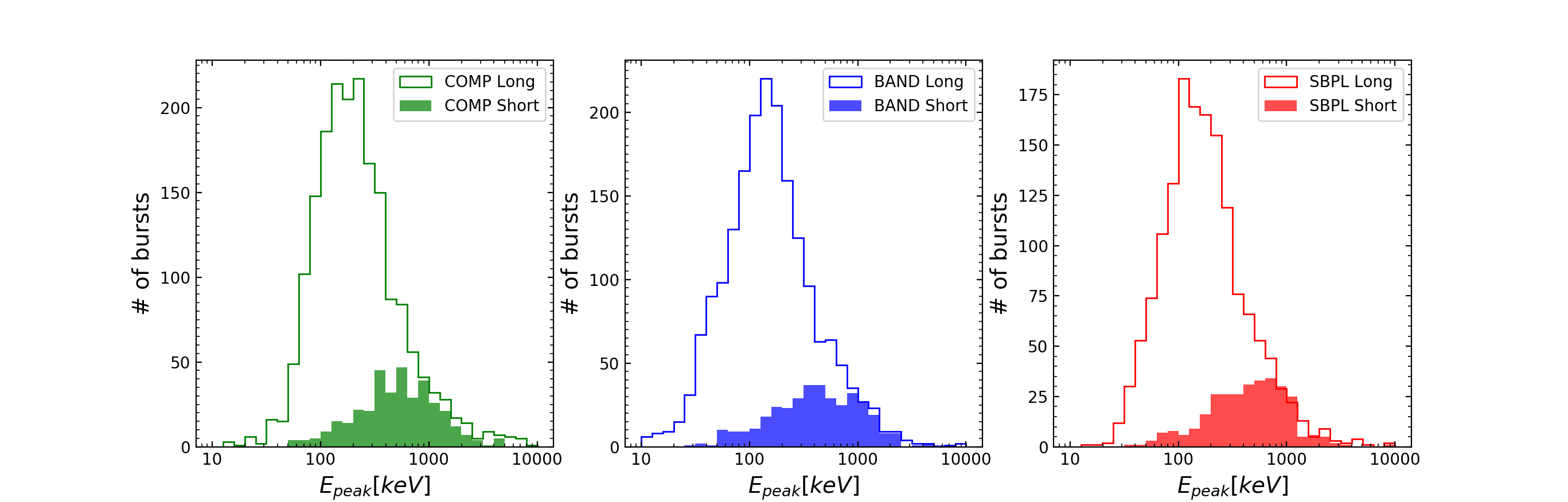}
    \caption{Comparison of $E_{peak}$ between long and short GRBs from \textit{F} spectral fits.}
    \label{fig:Epeak_Long_Short_F}
\end{figure}

\begin{figure}
    \centering
    \includegraphics[width=1.0\textwidth]{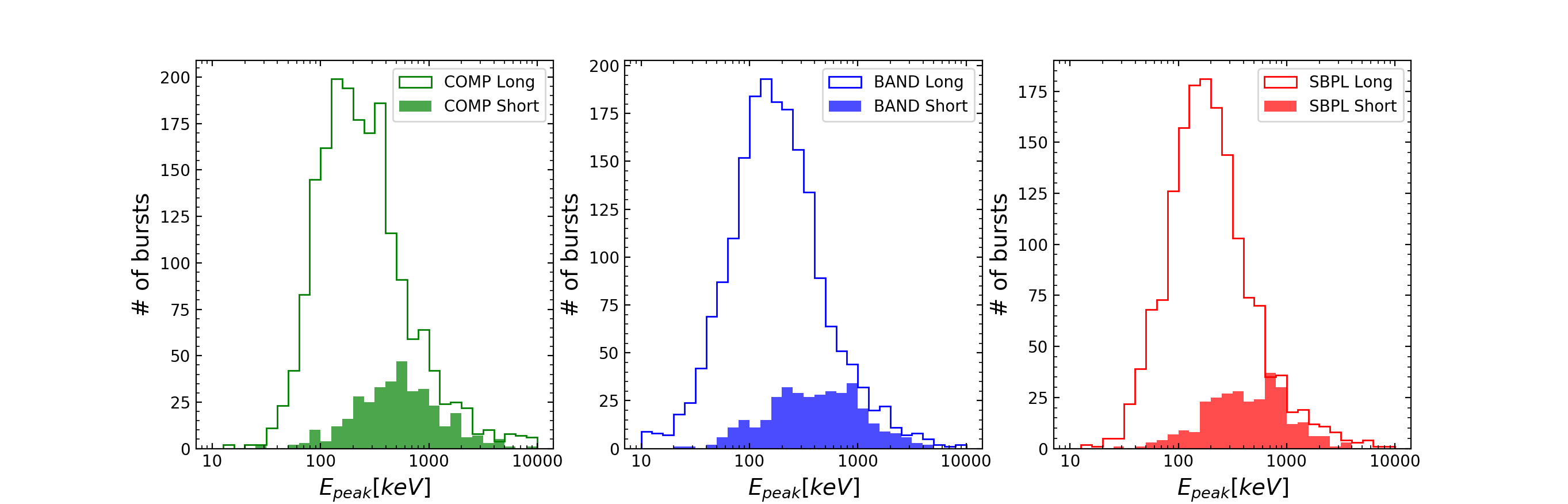}
    \caption{Comparison of $E_{peak}$ between long and short GRBs from \textit{P} spectral fits.}
    \label{fig:Epeak_Long_Short_P}
\end{figure}

\begin{figure}
    \centering
    \includegraphics[width=1.0\textwidth]{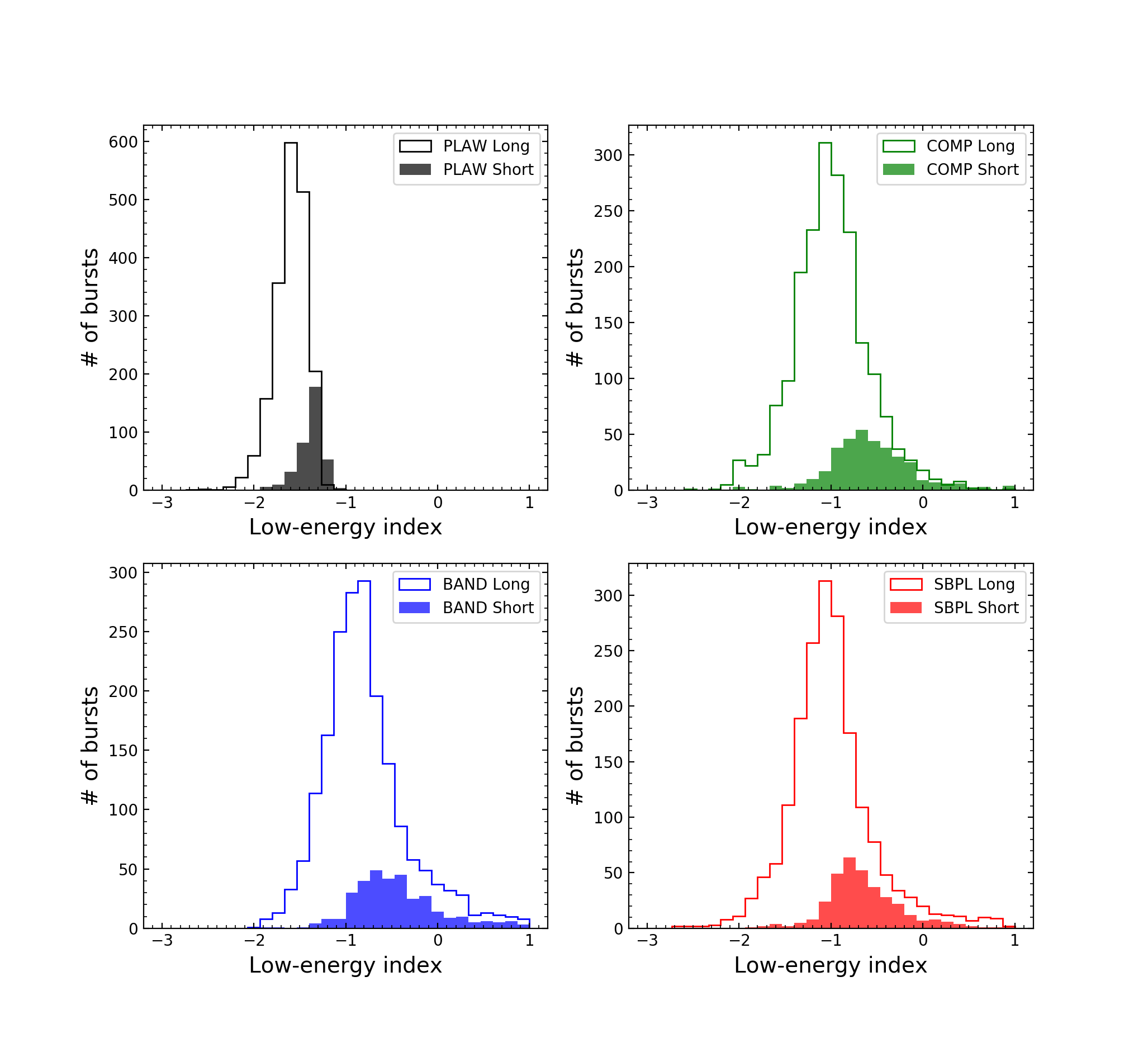}
    \caption{Comparison of low-energy indices between long and short GRBs from \textit{F} spectral fits.}
    \label{fig:LE_Long_Short_F}
\end{figure}

\begin{figure}
    \centering
    \includegraphics[width=1.0\textwidth]{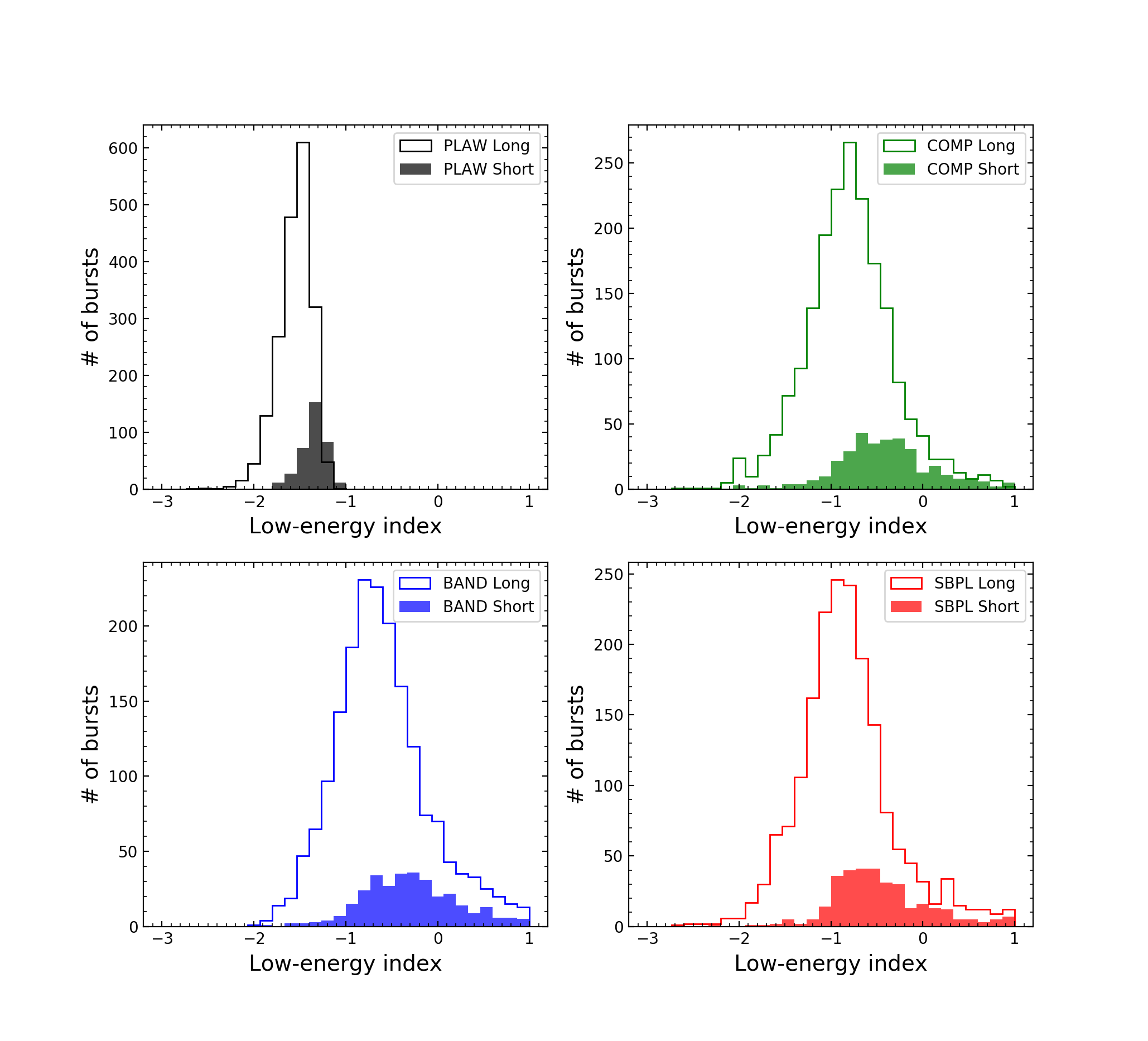}
    \caption{Comparison of low-energy indices between long and short GRBs from \textit{P} spectral fits.}
    \label{fig:LE_Long_Short_P}
\end{figure}

\begin{figure}
    \centering
    \includegraphics[width=1.0\textwidth]{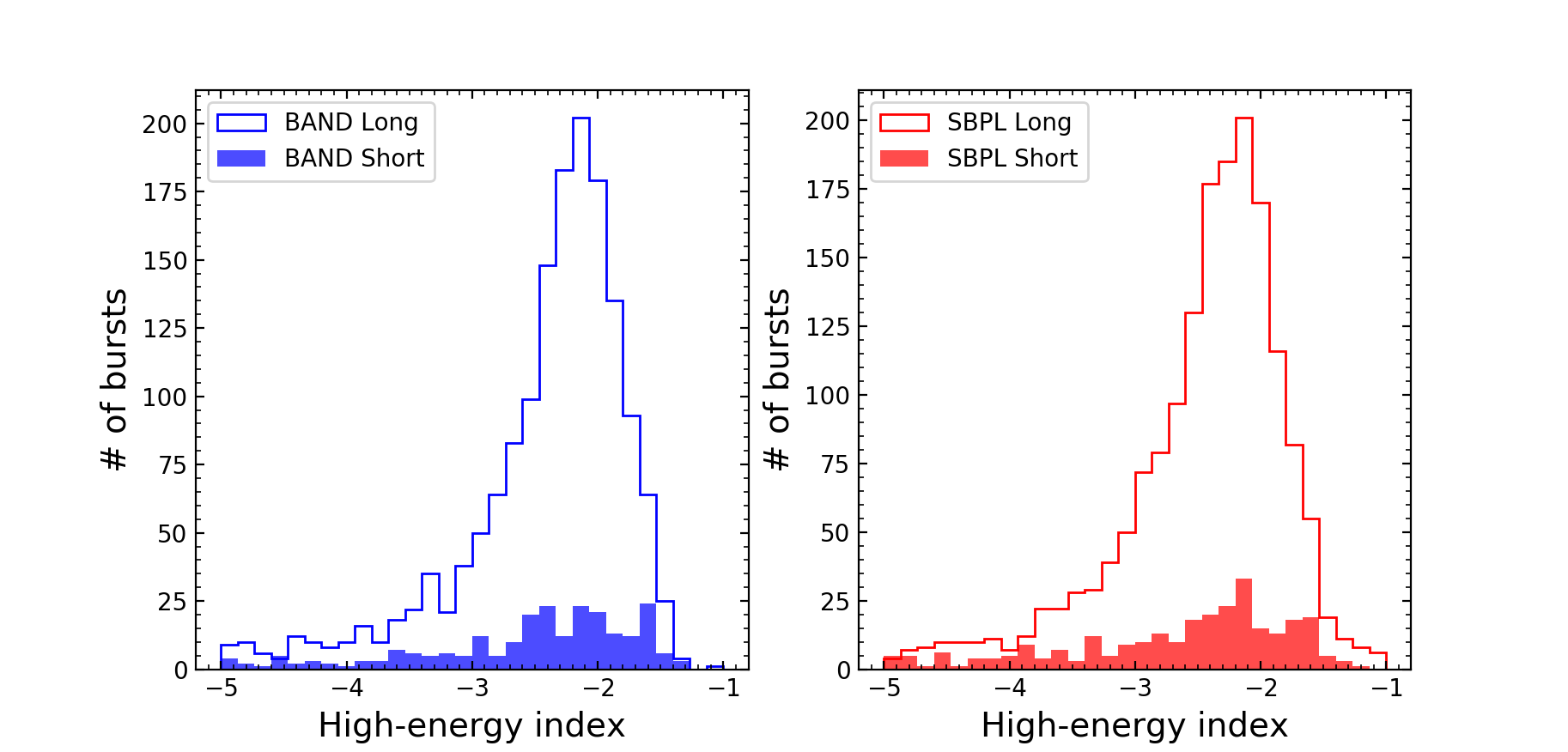}
    \caption{Comparison of high-energy indices between long and short GRBs from \textit{F} spectral fits.}
    \label{fig:HE_Long_Short_F}
\end{figure}

\begin{figure}
    \centering
    \includegraphics[width=1.0\textwidth]{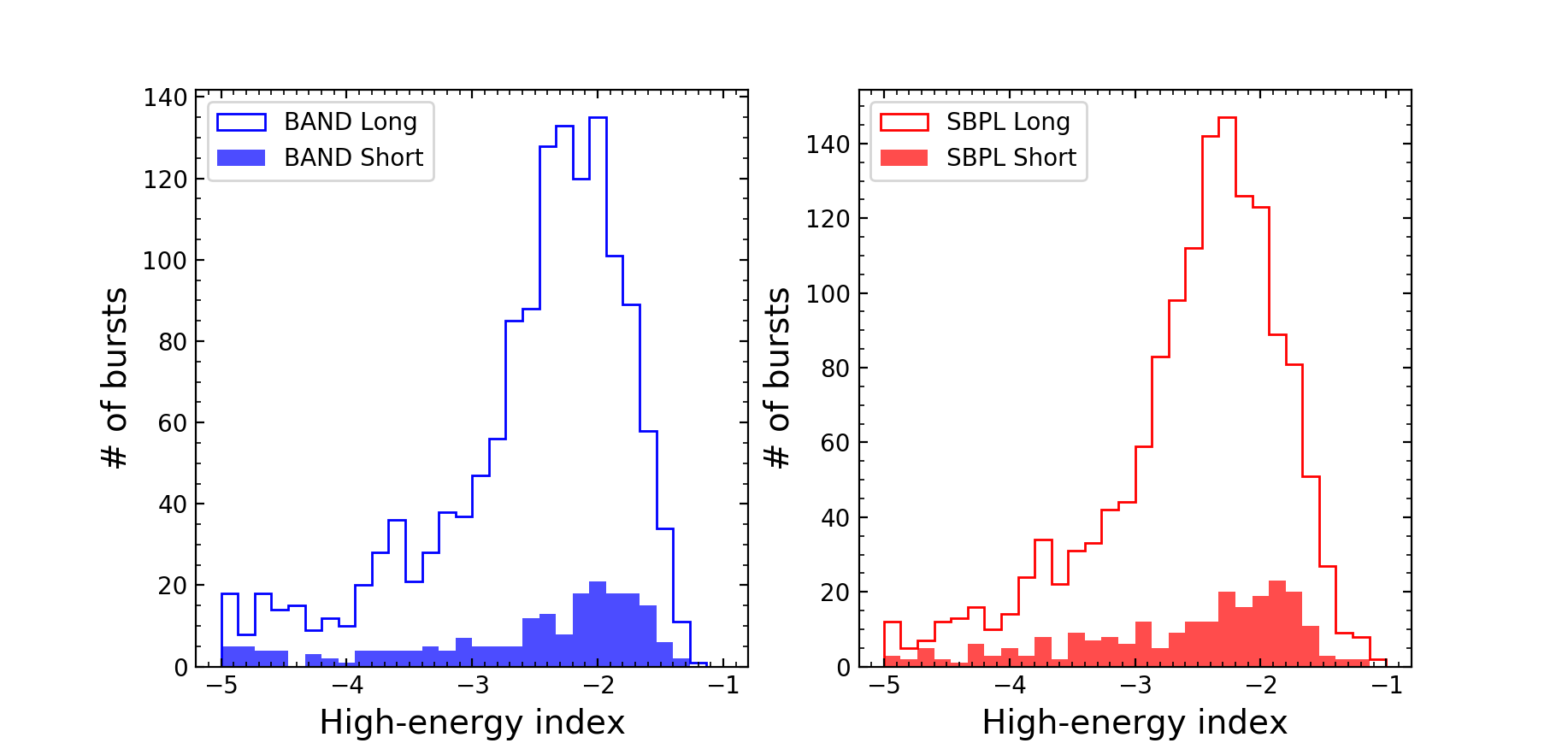}
    \caption{Comparison of high-energy indices between long and short GRBs from \textit{P} spectral fits.}
    \label{fig:HE_Long_Short_P}
\end{figure}

The hard nature of short bursts is even more dramatic when considering the distributions of the fitted parameters. 
Figures~\ref{fig:Epeak_Long_Short_F} and~\ref{fig:Epeak_Long_Short_P} compare $E_{\rm peak}$ between long and short GRBs for 
the fluence and peak-flux spectral fits respectively. In order to improve the sample size of the short burst population, 
we present fits from the total ensemble of bursts; one for each of the 
three models that have an energy-related parameter (COMP, BAND and SBPL).
Similarly, Figures~\ref{fig:LE_Long_Short_F} and~\ref{fig:LE_Long_Short_P} compare the low-energy indices between long and short 
GRBs for all four models (including PLAW) from the fluence and peak-flux spectral fits respectively. Clearly, it is highly 
improbable that the long and short distributions are the same for the $E_{\rm peak}$ and low-energy index distributions. 
Figures~\ref{fig:HE_Long_Short_F} and~\ref{fig:HE_Long_Short_P} compare the high-energy indices between long and short GRBs for 
BAND and SBPL from the fluence and peak-flux spectral fits respectively.  
Although the {\em presence} of a high-energy power-law component is a clear signal of `hardness', the fitted index itself 
seems to be invariant between the two classes of GRBs.

\begingroup
\def\arraystretch{1.3}
\begin{table}[h]
    \movetableright=-0.75in
    \caption{The median parameter values and the 68\% CL for all long and short GRBs}
    \centering
    \begin{tabular}{l|ccc|ccc}
	\hline \hline
	~    & ~                       & Long GRBs               & ~                  & ~                       & Short GRBs              & ~                  \\ \hline
	Model & Low-energy Index & High-energy Index& $E_{peak}$ (keV) & Low-energy Index & High-energy Index & $E_{peak}$ (keV) \\
	 \hline
	 \multicolumn{7}{c}{Fluence Spectra}\\
	 \hline
	PLAW   & $-1.58_{-0.18}^{+0.15}$ & ...                     & ...                & $-1.35_{-0.16}^{+0.09}$ & ...                     & ...                  \\
	COMP & $-1.01_{-0.35}^{+0.39}$ & ...                     & $205_{-109}^{+374}$& $-0.59_{-0.34}^{+0.49}$ & ...                     & $534_{-312}^{+660}$  \\ 
	BAND & $-0.84_{-0.35}^{+0.48}$ & $-2.41_{-3.96}^{+0.49}$ & $144_{-85}^{+229}$ & $-0.46_{-0.37}^{+0.68}$ & $-2.98_{-6.56}^{+1.04}$ & $413_{-263}^{+651}$  \\
	SBPL & $-1.03_{-0.34}^{+0.47}$ & $-2.40_{-1.49}^{+0.48}$ & $160_{-86}^{+255}$ & $-0.66_{-0.30}^{+0.53}$ & $-2.79_{-6.04}^{+0.87}$ & $476_{-280}^{+480}$  \\
	\hline
	\multicolumn{7}{c}{Peak Flux Spectra}\\
	\hline
	PLAW   & $-1.52_{-0.20}^{+0.14}$ & ...                     & ...                 & $-1.33_{-0.15}^{+0.11}$ & ...                     & ...                  \\
	COMP & $-0.80_{-0.43}^{+0.45}$ & ...                     & $224_{-124}^{+435}$& $-0.39_{-0.46}^{+0.63}$ & ...                     & $532_{-316}^{+732}$  \\ 
	BAND & $-0.64_{-0.42}^{+0.63}$ & $-2.69_{-5.23}^{+0.75}$ & $166_{-98}^{+295}$  & $-0.22_{-0.51}^{+0.92}$ & $-4.18_{-8.54}^{+2.26}$ & $426_{-279}^{+635}$  \\
	SBPL & $-0.85_{-0.42}^{+0.55}$ & $-2.62_{-5.84}^{+0.68}$ & $181_{-96}^{+272}$  & $-0.49_{-0.40}^{+0.81}$ & $-3.26_{-12.9}^{+1.34}$ & $415_{-236}^{+552}$  \\ 
    \hline 
    \end{tabular}
    \label{tab:Short and Long.}
\end{table}
\endgroup

\section{Rest-Frame Properties}
Calculating the rest-frame energetics is key to understanding the central engine and
emission physics of a GRB. Using 10 years of GBM data, and the known
redshift for $\sim$130 GRBs, we provide one of the largest samples of GRB energetics to date.

\begin{figure}
    \centering
    \includegraphics[width=0.5\textwidth]{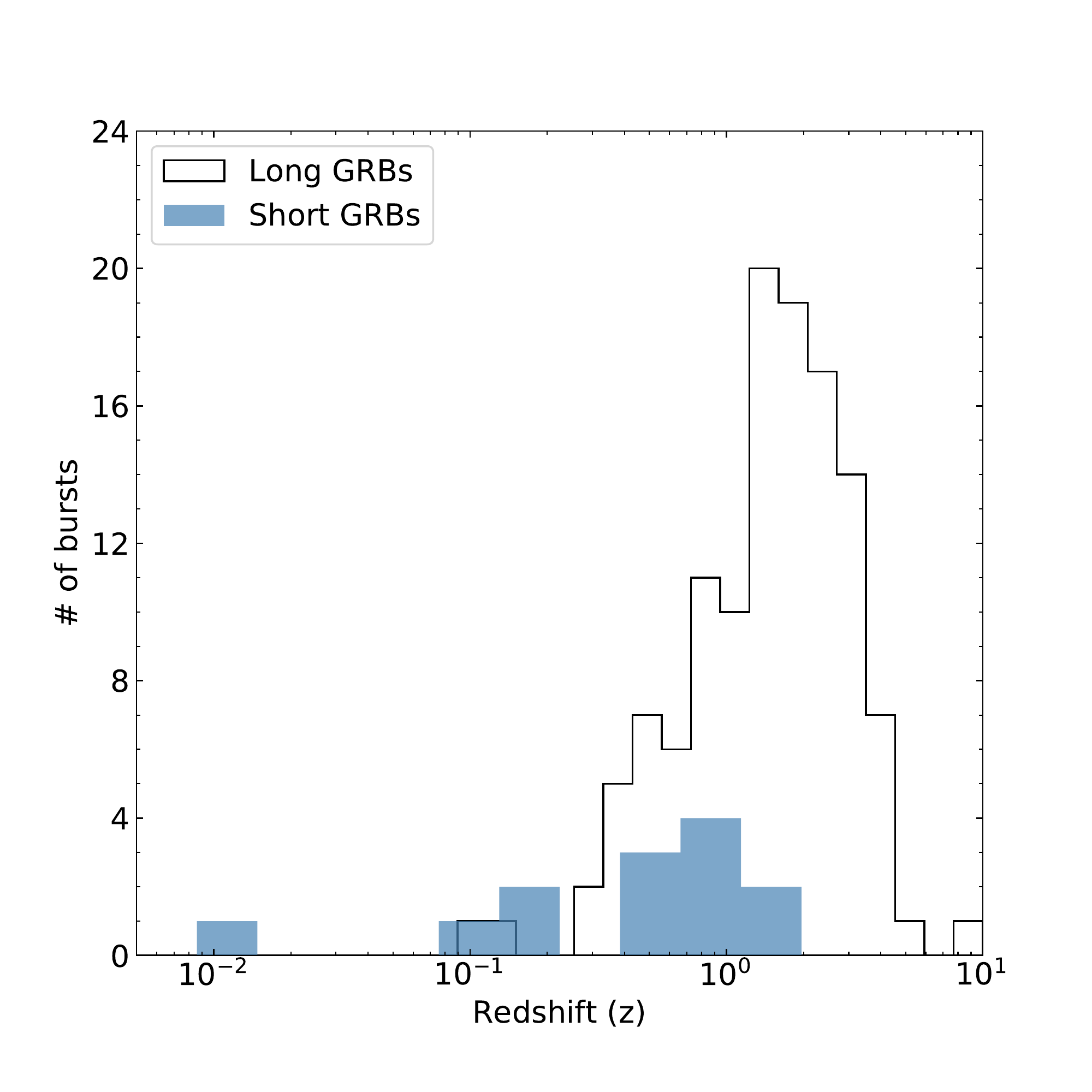}
    \caption{The distribution of long and short GRBs with respect to redshift. We find that there are more long GRBs in our sample. GRB 170817A is the outlier in this plot with a much lower redshift of $0.009$. }
    \label{fig:redshift_hist}
\end{figure}

GBM is found to detect more long GRBs than short, which is reflected in our redshift sample with 13 
short GRBs and 122 long GRBs. The distribution of the redshift for both short GRBs (black) and long GRBs (blue) are shown 
in Figure~\ref{fig:redshift_hist}. GRB 170817A, which was determined to be in coincidence with the neutron star merger event \citep{GBMLVC}, GW170817, 
is found to have a much lower redshift ($z=0.009$) than the rest of the sample.

\subsection{Rest-frame Properties using the BAND Spectral model} \label{sec:RestFrameBand}

\begin{figure}
\centering
    \includegraphics[width=\textwidth]{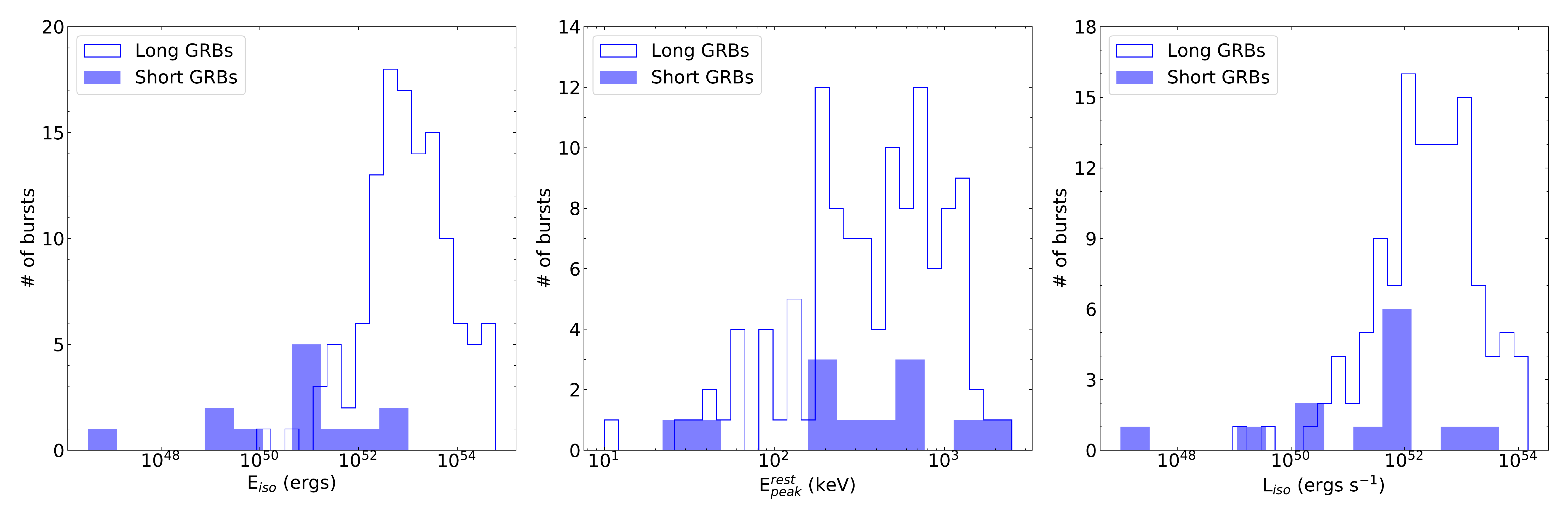}
\caption{Distribution of the k-corrected isotropic energy (\textit{left}), rest peak energy (\textit{middle}) and isotropic luminosity (\textit{right}) using the parameters of the Band spectral model. }
\label{fig:band_hist}
\end{figure}

We use the BAND model as discussed in Sec.~\ref{sec:Models}, which provides the peak energy ($E_{peak}$), amplitude, and the indices of $\alpha$ and $\beta$ as well as the measured fluence in the \textit{Fermi}-GBM bandpass (10-1000 keV). For the GRBs with known redshift ($z$), we calculated the k-corrected, isotropic-equivalent gamma-ray energy of the GRB in the co-moving bolometric bandpass of 1 keV to 10 MeV and the estimated uncertainty on the prompt energy release.\\
\indent In order to calculate the \textit{k}-correction, we use the fluence in the \textit{Fermi} GBM bandpass and expand the spectral model given in \citep{Band_1993} to the co-moving bolometric bandpass of 1 keV to 10 MeV \citep{Bloom+01kcorr}. The \textit{k}-value is defined as 
\begin{equation}
    k=\frac{S_{[\frac{E_1}{1+z},\frac{E_2}{1+z}]}}{S_{[e_1, e_2]}}    
\end{equation} 
where $S$ is the fluence for a range of given energies, $E_1=1$ keV, $E_2=10$ MeV, $e_1=10$ keV, $e_2=1000$ keV. The isotropic energy can then be calculated by 
\begin{equation}
    E_{iso}=\frac{4 \pi D_\ell^2}{1+z}S_{obs}k
\end{equation}
with $D_\ell$ being the luminosity distance and $S_{obs}$ being the observed fluence. The distribution of the $E_{iso}$ is presented in  Figure~\ref{fig:band_hist} and shows that the long GRBs have $E_{iso}$ centered around $10^{53}$ erg. The limited number of short GRBs with redshift does not provide much insight into their distribution but the outlier in this plot is GRB 170817A which was found to have a lower redshift than the other GRBs.  \\
\indent  The peak rest frame energy ($E^{rest}_{peak}$) is determined by $E^{rest}_{peak}=E^{obs}_{peak} (1+z)$, where $E^{obs}_{peak}$ is the observed peak energy from the F spectra. The distribution of $E^{rest}_{peak}$ is shown in Figure~\ref{fig:band_hist} to span from 10 keV to 3000 keV. Neither short nor long GRBs appear to exhibit a preference for a particular rest frame $E_{peak}$ but there does appear to be a separate group of GRBs with $E_{peak}>1$ MeV. 
Using the spectra of the brightest time bin, we can determine the isotropic luminosity for the GRBs with redshift using
\begin{equation}
    L_{iso}=4 \pi D_\ell^2 S_{[E1,E2]}
\end{equation}
The distribution of $L_{iso}$, shown in Figure~\ref{fig:band_hist}, shows that neither the short nor the long
GRBs have preference for $L_{iso}$. However, GRB 170817A has a $L_{iso}$ few order of magnitudes lower than the
rest. The $E_{iso}$, $E_{rest}^{peak}$, and $L_{iso}$ with their uncertainties are shown in Table
\ref{tab:comp_results}. 

\subsection{Rest-frame Properties using the COMP Spectral model}

\begin{figure}
\centering
    \includegraphics[width=\textwidth]{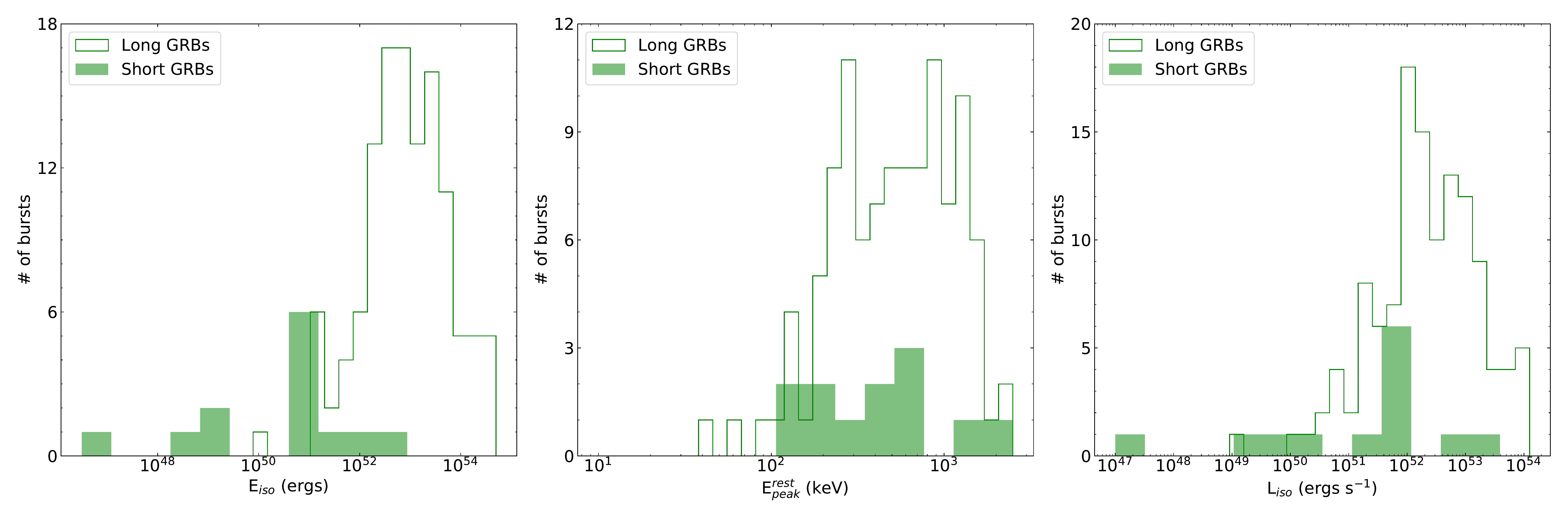}
\caption{Distribution of the isotropic energy (\textit{left}), rest peak energy  (\textit{middle}) and isotropic luminosity (\textit{right}) for the Comptonized spectral model.}
\label{fig:comp_hist}
\end{figure}

The COMP spectral model (Sec.~\ref{sec:Models}) is also used to determine the $E_{iso}$, $E^{rest}_{peak}$ and $L_{iso}$ in a similar way to Sec.~\ref{sec:RestFrameBand} and using the spectral shape from Eq.\ \ref{eq:comp}. The values and uncertainties for $E_{iso}$, $E^{rest}_{peak}$ and $L_{iso}$ are presented in Table \ref{tab:comp_results}. \\
\indent The distribution of the $E_{iso}$ in Figure~\ref{fig:comp_hist}, shows that the short and long GRBs do not have a preference for the $E_{iso}$. However, GRB 170817A is again at the low end of the $E_{iso}$ distribution. The distribution of $E^{rest}_{peak}$ using the COMP model spans from 10 kev to 3000 keV with a group of GRBs with $E_{peak}$ $>1$ MeV, similar to the BAND model. The $L_{iso}$ distribution for the COMP model also shows no preference for a value and GRB 170817A has a lower value than the rest of the GRBs.

\section{Comparison to Previous Analysis}
The number of GRBs with GOOD spectral fits presented in this catalog is similar to the total contained in the BATSE 5B spectroscopy catalog \citep{Goldstein_2013}.  Additionally, the methodology and instrument characteristics are similar, therefore these two data sets can be easily compared.  

\begin{figure}
    \centering
        \includegraphics[width=0.48\textwidth]{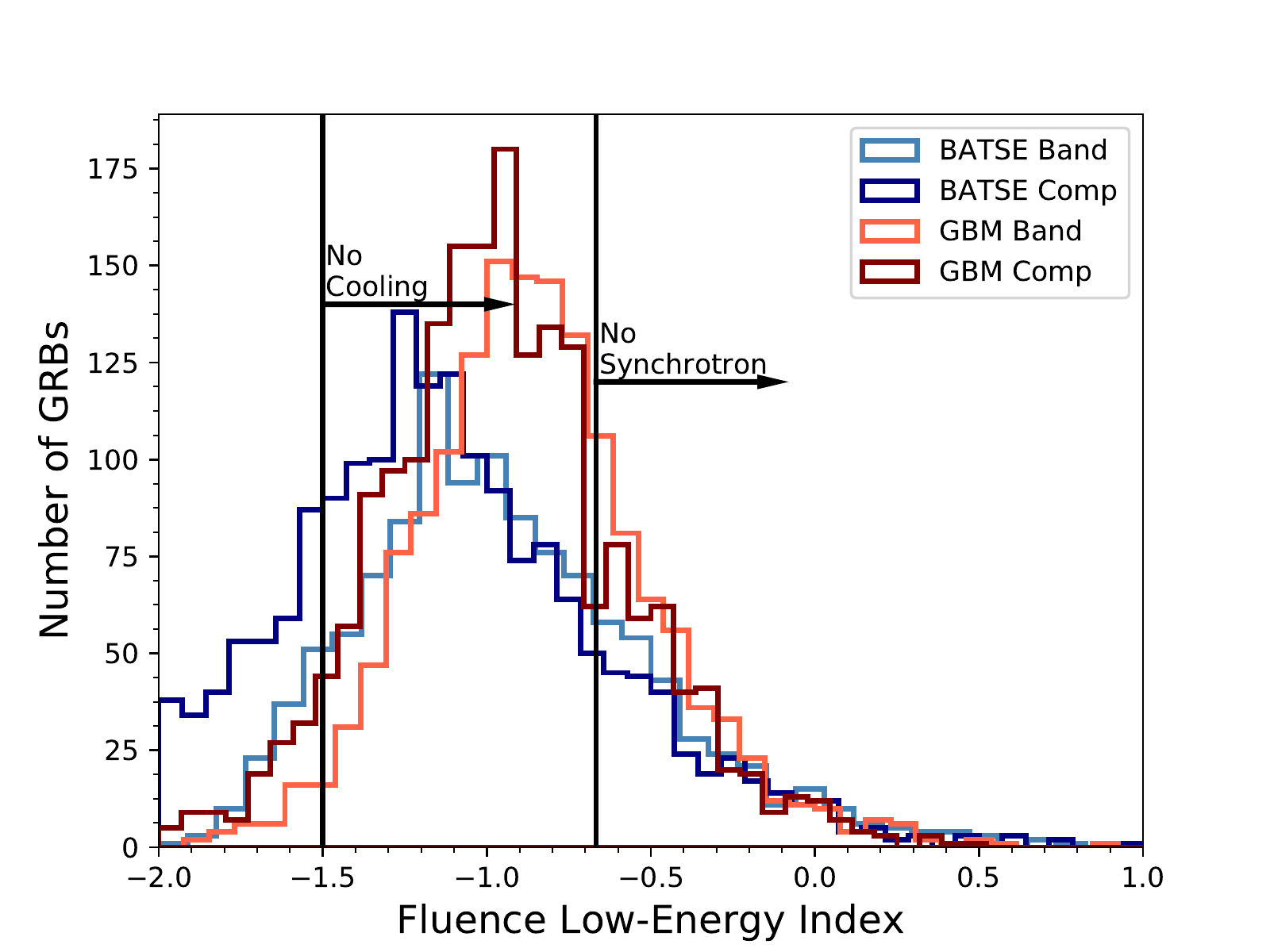}
        \includegraphics[width=0.48\textwidth]{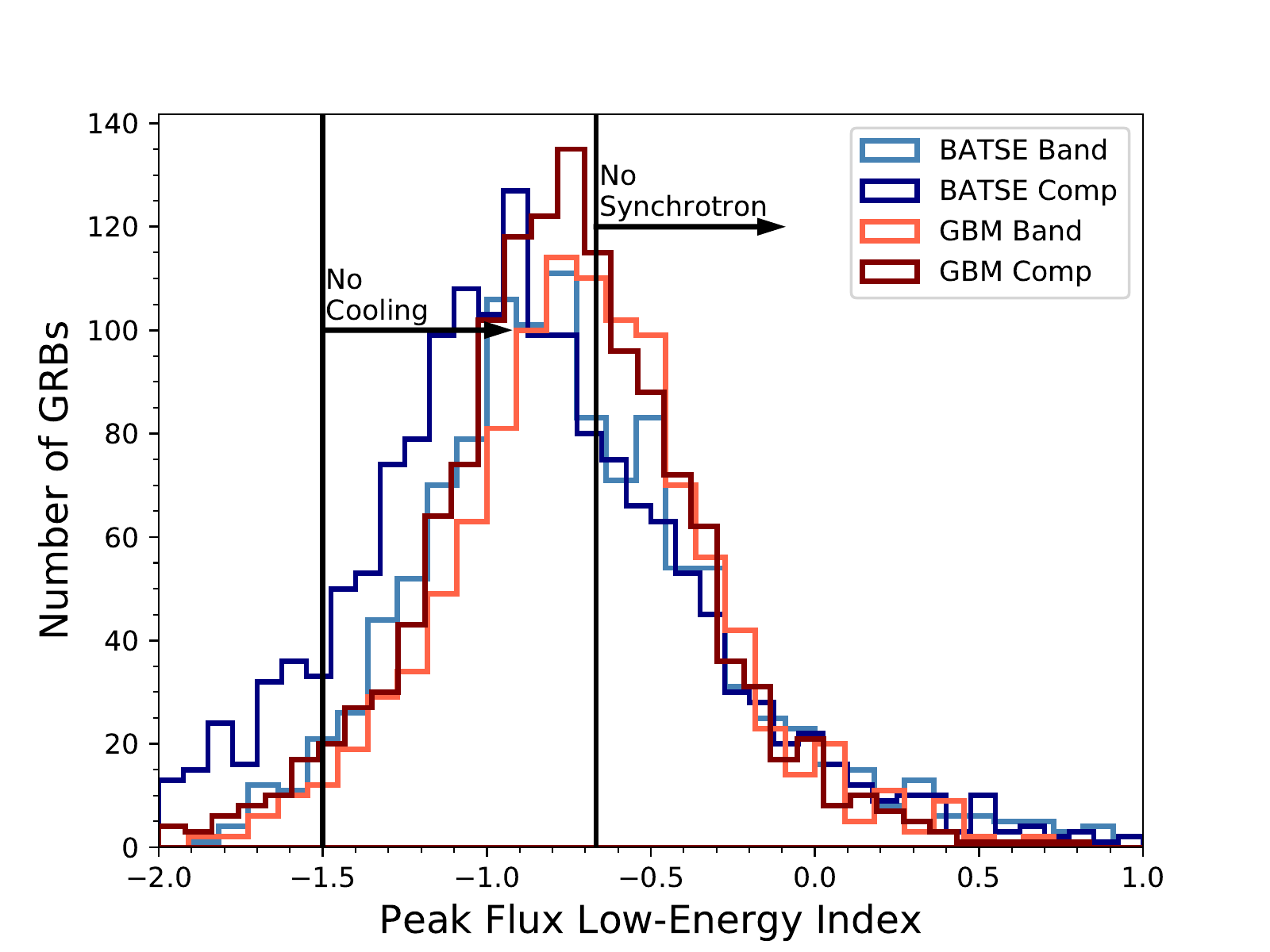}
    \caption{Comparison of the low-energy power-law index as measured by GBM to that measured by BATSE.  GBM-observed GRBs tend to have a slightly harder alpha, leading to an even larger violation of the synchrotron line-of-death.
    \label{fig:AlphaComparison}}
\end{figure}

Of particular interest for determining the emission mechanism that converts the bulk relativistic outflow into radiation is the low-energy power-law index.  Under the assumption that the emission is dominated by synchrotron radiation, the low-energy photon index should be no harder than $-2/3$ in the case of non-adiabatic cooling, and no harder than $-3/2$ in the case of adiabatic cooling~\citep{RybickiLightman79, Katz94}.  As has been widely noted previously~\citep[e.g.][]{Preece_1998a, Gonzalez03,Medvedev06}, BATSE-detected GRBs had a significant fraction of events that have measured low-energy indices that violate these conditions, and have been termed the synchrotron `line-of-death' problem.  A comparison of the GBM data with the BATSE data, shown in Figure~\ref{fig:AlphaComparison}, indicates some overall agreement in the distribution of the low-energy index, however GBM, on the average, measures a slightly harder index than what was measured in BATSE.  The GBM response extends down to $\sim$8 keV, whereas the BATSE response extended to $\sim$20 keV, therefore, the measurement of the low-energy index by GBM is likely more conclusive in most cases.  This leads to an increasingly worrisome case for synchrotron radiation as the primary emission mechanism because fewer GBM bursts are compatible with that interpretation.

\begin{figure}
    \centering
        \includegraphics[width=0.48\textwidth]{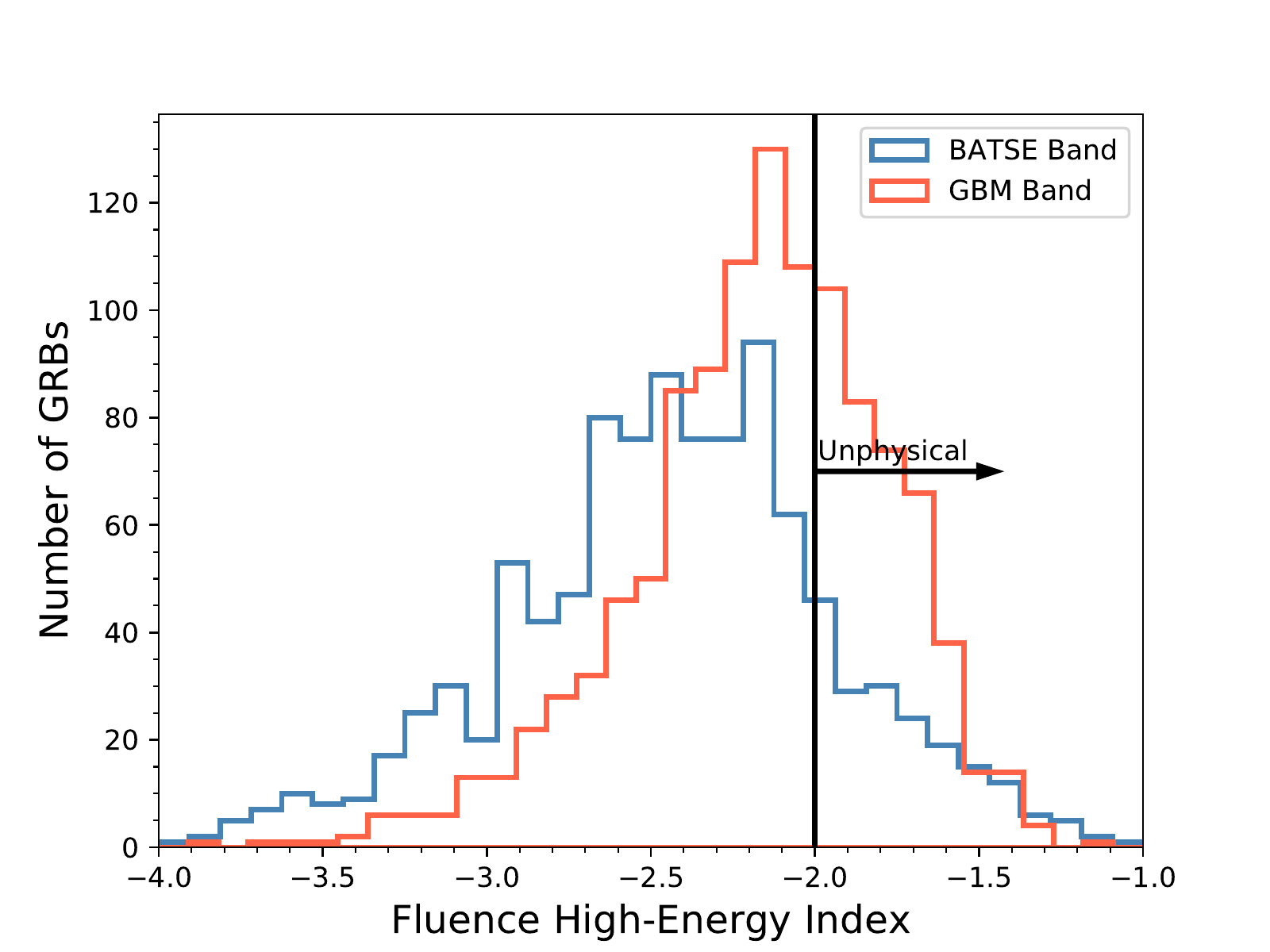}
        \includegraphics[width=0.48\textwidth]{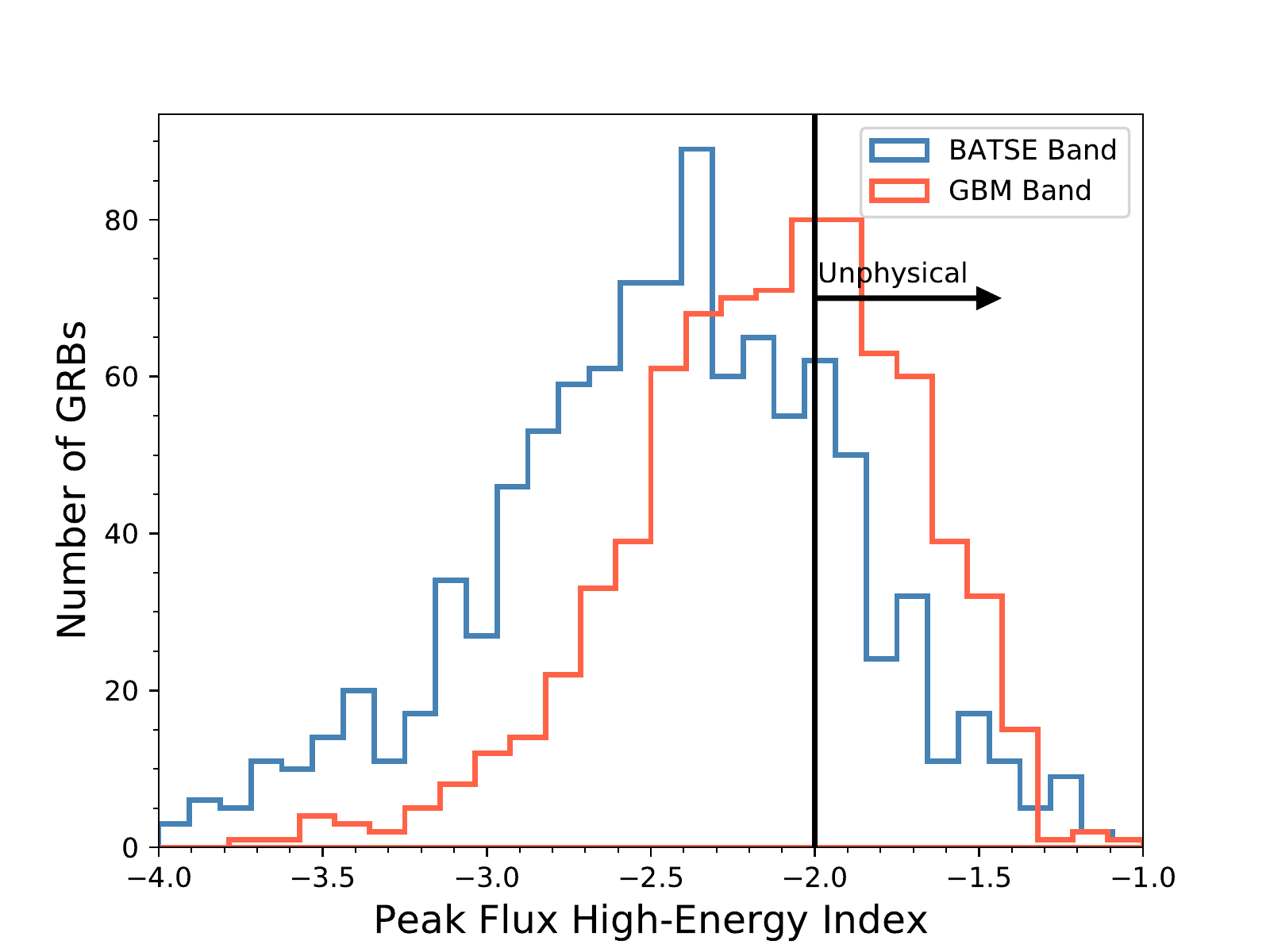}
    \caption{Comparison of the high-energy power-law index as measured by GBM to that measured by BATSE.  The high-energy index measured by GBM appears to be generally harder and thus a larger fraction represent an unphysical power law.
    \label{fig:BetaComparison}}
\end{figure}

Another interesting comparison between the BATSE and GBM bursts is the high-energy power-law index. Figure~\ref{fig:BetaComparison} shows that the GBM measurement of the high-energy index is generally shifted towards harder spectra compared to BATSE.  This creates an issue for a sizable fraction of GRBs because the Band function becomes unphysical at a $\beta \geq -2$ and leads to an infinite flux if extrapolated in energy.  Previous studies of early GBM GRBs also detected by the \textit{Fermi} LAT have shown that in many cases, the high-energy index is biased toward harder values for GBM data~\citep{Ackermann12}. Therefore, the shift in the GBM distribution may result in an issue with the fitting of the spectrum rather than an insight into the true spectrum.  Although GBM has an energy range that extends far beyond the data used in the BATSE 5B catalog (40 MeV vs. 2 MeV), the much smaller effective area of GBM may contribute to the bias in fitting the spectral indices.

\begin{figure}
    \centering
        \includegraphics[width=0.48\textwidth]{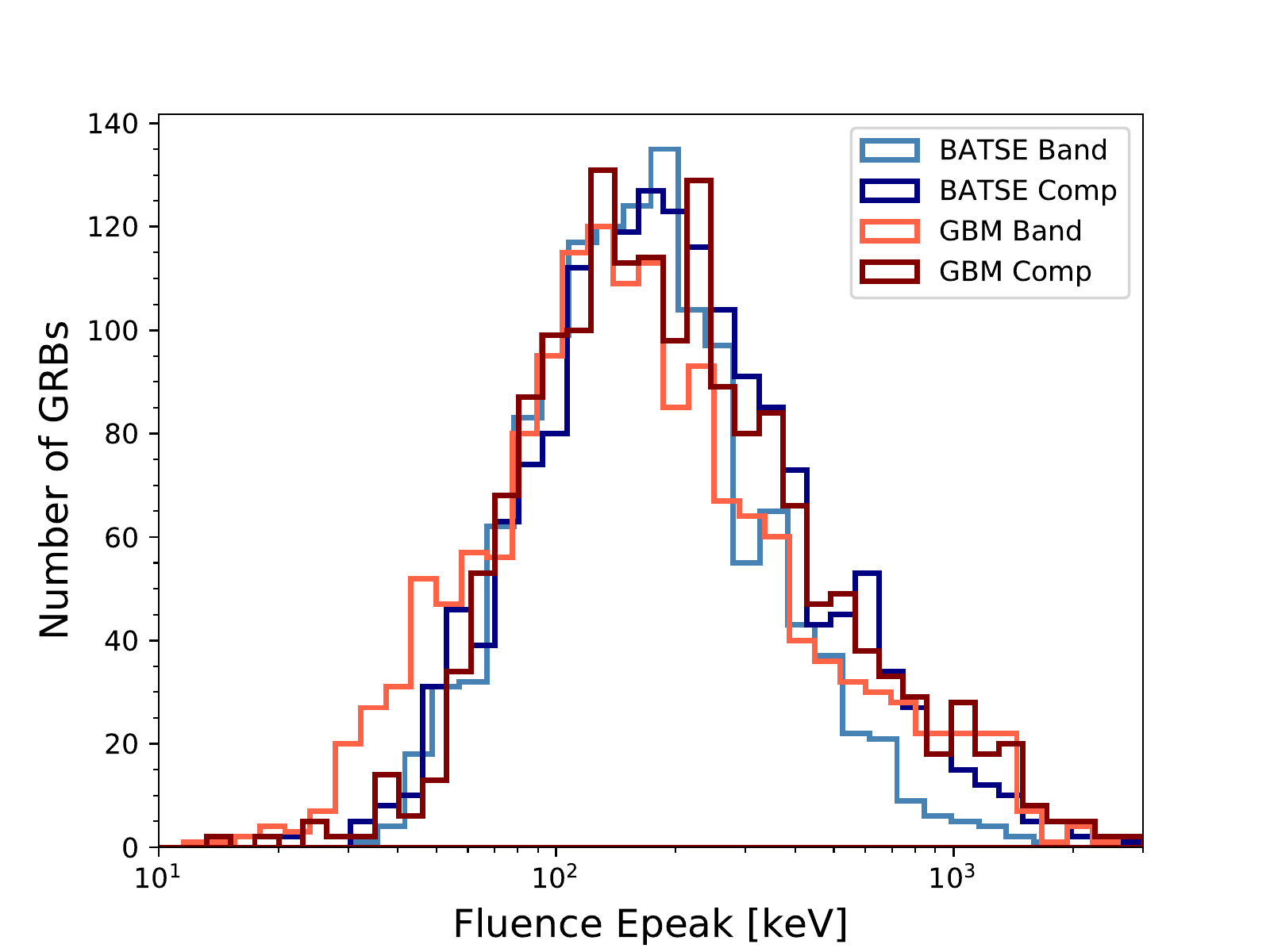}
        \includegraphics[width=0.48\textwidth]{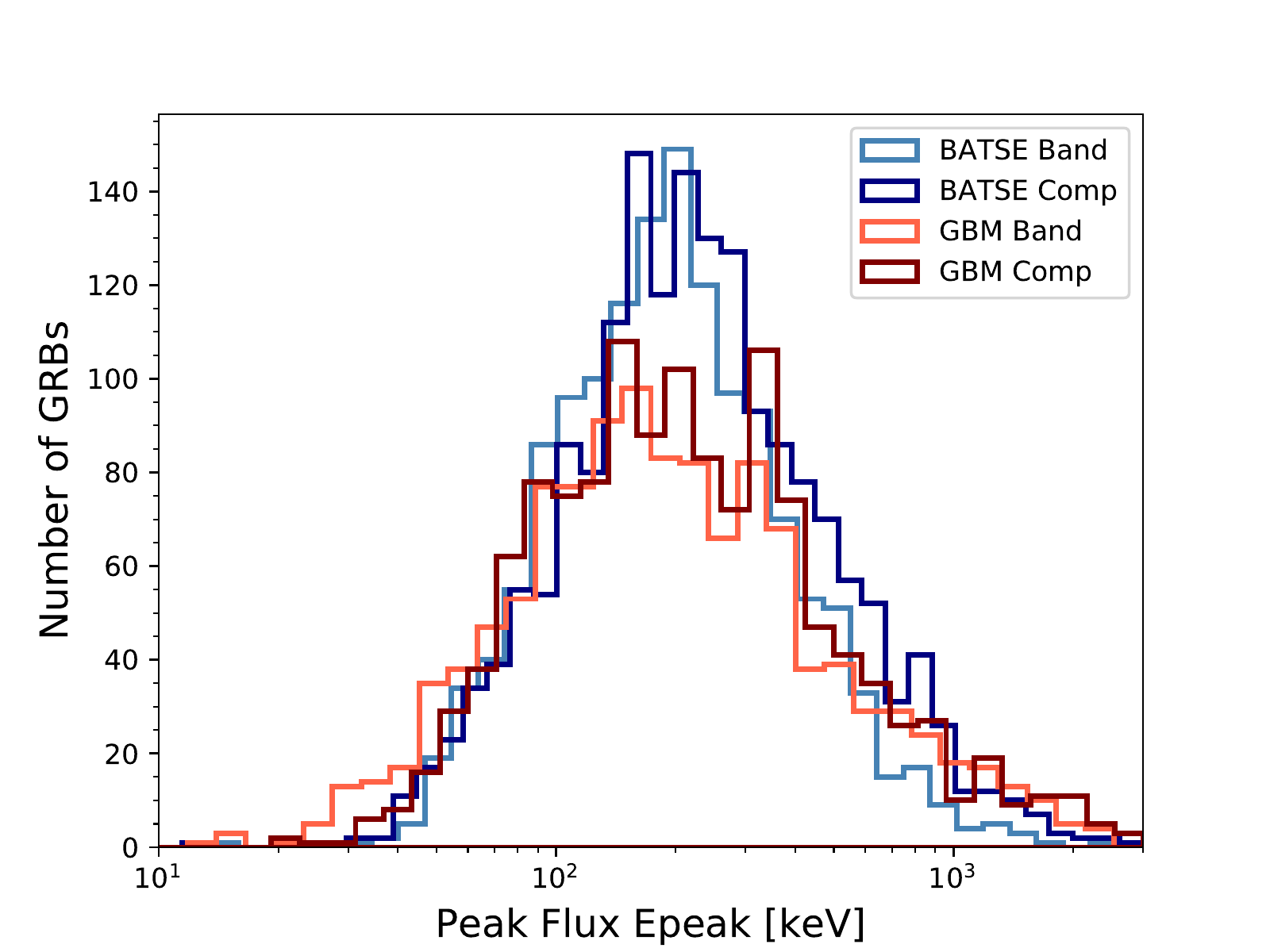}
    \caption{Comparison of $E_{\rm peak}$ as measured by GBM to that measured by BATSE.  The Band function results in a broader distribution of $E_{\rm peak}$ for GBM, expanding toward lower energies.
    \label{fig:EpeakComparison}}
\end{figure}

A spectral feature of GRBs that has been of particular interest to the community is the $E_{\rm peak}$, since it was previously thought to be an indicator used to standardize GRB energetics for purposes of studying cosmology~\citep{Lloyd00, Yonetoku04, Amati06, Ghirlanda07}.  In Figure~\ref{fig:EpeakComparison} we show the comparison of $E_{\rm peak}$ measurements between GBM and BATSE.  These distributions broadly agree, although it is clear that the larger energy range of the GBM allows the measurement of $E_{\rm peak}$ down to $\sim 10$ keV and an expanded population of GRBs with $E_{peak}$ values in the MeV range.

\section{Summary}

The third GBM spectral catalog includes 2297 GRBs detected by GBM during its first 10 years of operation. 
The spectral properties presented here are from time-integrated and peak-flux analysis,
produced using four photon models which were chosen based on their empirical importance to the shape of many GRB spectra. 
The analysis of each burst was performed as objectively as possible, in an attempt to minimize biased systematic 
errors inherent in a subjective analyses. We have described subsets of the full results in the form of data cuts 
based on parameter uncertainties (\textit{GOOD} models), as well as employing model comparison techniques to select the most 
statistically preferred model for each GRB (\textit{BEST} models).

We have illustrated alternative means to classify bursts as long or short; based on their accumulation times (Figure \ref{fig:Accum_times})
and using the \epeak/Fluence ratio (Figure \ref{fig:Epeak_Fluence_Comp}). These plots, alongside the classical $T_{90}$ distribution \citep{vonKienlin+20GBM10yrcat}, highlight the robustness of bimodalilty observed in GRB distribution.
The parameter distributions shown here are largely similar to those in previous studies \citep{Goldstein_2012, Gruber_2014} 
yet contain some important differences. Importantly, the energy ratio technique can be implemented 
solely from parameter values found in this Catalog. Bursts with energy ratios $> 1$ are very likely to 
belong to the short class of bursts. It is also not tied up with issues derived from 
$T_{90}$, which omits $10\%$ of the burst fluence. This is seen clearly in Figure 
\ref{fig:Accum_t90_scatterplot} for those values where $T_{90}$ is {\em less} than the accumulation 
time ($T_{90} > $ accumulation time is expected, as the latter doesn't count quiescent portions of 
the burst).

The \textit{Probability Density Histograms} presented in section \ref{sec:Parameter Distributions}
capture the two-tailed uncertainties of the fitted parameters; Figures \ref{fig:Epeak_Error}, \ref{fig:Band_Beta_Error} and
\ref{fig:SBPL_Index2_Error}, 
depicting the CDF for errors of $E_{peak}$, BAND $\beta$ and SBPL $\lambda_{2}$ respectively, showcase the 
differences in distribution of positive and negative uncertainties for these parameters. The \textit{GOOD} criteria cutoffs
(sec. \ref{sec:Good_Best}) has been altered, accounting for the introduction of asymmetric uncertainties.

The current models for GRB prompt emission can be split into two categories: magnetic (e.g., \citet{Lee_2000}) or internal/external
shock driven (e.g., \citet{Rees&Meszaros_1992}. The $\Delta S$ distribution is hence an interesting result; as comparing it to
predictions made by the SSM provides useful insights into the emission mechanisms of GRBs.
The results obtained here are compatible and in line with results obtained by \citet{Preece_2002} and previous GBM catalogs
\citep{Goldstein_2012, Gruber_2014}. Thus, we conclude that the predictions of the SSM model, in its simplest 
form, are not reconcilable by observations made by GBM.
In its 10 years of operation, GBM has observed 130 GRBs with known redshifts, hence providing one of the largest samples of rest-frame
properties ($E^{rest}_{peak}$, $E_{iso}$ and $L_{iso}$) to date. This helps us assess our current understanding of the central engine
and emission physics of a GRB. 

This catalog should be treated as a starting point for future research on 
interesting bursts and ideas. As has been the case in previous GRB spectral catalogs, we 
hope this catalog will be of great assistance and importance to the search for the
physical properties of GRBs and other related studies.

\section{Acknowledgments}
The UAH coauthors gratefully acknowledge NASA funding from cooperative agreement NNM11AA01A.
The USRA coauthors acknowledge NASA funding through co-operative agreement NNM13AA43C.
C.M. is supported by an appointment to the NASA Postdoctoral Program at the Marshall
Space Flight Center, administered by Universities Space Research Association under
contract with NASA.
P.V. acknowledges support from NASA grant 80NSSC19K0595.
Support for the German contribution to GBM was provided by the Bundesministerium
f{\"u}r Bildung und Forschung (BMBF) via the Deutsches Zentrum f{\"u}r Luft und
Raumfahrt (DLR) under grant number 50 QV 0301.

\newpage
\begin{longrotatetable}
\begin{deluxetable}{lccccccccc}
\tabletypesize{\footnotesize}
\tablewidth{550pt}
\movetabledown=.5in
\tablecaption{ \textit{k}-correction, isotropic energy, isotropic luminosity and rest peak energy for those GRBs with redshift and the parameters from the BAND and COMP spectral models. \label{tab:comp_results}}
 
\tablehead{  &\multicolumn{4}{c}{COMP} & & \multicolumn{4}{c}{BAND}\\ \cline{2-5} \cline{7-10}
\colhead{ID} & \colhead{\textit{k}} & \colhead{E$_{iso}$} & \colhead{L$_{iso}$} & \colhead{E$^{rest}_{peak}$} & & \colhead{\textit{k}} & \colhead{E$_{iso}$} & \colhead{ L$_{iso}$} & \colhead{E$^{rest}_{peak}$} \\
 &  &  \colhead{(erg)} & \colhead{(erg s$^{-1}$)} & \colhead{(keV)} & & & \colhead{(erg)} & \colhead{(erg s$^{-1}$)}  & \colhead{(keV)}}

\startdata
GRB 080804972& 1.546& 2.25e+53 $\pm$ 1.19e+52& 3.70e+52 $\pm$ 6.15e+51&7.08e+02 $\pm$ 2.30e+01 &&1.041& 1.43e+53 $\pm$ 6.08e+51&2.24e+52 $\pm$ 2.35e+51&9.91e+02 $\pm$ 2.42e+01\\ 
GRB 080810549& 1.204& 4.67e+53 $\pm$ 1.57e+52&1.12e+53 $\pm$ 1.09e+52&2.56e+03 $\pm$ 8.58e+01 &&1.235& 4.79e+53 $\pm$ 1.56e+52&1.14e+53 $\pm$ 1.11e+52&2.57e+03 $\pm$ 7.52e+01\\ 
GRB 080905499& 1.976& 3.84e+49 $\pm$ 1.17e+51&2.54e+50 $\pm$ 7.74e+51&3.56e+02 $\pm$ 5.25e+01 &&1.019& 1.83e+49 $\pm$ 2.40e+48&1.32e+50 $\pm$ 2.08e+49&3.92e+02 $\pm$ 5.53e+01\\ 
GRB 080905705& 1.240& 4.03e+52 $\pm$ 1.77e+53&1.49e+52 $\pm$ 6.58e+52&6.15e+02 $\pm$ 5.15e+01 &&1.042& 3.08e+52 $\pm$ 3.44e+51&1.23e+52 $\pm$ 3.18e+51&6.71e+02 $\pm$ 3.14e+01\\ 
GRB 080916009& 1.279& 4.46e+54 $\pm$ 7.66e+52&1.01e+54 $\pm$ 4.87e+52&3.57e+03 $\pm$ 4.17e+01 &&1.275& 4.60e+54 $\pm$ 5.18e+52&1.03e+54 $\pm$ 4.29e+52&4.11e+03 $\pm$ 4.13e+01\\ 
GRB 080916406& 1.904& 3.51e+52 $\pm$ 2.05e+51&3.92e+51 $\pm$ 4.89e+50&1.79e+02 $\pm$ 2.04e+01 &&1.085& 1.75e+52 $\pm$ 7.83e+50&2.21e+51 $\pm$ 2.19e+50&4.19e+02 $\pm$ 2.72e+01\\ 
GRB 080928628& 1.719& 2.76e+52 $\pm$ 3.96e+51&7.86e+51 $\pm$ 2.76e+51&1.33e+02 $\pm$ 1.68e+02 &&1.669& 2.61e+52 $\pm$ 2.34e+51&7.12e+51 $\pm$ 1.12e+51&2.90e+03 $\pm$ 3.19e+03\\ 
GRB 081007224& 1.561& 1.42e+51 $\pm$ 5.15e+50&4.29e+50 $\pm$ 1.65e+50&5.76e+01 $\pm$ 5.47e+00 &&1.391& 1.25e+51 $\pm$ 1.01e+50&3.82e+50 $\pm$ 5.78e+49&5.72e+01 $\pm$ 5.19e+00\\ 
GRB 081008832& 1.352& 1.14e+53 $\pm$ 1.65e+52&1.44e+52 $\pm$ 5.03e+51&4.96e+02 $\pm$ 3.05e+01 &&1.079& 8.62e+52 $\pm$ 6.14e+51&1.11e+52 $\pm$ 2.04e+51&6.82e+02 $\pm$ 3.30e+01\\ 
GRB 081109293& 2.016& 3.90e+52 $\pm$ 2.88e+51&4.15e+51 $\pm$ 6.27e+50&5.66e+01 $\pm$ 1.10e+01 &&2.150& 4.13e+52 $\pm$ 1.87e+51&3.90e+51 $\pm$ 6.12e+50&9.21e+06 $\pm$ 1.66e+09\\ 
GRB 081121858& 1.350& 3.32e+53 $\pm$ 1.58e+52&1.22e+53 $\pm$ 8.76e+51&5.65e+02 $\pm$ 1.44e+01 &&1.034& 2.41e+53 $\pm$ 8.14e+51&9.13e+52 $\pm$ 7.60e+51&8.61e+02 $\pm$ 1.50e+01\\ 
GRB 081221681& 1.083& 4.44e+53 $\pm$ 7.83e+51&1.20e+53 $\pm$ 4.40e+51&2.83e+02 $\pm$ 1.33e+00 &&1.090& 4.33e+53 $\pm$ 3.51e+51&1.18e+53 $\pm$ 2.89e+51&2.88e+02 $\pm$ 1.11e+00\\ 
GRB 081222204& 1.213& 2.87e+53 $\pm$ 1.17e+52&1.53e+53 $\pm$ 9.20e+51&5.55e+02 $\pm$ 8.43e+00 &&1.053& 2.27e+53 $\pm$ 5.61e+51&1.27e+53 $\pm$ 5.88e+51&6.71e+02 $\pm$ 8.14e+00\\ 
GRB 090102122& 1.139& 2.65e+53 $\pm$ 1.88e+52&6.05e+52 $\pm$ 4.80e+51&1.07e+03 $\pm$ 1.80e+01 &&1.119& 2.59e+53 $\pm$ 4.02e+51&5.95e+52 $\pm$ 2.23e+51&1.06e+03 $\pm$ 1.72e+01\\ 
GRB 090113778& 1.390& 1.88e+52 $\pm$ 4.57e+51&1.61e+52 $\pm$ 4.59e+51&3.92e+02 $\pm$ 3.08e+01 &&1.111& 1.35e+52 $\pm$ 1.42e+51&1.27e+52 $\pm$ 1.57e+51&5.03e+02 $\pm$ 3.73e+01\\ 
GRB 090323002& 1.240& 4.86e+54 $\pm$ 1.15e+53&5.69e+53 $\pm$ 2.70e+52&2.07e+03 $\pm$ 2.36e+01 &&1.189& 4.73e+54 $\pm$ 5.09e+52&5.43e+53 $\pm$ 2.24e+52&2.20e+03 $\pm$ 1.86e+01\\ 
GRB 090328401& 1.521& 1.30e+53 $\pm$ 9.39e+51&1.76e+52 $\pm$ 1.36e+51&1.13e+03 $\pm$ 4.38e+01 &&1.358& 1.18e+53 $\pm$ 1.34e+51&1.58e+52 $\pm$ 4.40e+50&1.24e+03 $\pm$ 3.93e+01\\ 
GRB 090423330& 1.047& 7.58e+52 $\pm$ 1.51e+52&1.63e+53 $\pm$ 8.66e+52&6.10e+02 $\pm$ 1.08e+01 &&1.085& 6.44e+52 $\pm$ 6.46e+51&1.58e+53 $\pm$ 2.66e+52&6.58e+02 $\pm$ 8.70e+00\\ 
GRB 090424592& 1.171& 4.79e+52 $\pm$ 1.49e+51&1.75e+52 $\pm$ 5.55e+50&2.46e+02 $\pm$ 3.98e+00 &&1.067& 4.21e+52 $\pm$ 3.77e+50&1.58e+52 $\pm$ 1.95e+50&2.63e+02 $\pm$ 2.97e+00\\ 
GRB 090510016& 4.247& 4.65e+52 $\pm$ 2.11e+51&3.64e+53 $\pm$ 2.06e+52&8.07e+03 $\pm$ 4.40e+02 &&4.116& 4.47e+52 $\pm$ 1.06e+51&3.52e+53 $\pm$ 1.49e+52&8.98e+03 $\pm$ 3.49e+02\\ 
GRB 090516353& 1.389& 1.18e+54 $\pm$ 6.35e+52&1.39e+53 $\pm$ 1.73e+52&7.26e+02 $\pm$ 2.65e+01 &&1.257& 1.04e+54 $\pm$ 4.00e+52&1.33e+53 $\pm$ 2.10e+52&8.37e+02 $\pm$ 1.95e+01\\ 
GRB 090519881& 1.639& 2.62e+53 $\pm$ 2.19e+52&1.02e+53 $\pm$ 2.08e+52&7.82e+03 $\pm$ 5.42e+02 &&1.606& 2.57e+53 $\pm$ 1.64e+52&1.00e+53 $\pm$ 1.94e+52&7.69e+03 $\pm$ 5.16e+02\\ 
GRB 090618353& 1.370& 3.27e+53 $\pm$ 3.59e+51&2.48e+52 $\pm$ 5.21e+50&2.30e+02 $\pm$ 3.29e+00 &&1.104& 2.59e+53 $\pm$ 1.61e+51&2.02e+52 $\pm$ 3.58e+50&3.23e+02 $\pm$ 3.30e+00\\ 
GRB 090902462& 1.593& 4.00e+54 $\pm$ 1.51e+52&7.88e+53 $\pm$ 7.45e+51&2.98e+03 $\pm$ 1.63e+01 &&1.599& 4.02e+54 $\pm$ 1.16e+52&7.91e+53 $\pm$ 7.22e+51&2.98e+03 $\pm$ 1.57e+01\\ 
GRB 090926181& 1.311& 2.47e+54 $\pm$ 3.11e+52&9.28e+53 $\pm$ 1.47e+52&1.04e+03 $\pm$ 5.84e+00 &&1.089& 2.09e+54 $\pm$ 1.08e+52&7.92e+53 $\pm$ 8.77e+51&1.17e+03 $\pm$ 4.57e+00\\ 
GRB 090926914& 1.052& 4.06e+52 $\pm$ 2.40e+51&4.15e+51 $\pm$ 9.12e+50&1.84e+02 $\pm$ 2.56e+00 &&1.019& 3.55e+52 $\pm$ 7.89e+50&3.94e+51 $\pm$ 3.53e+50&1.90e+02 $\pm$ 2.05e+00\\ 
GRB 090927422& 1.052& 4.91e+52 $\pm$ 2.89e+51&5.31e+51 $\pm$ 1.17e+51&1.95e+02 $\pm$ 2.56e+00 &&1.026& 1.21e+51 $\pm$ 2.29e+50&7.29e+51 $\pm$ 1.45e+51&4.17e+02 $\pm$ 4.14e+01\\ 
GRB 091003191& 1.459& 1.25e+53 $\pm$ 8.38e+51&5.70e+52 $\pm$ 3.88e+51&7.03e+02 $\pm$ 2.66e+01 &&1.156& 1.02e+53 $\pm$ 1.58e+51&4.77e+52 $\pm$ 8.85e+50&8.21e+02 $\pm$ 2.11e+01\\ 
GRB 091020900& 1.255& 8.59e+52 $\pm$ 6.67e+53&3.54e+52 $\pm$ 2.75e+53&6.19e+02 $\pm$ 4.88e+01 &&1.112& 7.50e+52 $\pm$ 3.77e+51&3.44e+52 $\pm$ 2.45e+51&6.61e+02 $\pm$ 2.75e+01\\ 
GRB 091024372& 1.662& 5.69e+52 $\pm$ 4.40e+51&9.49e+51 $\pm$ 2.20e+51&3.63e+03 $\pm$ 9.02e+02 &&1.909& 6.52e+52 $\pm$ 3.71e+51&1.09e+52 $\pm$ 1.89e+51&3.64e+03 $\pm$ 8.96e+02\\ 
GRB 091127976& 1.510& 1.97e+52 $\pm$ 5.40e+50&8.91e+51 $\pm$ 2.61e+50&5.28e+01 $\pm$ 1.55e+00 &&1.467& 1.70e+52 $\pm$ 1.99e+50&7.51e+51 $\pm$ 9.91e+49&8.71e+01 $\pm$ 1.61e+00\\ 
GRB 091208410& 1.533& 4.00e+52 $\pm$ 2.17e+51&2.84e+52 $\pm$ 1.59e+51&9.21e+01 $\pm$ 1.28e+01 &&1.174& 2.48e+52 $\pm$ 1.17e+51&2.04e+52 $\pm$ 8.68e+50&2.56e+02 $\pm$ 1.31e+01\\ 
GRB 100117879& 1.212& 1.16e+51 $\pm$ 7.20e+50&1.09e+52 $\pm$ 7.07e+51&6.28e+02 $\pm$ 5.29e+01 &&1.020& 9.75e+50 $\pm$ 1.49e+50&8.70e+51 $\pm$ 1.95e+51&6.25e+02 $\pm$ 5.11e+01\\ 
GRB 100206563& 1.977& 8.83e+50 $\pm$ 3.00e+50&1.27e+52 $\pm$ 4.33e+51&6.39e+02 $\pm$ 6.36e+01 &&1.148& 5.40e+50 $\pm$ 3.42e+49&7.77e+51 $\pm$ 4.91e+50&7.48e+02 $\pm$ 7.17e+01\\ 
GRB 100414097& 1.567& 7.99e+53 $\pm$ 3.82e+52&1.10e+53 $\pm$ 5.98e+51&1.57e+03 $\pm$ 1.54e+01 &&1.282& 6.56e+53 $\pm$ 4.69e+51&9.24e+52 $\pm$ 2.23e+51&1.58e+03 $\pm$ 1.46e+01\\ 
GRB 100615083& 1.639& 1.26e+53 $\pm$ 4.42e+51&1.81e+52 $\pm$ 1.35e+51&1.28e+02 $\pm$ 7.50e+00 &&1.157& 6.77e+52 $\pm$ 2.25e+51&9.71e+51 $\pm$ 5.30e+50&3.43e+02 $\pm$ 1.04e+01\\ 
GRB 100625773& 1.290& 9.87e+50 $\pm$ 1.30e+50&5.88e+51 $\pm$ 9.92e+50&7.00e+02 $\pm$ 6.19e+01 &&1.126& 8.61e+50 $\pm$ 6.47e+49&5.88e+51 $\pm$ 7.79e+50&7.02e+02 $\pm$ 6.33e+01\\ 
GRB 100724029& 1.664& 2.01e+54 $\pm$ 2.40e+52&1.21e+53 $\pm$ 2.88e+51&8.20e+02 $\pm$ 8.16e+00 &&1.159& 1.46e+54 $\pm$ 8.38e+51&8.99e+52 $\pm$ 1.78e+51&1.11e+03 $\pm$ 7.70e+00\\ 
GRB 100728095& 1.257& 1.08e+54 $\pm$ 3.44e+52&6.89e+52 $\pm$ 3.64e+51&7.45e+02 $\pm$ 7.82e+00 &&1.041& 8.95e+53 $\pm$ 9.80e+51&5.95e+52 $\pm$ 2.44e+51&7.99e+02 $\pm$ 6.11e+00\\ 
GRB 100728439& 1.892& 9.35e+52 $\pm$ 2.57e+53&6.13e+52 $\pm$ 1.68e+53&6.26e+01 $\pm$ 6.03e+04 &&1.069& 3.50e+52 $\pm$ 2.42e+51&2.05e+52 $\pm$ 2.09e+51&4.76e+02 $\pm$ 1.85e+01\\ 
GRB 100814160& 1.249& 1.07e+53 $\pm$ 9.32e+51&1.87e+52 $\pm$ 1.96e+51&3.31e+02 $\pm$ 1.05e+01 &&1.022& 7.60e+52 $\pm$ 2.71e+51&1.14e+52 $\pm$ 1.15e+51&3.81e+02 $\pm$ 8.03e+00\\ 
GRB 100816026& 1.175& 8.93e+51 $\pm$ 7.52e+50&7.16e+51 $\pm$ 6.09e+50&2.41e+02 $\pm$ 7.08e+00 &&1.019& 6.79e+51 $\pm$ 2.24e+50&5.44e+51 $\pm$ 2.06e+50&2.54e+02 $\pm$ 6.07e+00\\ 
GRB 100906576& 1.173& 2.43e+53 $\pm$ 5.28e+52&6.16e+52 $\pm$ 1.36e+52&6.11e+02 $\pm$ 2.28e+01 &&1.155& 2.39e+53 $\pm$ 8.52e+51&5.58e+52 $\pm$ 3.40e+51&6.11e+02 $\pm$ 2.14e+01\\ 
GRB 101213451& 1.423& 9.57e+51 $\pm$ 3.12e+51&7.98e+50 $\pm$ 2.88e+50&4.93e+02 $\pm$ 3.38e+01 &&1.081& 7.10e+51 $\pm$ 3.26e+50&5.96e+50 $\pm$ 8.87e+49&4.85e+02 $\pm$ 3.03e+01\\ 
GRB 101219686& 1.420& 3.90e+51 $\pm$ 5.15e+50&4.61e+50 $\pm$ 1.31e+50&9.00e+01 $\pm$ 6.81e+00 &&1.023& 2.01e+51 $\pm$ 1.04e+50&2.70e+50 $\pm$ 4.05e+49&1.28e+02 $\pm$ 4.63e+00\\ 
GRB 110106893& 1.063& 2.43e+51 $\pm$ 1.22e+51&5.50e+50 $\pm$ 3.21e+50&2.12e+02 $\pm$ 1.49e+01 &&1.067& 2.43e+51 $\pm$ 2.07e+50&6.27e+50 $\pm$ 1.76e+50&2.12e+02 $\pm$ 1.45e+01\\ 
GRB 110128073& 1.715& 1.73e+52 $\pm$ 2.59e+53&2.35e+52 $\pm$ 3.53e+53&3.71e+01 $\pm$ 1.55e+03 &&1.794& 1.92e+52 $\pm$ 2.69e+51&2.30e+52 $\pm$ 7.11e+51&1.11e+05 $\pm$ 3.49e+05\\ 
GRB 110213220& 1.461& 9.56e+52 $\pm$ 5.74e+51&2.83e+52 $\pm$ 1.84e+51&2.77e+02 $\pm$ 1.20e+01 &&1.285& 8.41e+52 $\pm$ 3.49e+51&2.49e+52 $\pm$ 1.14e+51&2.77e+02 $\pm$ 1.18e+01\\ 
GRB 110731465& 1.245& 5.72e+53 $\pm$ 3.13e+52&2.37e+53 $\pm$ 1.68e+52&1.23e+03 $\pm$ 1.69e+01 &&1.075& 4.96e+53 $\pm$ 9.07e+51&2.03e+53 $\pm$ 8.08e+51&1.33e+03 $\pm$ 1.40e+01\\ 
GRB 110818860& 1.418& 2.43e+53 $\pm$ 2.00e+52&7.65e+52 $\pm$ 1.00e+52&7.94e+02 $\pm$ 5.79e+01 &&1.152& 1.92e+53 $\pm$ 1.33e+52&7.48e+52 $\pm$ 8.41e+51&1.57e+03 $\pm$ 7.45e+01\\ 
GRB 111107035& 1.474& 1.97e+53 $\pm$ 1.80e+52&5.53e+52 $\pm$ 7.57e+51&7.09e+02 $\pm$ 5.79e+01 &&1.124& 4.40e+52 $\pm$ 6.45e+51&2.06e+52 $\pm$ 5.14e+51&1.03e+03 $\pm$ 7.68e+01\\ 
GRB 111117510& 1.608& 5.35e+51 $\pm$ 1.49e+53&5.97e+52 $\pm$ 1.66e+54&1.16e+03 $\pm$ 1.11e+02 &&1.164& 4.01e+51 $\pm$ 3.53e+50&4.02e+52 $\pm$ 5.67e+51&1.25e+03 $\pm$ 1.03e+02\\ 
GRB 111228657& 1.933& 2.05e+52 $\pm$ 1.44e+51&2.67e+51 $\pm$ 2.42e+50&3.92e+01 $\pm$ 1.25e+00 &&1.722& 1.60e+52 $\pm$ 2.92e+50&2.26e+51 $\pm$ 9.76e+49&4.01e+01 $\pm$ 2.05e+00\\ 
GRB 120118709& 1.108& 6.54e+52 $\pm$ 4.23e+51&1.96e+52 $\pm$ 4.08e+51&1.70e+02 $\pm$ 3.06e+00 &&1.169& 5.76e+52 $\pm$ 2.14e+51&1.82e+52 $\pm$ 2.09e+51&2.19e+02 $\pm$ 3.02e+00\\ 
GRB 120119170& 1.250& 4.24e+53 $\pm$ 2.01e+52&8.54e+52 $\pm$ 4.65e+51&4.99e+02 $\pm$ 1.05e+01 &&1.060& 3.45e+53 $\pm$ 4.72e+51&7.35e+52 $\pm$ 2.57e+51&5.86e+02 $\pm$ 6.07e+00\\ 
GRB 120326056& 1.212& 4.35e+52 $\pm$ 2.68e+51&1.91e+52 $\pm$ 1.84e+51&1.24e+02 $\pm$ 5.59e+00 &&1.215& 3.54e+52 $\pm$ 1.08e+51&1.62e+52 $\pm$ 9.85e+50&1.75e+02 $\pm$ 3.09e+00\\ 
GRB 120624933& 1.498& 3.91e+54 $\pm$ 7.15e+52&3.11e+53 $\pm$ 1.08e+52&2.04e+03 $\pm$ 2.45e+01 &&1.336& 3.61e+54 $\pm$ 2.35e+52&2.87e+53 $\pm$ 6.60e+51&2.31e+03 $\pm$ 2.17e+01\\ 
GRB 120711115& 1.869& 2.17e+54 $\pm$ 1.89e+52&2.07e+53 $\pm$ 4.76e+51&3.18e+03 $\pm$ 4.23e+01 &&1.940& 2.26e+54 $\pm$ 1.14e+52&2.11e+53 $\pm$ 4.52e+51&3.59e+03 $\pm$ 4.59e+01\\ 
GRB 120712571& 1.243& 2.14e+53 $\pm$ 1.29e+52&1.18e+53 $\pm$ 1.44e+52&7.85e+02 $\pm$ 1.96e+01 &&1.060& 1.75e+53 $\pm$ 1.04e+52&8.77e+52 $\pm$ 1.16e+52&1.34e+03 $\pm$ 3.37e+01\\ 
GRB 120716712& 1.144& 2.64e+53 $\pm$ 1.20e+52&7.19e+52 $\pm$ 4.79e+51&4.19e+02 $\pm$ 6.46e+00 &&1.071& 2.26e+53 $\pm$ 5.27e+51&5.92e+52 $\pm$ 3.46e+51&4.59e+02 $\pm$ 5.29e+00\\ 
GRB 120729456& 2.190& 2.16e+52 $\pm$ 2.40e+51&5.46e+51 $\pm$ 7.67e+50&1.24e+03 $\pm$ 4.57e+03 &&2.247& 2.23e+52 $\pm$ 1.11e+51&5.76e+51 $\pm$ 5.60e+50&5.76e+05 $\pm$ 7.78e+06\\ 
GRB 120811649& 1.114& 7.57e+52 $\pm$ 6.20e+51&3.95e+52 $\pm$ 5.27e+51&2.03e+02 $\pm$ 3.93e+00 &&1.156& 7.01e+52 $\pm$ 3.20e+51&3.12e+52 $\pm$ 3.07e+51&2.23e+02 $\pm$ 3.73e+00\\ 
GRB 120907017& 1.054& 2.15e+51 $\pm$ 9.08e+50&2.44e+51 $\pm$ 1.02e+51&2.53e+02 $\pm$ 3.03e+01 &&1.054& 2.08e+51 $\pm$ 3.27e+50&2.42e+51 $\pm$ 4.43e+50&2.38e+02 $\pm$ 2.21e+01\\ 
GRB 120909070& 1.287& 8.09e+53 $\pm$ 3.11e+52&1.67e+53 $\pm$ 2.16e+52&9.84e+02 $\pm$ 2.43e+01 &&1.074& 6.53e+53 $\pm$ 2.59e+52&1.42e+53 $\pm$ 1.49e+52&1.50e+03 $\pm$ 2.86e+01\\ 
GRB 120922393& 1.357& 5.83e+53 $\pm$ 2.46e+52&9.99e+52 $\pm$ 1.30e+52&8.19e+02 $\pm$ 2.43e+01 &&1.075& 4.46e+53 $\pm$ 1.77e+52&8.05e+52 $\pm$ 8.44e+51&1.25e+03 $\pm$ 2.86e+01\\ 
GRB 121011469& 1.982& 1.59e+53 $\pm$ 1.00e+52&4.15e+52 $\pm$ 5.65e+51&1.13e+03 $\pm$ 1.38e+02 &&2.364& 2.11e+53 $\pm$ 7.90e+51&5.40e+52 $\pm$ 5.30e+51&1.53e+04 $\pm$ 1.20e+03\\ 
GRB 121128212& 1.170& 1.55e+53 $\pm$ 6.30e+51&7.84e+52 $\pm$ 7.89e+51&1.92e+02 $\pm$ 3.85e+00 &&1.136& 1.25e+53 $\pm$ 2.81e+51&7.48e+52 $\pm$ 2.94e+51&2.46e+02 $\pm$ 2.87e+00\\ 
GRB 121211574& 1.017& 1.63e+51 $\pm$ 4.62e+50&1.21e+51 $\pm$ 4.79e+50&2.04e+02 $\pm$ 1.39e+01 &&1.026& 1.59e+51 $\pm$ 1.63e+50&1.12e+51 $\pm$ 2.09e+50&1.98e+02 $\pm$ 1.03e+01\\ 
GRB 130215063& 2.434& 4.18e+52 $\pm$ 2.43e+51&2.08e+51 $\pm$ 2.59e+50&4.11e+02 $\pm$ 1.30e+02 &&2.369& 4.42e+52 $\pm$ 1.03e+51&2.11e+51 $\pm$ 2.37e+50&5.39e+03 $\pm$ 1.13e+03\\ 
GRB 130420313& 1.252& 3.64e+52 $\pm$ 6.52e+51&4.31e+51 $\pm$ 5.27e+50&1.32e+02 $\pm$ 3.14e+00 &&1.216& 3.53e+52 $\pm$ 1.27e+51&4.18e+51 $\pm$ 4.36e+50&1.31e+02 $\pm$ 3.04e+00\\ 
GRB 130427324& 1.576& 7.43e+53 $\pm$ 3.76e+51&1.38e+53 $\pm$ 8.06e+50&1.11e+03 $\pm$ 5.45e+00 &&1.485& 7.05e+53 $\pm$ 8.28e+50&1.32e+53 $\pm$ 3.95e+50&1.16e+03 $\pm$ 4.98e+00\\ 
GRB 130518580& 1.375& 2.17e+54 $\pm$ 4.59e+52&8.83e+53 $\pm$ 2.23e+52&1.33e+03 $\pm$ 1.46e+01 &&1.136& 1.85e+54 $\pm$ 1.76e+52&7.86e+53 $\pm$ 1.10e+52&1.58e+03 $\pm$ 1.20e+01\\ 
GRB 130528695& 1.103& 4.67e+52 $\pm$ 1.25e+52&1.33e+52 $\pm$ 3.69e+51&2.73e+02 $\pm$ 6.94e+00 &&1.095& 4.62e+52 $\pm$ 1.47e+51&1.25e+52 $\pm$ 8.00e+50&2.71e+02 $\pm$ 6.65e+00\\ 
GRB 130610133& 1.304& 9.09e+52 $\pm$ 1.06e+52&1.49e+52 $\pm$ 3.17e+51&9.45e+02 $\pm$ 1.60e+02 &&1.289& 8.90e+52 $\pm$ 7.37e+51&1.48e+52 $\pm$ 2.88e+51&8.81e+02 $\pm$ 6.16e+01\\ 
GRB 130612141& 1.194& 7.59e+51 $\pm$ 1.03e+51&7.69e+51 $\pm$ 1.49e+51&8.71e+01 $\pm$ 8.23e+00 &&1.315& 6.86e+51 $\pm$ 6.90e+50&6.88e+51 $\pm$ 1.01e+51&1.74e+02 $\pm$ 1.00e+01\\ 
GRB 130702004& 1.133& 4.75e+50 $\pm$ 1.82e+49&5.21e+49 $\pm$ 3.83e+48&1.20e+01 $\pm$ 1.13e+00 &&3.559& 1.50e+51 $\pm$ 4.26e+49&1.64e+50 $\pm$ 1.21e+49&7.80e+03 $\pm$ 2.75e+02\\ 
GRB 130925173& 1.520& 5.50e+52 $\pm$ 1.64e+51&5.79e+50 $\pm$ 6.35e+49&3.13e+01 $\pm$ 8.09e-01 &&1.353& 4.15e+52 $\pm$ 3.44e+50&5.20e+50 $\pm$ 2.68e+49&1.14e+02 $\pm$ 1.61e+00\\ 
GRB 131004904& 1.241& 8.57e+50 $\pm$ 1.29e+50&2.78e+51 $\pm$ 8.32e+50&2.03e+02 $\pm$ 2.44e+01 &&1.173& 8.10e+50 $\pm$ 9.41e+49&2.65e+51 $\pm$ 7.12e+50&2.03e+02 $\pm$ 2.42e+01\\ 
GRB 131011741& 1.396& 1.63e+53 $\pm$ 1.73e+52&3.39e+52 $\pm$ 4.10e+51&6.25e+02 $\pm$ 4.09e+01 &&1.059& 1.18e+53 $\pm$ 5.05e+51&2.53e+52 $\pm$ 2.77e+51&7.87e+02 $\pm$ 2.43e+01\\ 
GRB 131105087& 1.112& 2.26e+53 $\pm$ 3.02e+52&3.96e+52 $\pm$ 5.64e+51&7.23e+02 $\pm$ 1.83e+01 &&1.119& 2.26e+53 $\pm$ 5.39e+51&3.98e+52 $\pm$ 2.19e+51&7.16e+02 $\pm$ 1.56e+01\\ 
GRB 131108862& 1.274& 7.14e+53 $\pm$ 2.80e+52&2.98e+53 $\pm$ 1.30e+52&1.25e+03 $\pm$ 1.63e+01 &&1.107& 6.31e+53 $\pm$ 8.38e+51&2.75e+53 $\pm$ 7.24e+51&1.35e+03 $\pm$ 1.37e+01\\ 
GRB 131231198& 1.343& 2.55e+53 $\pm$ 3.47e+51&4.00e+52 $\pm$ 6.77e+50&2.92e+02 $\pm$ 4.03e+00 &&1.114& 2.08e+53 $\pm$ 1.01e+51&3.39e+52 $\pm$ 3.82e+50&3.63e+02 $\pm$ 2.90e+00\\ 
GRB 140206304& 1.220& 3.49e+53 $\pm$ 1.27e+52&2.29e+53 $\pm$ 1.03e+52&4.52e+02 $\pm$ 5.83e+00 &&1.014& 2.50e+53 $\pm$ 5.33e+51&1.67e+53 $\pm$ 5.12e+51&5.69e+02 $\pm$ 4.61e+00\\ 
GRB 140213807& 1.320& 1.39e+53 $\pm$ 3.53e+51&3.75e+52 $\pm$ 1.25e+51&1.90e+02 $\pm$ 4.10e+00 &&1.154& 1.06e+53 $\pm$ 1.45e+51&3.04e+52 $\pm$ 5.60e+50&2.51e+02 $\pm$ 3.27e+00\\ 
GRB 140304557& 1.107& 1.32e+53 $\pm$ 1.72e+52&1.37e+53 $\pm$ 2.74e+52&7.71e+02 $\pm$ 3.14e+01 &&1.058& 1.12e+53 $\pm$ 9.40e+51&1.37e+53 $\pm$ 2.87e+52&8.91e+02 $\pm$ 1.85e+01\\ 
GRB 140423356& 1.411& 7.79e+53 $\pm$ 2.87e+52&7.90e+52 $\pm$ 1.03e+52&4.95e+02 $\pm$ 1.59e+01 &&1.054& 5.03e+53 $\pm$ 2.05e+52&4.94e+52 $\pm$ 9.58e+51&1.03e+03 $\pm$ 1.98e+01\\ 
GRB 140506880& 1.095& 1.21e+52 $\pm$ 3.88e+51&1.03e+52 $\pm$ 3.25e+51&3.72e+02 $\pm$ 2.53e+01 &&1.088& 1.21e+52 $\pm$ 8.63e+50&9.39e+51 $\pm$ 5.15e+50&3.73e+02 $\pm$ 2.63e+01\\ 
GRB 140508128& 1.332& 2.70e+53 $\pm$ 9.35e+51&1.07e+53 $\pm$ 3.96e+51&5.23e+02 $\pm$ 1.21e+01 &&1.110& 2.24e+53 $\pm$ 3.24e+51&9.23e+52 $\pm$ 1.68e+51&5.88e+02 $\pm$ 1.09e+01\\ 
GRB 140512814& 1.282& 9.41e+52 $\pm$ 5.87e+52&8.69e+51 $\pm$ 5.43e+51&1.20e+03 $\pm$ 5.82e+01 &&1.363& 1.00e+53 $\pm$ 1.61e+51&9.24e+51 $\pm$ 3.86e+50&1.20e+03 $\pm$ 5.71e+01\\ 
GRB 140606133& 1.713& 6.76e+51 $\pm$ 2.11e+51&2.32e+51 $\pm$ 7.31e+50&7.37e+02 $\pm$ 1.13e+02 &&1.283& 5.17e+51 $\pm$ 2.29e+50&1.76e+51 $\pm$ 8.41e+49&7.97e+02 $\pm$ 1.03e+02\\ 
GRB 140620219& 1.313& 1.07e+53 $\pm$ 6.07e+51&2.82e+52 $\pm$ 3.25e+51&2.11e+02 $\pm$ 1.07e+01 &&1.146& 8.19e+52 $\pm$ 3.78e+51&1.97e+52 $\pm$ 1.68e+51&3.96e+02 $\pm$ 1.22e+01\\ 
GRB 140623224& 1.172& 4.13e+52 $\pm$ 4.61e+51&9.11e+51 $\pm$ 1.75e+51&9.54e+02 $\pm$ 1.38e+02 &&1.198& 4.18e+52 $\pm$ 4.34e+51&8.65e+51 $\pm$ 1.66e+51&9.26e+02 $\pm$ 8.26e+01\\ 
GRB 140703026& 1.193& 2.63e+53 $\pm$ 2.49e+52&8.04e+52 $\pm$ 9.89e+51&8.65e+02 $\pm$ 3.47e+01 &&1.118& 2.44e+53 $\pm$ 1.12e+52&7.38e+52 $\pm$ 7.54e+51&9.05e+02 $\pm$ 2.32e+01\\ 
GRB 140801792& 1.037& 6.67e+52 $\pm$ 2.98e+51&3.28e+52 $\pm$ 1.66e+51&2.77e+02 $\pm$ 2.64e+00 &&1.026& 6.40e+52 $\pm$ 8.24e+50&3.11e+52 $\pm$ 7.36e+50&2.81e+02 $\pm$ 2.11e+00\\ 
GRB 140808038& 1.087& 1.01e+53 $\pm$ 7.58e+51&1.43e+53 $\pm$ 1.26e+52&5.04e+02 $\pm$ 6.46e+00 &&1.029& 8.65e+52 $\pm$ 3.04e+51&1.26e+53 $\pm$ 6.68e+51&5.38e+02 $\pm$ 6.15e+00\\ 
GRB 140907672& 1.076& 2.97e+52 $\pm$ 8.09e+51&4.23e+51 $\pm$ 1.29e+51&3.08e+02 $\pm$ 1.03e+01 &&1.074& 2.94e+52 $\pm$ 1.04e+51&3.69e+51 $\pm$ 3.28e+50&3.03e+02 $\pm$ 7.81e+00\\ 
GRB 141004973& 1.663& 2.51e+51 $\pm$ 2.80e+50&2.35e+51 $\pm$ 2.53e+50&4.37e+01 $\pm$ 6.64e+00 &&1.259& 1.74e+51 $\pm$ 2.21e+50&1.49e+51 $\pm$ 1.52e+50&2.85e+02 $\pm$ 5.56e+01\\ 
GRB 141005217& 1.197& 1.91e+52 $\pm$ 1.82e+52&1.72e+52 $\pm$ 1.65e+52&2.93e+02 $\pm$ 1.10e+01 &&1.022& 1.53e+52 $\pm$ 7.71e+50&1.46e+52 $\pm$ 9.87e+50&3.55e+02 $\pm$ 1.02e+01\\ 
GRB 141028455& 1.462& 8.52e+53 $\pm$ 2.11e+52&3.01e+53 $\pm$ 1.04e+52&9.76e+02 $\pm$ 1.80e+01 &&1.122& 6.75e+53 $\pm$ 8.70e+51&2.41e+53 $\pm$ 6.00e+51&1.39e+03 $\pm$ 1.53e+01\\ 
GRB 141220252& 1.037& 2.98e+52 $\pm$ 5.96e+51&2.64e+52 $\pm$ 5.30e+51&4.15e+02 $\pm$ 1.01e+01 &&1.039& 2.99e+52 $\pm$ 9.56e+50&2.65e+52 $\pm$ 1.01e+51&4.14e+02 $\pm$ 9.11e+00\\ 
GRB 141221338& 1.460& 2.92e+52 $\pm$ 4.84e+51&1.75e+52 $\pm$ 3.52e+51&2.26e+02 $\pm$ 2.87e+01 &&1.093& 1.94e+52 $\pm$ 1.78e+51&1.32e+52 $\pm$ 1.86e+51&4.46e+02 $\pm$ 3.20e+01\\ 
GRB 141225959& 1.626& 2.66e+52 $\pm$ 3.27e+51&4.32e+51 $\pm$ 7.29e+50&3.42e+02 $\pm$ 1.93e+01 &&1.024& 1.55e+52 $\pm$ 9.67e+50&2.42e+51 $\pm$ 3.81e+50&4.95e+02 $\pm$ 2.70e+01\\ 
GRB 150101641& 1.837& 8.02e+48 $\pm$ 2.36e+48&3.53e+49 $\pm$ 1.04e+49&3.24e+01 $\pm$ 6.74e+00 &&1.157& 4.20e+48 $\pm$ 1.01e+48&2.22e+49 $\pm$ 3.32e+48&1.41e+02 $\pm$ 4.86e+01\\ 
GRB 150301818& 1.322& 3.77e+52 $\pm$ 6.51e+51&1.15e+52 $\pm$ 2.34e+51&4.61e+02 $\pm$ 2.87e+01 &&1.078& 2.95e+52 $\pm$ 1.79e+51&9.32e+51 $\pm$ 1.08e+51&5.68e+02 $\pm$ 2.75e+01\\ 
GRB 150314205& 1.310& 1.03e+54 $\pm$ 2.39e+52&5.10e+53 $\pm$ 1.29e+52&9.57e+02 $\pm$ 7.90e+00 &&1.073& 8.59e+53 $\pm$ 6.37e+51&4.26e+53 $\pm$ 5.72e+51&1.05e+03 $\pm$ 5.76e+00\\ 
GRB 150403913& 1.473& 1.11e+54 $\pm$ 2.54e+52&4.74e+53 $\pm$ 1.35e+52&1.31e+03 $\pm$ 2.11e+01 &&1.202& 9.56e+53 $\pm$ 9.41e+51&4.09e+53 $\pm$ 7.27e+51&1.66e+03 $\pm$ 1.70e+01\\ 
GRB 150514774& 1.321& 1.35e+52 $\pm$ 8.30e+50&6.07e+51 $\pm$ 3.93e+50&1.17e+02 $\pm$ 5.91e+00 &&1.233& 1.11e+52 $\pm$ 3.16e+50&5.04e+51 $\pm$ 1.43e+50&1.42e+02 $\pm$ 4.22e+00\\ 
GRB 150727793& 1.690& 3.18e+51 $\pm$ 5.11e+50&2.31e+50 $\pm$ 4.79e+49&1.95e+02 $\pm$ 1.83e+01 &&1.011& 1.68e+51 $\pm$ 9.20e+49&1.19e+50 $\pm$ 2.70e+49&2.74e+02 $\pm$ 1.71e+01\\ 
GRB 150821406& 1.474& 1.81e+53 $\pm$ 9.04e+51&8.74e+51 $\pm$ 5.74e+50&4.94e+02 $\pm$ 1.71e+01 &&1.148& 1.41e+53 $\pm$ 1.96e+51&7.04e+51 $\pm$ 3.68e+50&5.94e+02 $\pm$ 1.48e+01\\ 
GRB 151027166& 1.608& 5.57e+52 $\pm$ 3.99e+51&9.85e+51 $\pm$ 7.95e+50&3.65e+02 $\pm$ 2.45e+01 &&1.138& 3.79e+52 $\pm$ 1.25e+51&6.96e+51 $\pm$ 4.02e+50&5.10e+02 $\pm$ 2.80e+01\\ 
GRB 151111356& 1.006& 5.97e+52 $\pm$ 2.49e+52&1.53e+52 $\pm$ 1.03e+52&5.34e+02 $\pm$ 5.03e+01 &&1.015& 6.01e+52 $\pm$ 3.87e+51&1.51e+52 $\pm$ 3.61e+51&5.33e+02 $\pm$ 9.65e+00\\ 
GRB 160509374& 1.426& 1.20e+54 $\pm$ 2.37e+52&1.90e+53 $\pm$ 4.16e+51&7.71e+02 $\pm$ 9.88e+00 &&1.146& 9.98e+53 $\pm$ 5.74e+51&1.60e+53 $\pm$ 1.77e+51&9.30e+02 $\pm$ 8.02e+00\\ 
GRB 160623209& 2.521& 3.97e+51 $\pm$ 1.73e+51&5.67e+50 $\pm$ 2.82e+50&1.36e+03 $\pm$ 4.63e+03 &&2.710& 4.32e+51 $\pm$ 3.38e+50&6.25e+50 $\pm$ 1.27e+50&5.90e+04 $\pm$ 1.58e+05\\ 
GRB 160624477& 2.435& 8.83e+50 $\pm$ 2.48e+50&7.37e+51 $\pm$ 2.27e+51&1.73e+03 $\pm$ 5.47e+02 &&1.822& 6.62e+50 $\pm$ 6.19e+49&5.50e+51 $\pm$ 8.99e+50&1.71e+03 $\pm$ 4.88e+02\\ 
GRB 160625945& 1.515& 6.01e+54 $\pm$ 5.06e+52&1.47e+54 $\pm$ 1.46e+52&1.13e+03 $\pm$ 6.45e+00 &&1.214& 5.00e+54 $\pm$ 1.62e+52&1.26e+54 $\pm$ 6.67e+51&1.32e+03 $\pm$ 5.37e+00\\ 
GRB 160629930& 1.163& 5.89e+53 $\pm$ 1.04e+53&1.11e+53 $\pm$ 2.22e+52&1.20e+03 $\pm$ 2.14e+01 &&1.074& 5.41e+53 $\pm$ 1.75e+52&1.01e+53 $\pm$ 9.16e+51&1.26e+03 $\pm$ 1.93e+01\\ 
GRB 160804065& 1.183& 2.65e+52 $\pm$ 2.41e+51&1.41e+51 $\pm$ 2.51e+50&1.24e+02 $\pm$ 4.18e+00 &&1.152& 2.34e+52 $\pm$ 5.39e+50&1.15e+51 $\pm$ 1.81e+50&1.32e+02 $\pm$ 2.81e+00\\ 
GRB 160821937& 1.576& 2.27e+49 $\pm$ 5.16e+48&1.34e+50 $\pm$ 4.26e+49&4.43e+01 $\pm$ 2.75e+01 &&1.208& 1.41e+49 $\pm$ 2.10e+48&9.34e+49 $\pm$ 2.09e+49&1.07e+02 $\pm$ 2.79e+01\\ 
GRB 161014522& 1.031& 1.06e+53 $\pm$ 5.66e+51&5.82e+52 $\pm$ 5.79e+51&6.46e+02 $\pm$ 1.44e+01 &&1.035& 1.07e+53 $\pm$ 5.48e+51&5.78e+52 $\pm$ 5.23e+51&6.49e+02 $\pm$ 1.28e+01\\ 
GRB 161017745& 1.255& 8.25e+52 $\pm$ 1.70e+52&3.41e+52 $\pm$ 8.16e+51&7.19e+02 $\pm$ 4.08e+01 &&1.080& 7.07e+52 $\pm$ 4.82e+51&2.90e+52 $\pm$ 3.91e+51&8.35e+02 $\pm$ 4.06e+01\\ 
GRB 161117066& 1.105& 2.58e+53 $\pm$ 1.09e+52&1.45e+52 $\pm$ 1.64e+51&2.06e+02 $\pm$ 3.05e+00 &&1.096& 2.36e+53 $\pm$ 3.05e+51&1.34e+52 $\pm$ 7.46e+50&2.17e+02 $\pm$ 1.72e+00\\ 
GRB 161129300& 1.623& 2.00e+52 $\pm$ 2.72e+51&2.81e+51 $\pm$ 4.71e+50&2.41e+02 $\pm$ 4.26e+01 &&1.083& 1.20e+52 $\pm$ 5.94e+50&1.86e+51 $\pm$ 1.94e+50&3.52e+02 $\pm$ 2.24e+01\\ 
GRB 161228553& 1.272& 1.43e+50 $\pm$ 3.83e+49&1.53e+49 $\pm$ 4.32e+48&1.86e+02 $\pm$ 1.66e+01 &&1.069& 1.13e+50 $\pm$ 6.53e+48&1.27e+49 $\pm$ 1.39e+48&2.00e+02 $\pm$ 1.91e+01\\ 
GRB 170113420& 1.717& 3.30e+52 $\pm$ 1.98e+52&2.13e+52 $\pm$ 1.32e+52&3.34e+02 $\pm$ 5.88e+01 &&1.402& 2.65e+52 $\pm$ 4.49e+51&1.54e+52 $\pm$ 4.84e+51&3.16e+02 $\pm$ 3.67e+01\\ 
GRB 170214649& 1.295& 4.43e+54 $\pm$ 7.84e+52&3.42e+53 $\pm$ 1.12e+52&1.70e+03 $\pm$ 1.12e+01 &&1.186& 4.14e+54 $\pm$ 2.30e+52&3.36e+53 $\pm$ 8.43e+51&1.81e+03 $\pm$ 9.64e+00\\ 
GRB 170405777& 1.215& 2.87e+54 $\pm$ 5.51e+52&5.40e+53 $\pm$ 1.94e+52&1.20e+03 $\pm$ 9.29e+00 &&1.059& 2.52e+54 $\pm$ 2.94e+52&4.80e+53 $\pm$ 1.76e+52&1.41e+03 $\pm$ 7.77e+00\\ 
GRB 170607971& 1.329& 1.20e+52 $\pm$ 3.47e+50&3.18e+51 $\pm$ 1.78e+50&1.74e+02 $\pm$ 9.03e+00 &&1.165& 1.02e+52 $\pm$ 3.49e+50&2.77e+51 $\pm$ 1.71e+50&2.26e+02 $\pm$ 1.19e+01\\ 
GRB 170705115& 1.293& 1.42e+53 $\pm$ 4.47e+51&7.28e+52 $\pm$ 4.16e+51&3.37e+02 $\pm$ 9.03e+00 &&1.177& 1.25e+53 $\pm$ 4.29e+51&6.59e+52 $\pm$ 4.05e+51&4.37e+02 $\pm$ 1.19e+01\\ 
GRB 170817529& 1.070& 3.32e+46 $\pm$ 2.79e+46&1.93e+47 $\pm$ 1.73e+47&2.17e+02 $\pm$ 5.66e+01 &&1.004& 3.12e+46 $\pm$ 6.72e+45&1.63e+47 $\pm$ 5.64e+46&2.17e+02 $\pm$ 5.42e+01\\ 
GRB 170903534& 1.249& 1.00e+52 $\pm$ 1.14e+52&3.14e+51 $\pm$ 3.60e+51&1.79e+02 $\pm$ 1.34e+01 &&1.185& 9.53e+51 $\pm$ 6.82e+50&2.95e+51 $\pm$ 5.53e+50&1.80e+02 $\pm$ 1.34e+01\\ 
GRB 171010792& 1.418& 2.99e+53 $\pm$ 1.65e+51&1.19e+52 $\pm$ 1.44e+50&1.83e+02 $\pm$ 1.43e+00 &&1.096& 2.21e+53 $\pm$ 5.96e+50&9.50e+51 $\pm$ 1.12e+50&2.59e+02 $\pm$ 1.30e+00\\ 
GRB 171222684& 1.278& 3.54e+52 $\pm$ 5.95e+51&1.08e+52 $\pm$ 3.24e+51&5.98e+01 $\pm$ 4.14e+00 &&2.029& 6.34e+52 $\pm$ 3.03e+51&1.20e+52 $\pm$ 2.00e+51&6.94e+02 $\pm$ 3.44e+00\\ 
GRB 180205184& 1.338& 9.62e+51 $\pm$ 1.54e+51&5.27e+51 $\pm$ 1.14e+51&8.48e+01 $\pm$ 1.70e+01 &&1.874& 1.53e+52 $\pm$ 1.31e+51&9.56e+51 $\pm$ 1.37e+51&3.40e+04 $\pm$ 4.26e+05\\ 
GRB 180314030& 1.054& 1.13e+53 $\pm$ 1.06e+52&1.49e+52 $\pm$ 1.93e+51&2.52e+02 $\pm$ 4.49e+00 &&1.034& 1.04e+53 $\pm$ 2.10e+51&1.24e+52 $\pm$ 7.43e+50&2.60e+02 $\pm$ 2.86e+00\\ 
GRB 180620660& 1.637& 6.69e+52 $\pm$ 3.79e+51&1.33e+52 $\pm$ 1.24e+51&2.72e+02 $\pm$ 1.70e+01 &&1.204& 5.09e+52 $\pm$ 2.05e+51&1.04e+52 $\pm$ 1.03e+51&8.80e+02 $\pm$ 8.12e+01\\ 
\enddata

\end{deluxetable}
\end{longrotatetable}


\bibliography{references}

\end{document}